\documentclass{emulateapj}

\usepackage{natbib,aas_macros}
\usepackage{multirow,color}

\slugcomment{Accepted for Publication in The Astrophysical Journal}

\shorttitle{Statistics of Ly$\alpha$ Emitters at $\lowercase{z}\sim 7$}
\shortauthors{Ouchi et al.}

\begin{document}

\title{Statistics of 207 Ly$\alpha$ Emitters at a Redshift Near 7:\\
Constraints on Reionization and Galaxy Formation Models 
\altaffilmark{\ddag}
}

\author{
Masami Ouchi        \altaffilmark{1,2},
Kazuhiro Shimasaku  \altaffilmark{3,4},
Hisanori Furusawa   \altaffilmark{5},
Tomoki Saito        \altaffilmark{6},\\
Makiko Yoshida      \altaffilmark{3}, 
Masayuki Akiyama    \altaffilmark{7},
Yoshiaki Ono        \altaffilmark{3},
Toru Yamada         \altaffilmark{7},\\
Kazuaki Ota         \altaffilmark{8},
Nobunari Kashikawa  \altaffilmark{5}, 
Masanori Iye        \altaffilmark{5},
Tadayuki Kodama     \altaffilmark{5},\\
Sadanori Okamura    \altaffilmark{3,4},
Chris Simpson       \altaffilmark{9},
Michitoshi Yoshida  \altaffilmark{10}
}

\altaffiltext{1}{Observatories of the Carnegie Institution of Washington,
        813 Santa Barbara St., Pasadena, CA 91101}
\altaffiltext{2}{Carnegie Fellow; ouchi \_at\_ obs.carnegiescience.edu}
\altaffiltext{3}{Department of Astronomy, School of Science, University of Tokyo, Tokyo 113-0033, Japan}
\altaffiltext{4}{Research center for the Early Universe, School of Science,
        University of Tokyo, Tokyo 113-0033, Japan}
\altaffiltext{5}{National Astronomical Observatory, Mitaka, Tokyo 181-8588, Japan}
\altaffiltext{6}{Institute for the Physics and Mathematics of the Universe (IPMU), University of Tokyo, Kashiwa 277-8568, Japan}
\altaffiltext{7}{Astronomical Institute, Graduate School of Science, Tohoku University, Aramaki, Aoba, Sendai 980-8578, Japan}
\altaffiltext{8}{Institute for Cosmic Ray Research, University of Tokyo, Kashiwa 277-8582, Japan}
\altaffiltext{9}{Astrophysics Research Institute, Liverpool John Moores University, Twelve Quays House, Egerton Wharf, Birkenhead CH41 1LD, UK}
\altaffiltext{10}{Okayama Astrophysical Observatory, National Astronomical Observatory, Kamogata, Okayama 719-0232, Japan}

\altaffiltext{\ddag}{Based on data obtained with 
the Subaru Telescope and the W.M. Keck Observatory.
The Subaru Telescope is 
operated by the National Astronomical Observatory of Japan.
The W.M. Keck Observatory is operated as a scientific partnership
among the California Institute of Technology,
the University of California and the National
Aeronautics and Space Administration.
}

\begin{abstract}
We present Ly$\alpha$ luminosity function (LF), 
clustering measurements, and Ly$\alpha$ line profiles
based on the largest sample, to date, of 207
Ly$\alpha$ emitters (LAEs) at $z=6.6$ 
on the 1-deg$^2$ sky of Subaru/XMM-Newton Deep Survey (SXDS) field.
Our $z=6.6$ Ly$\alpha$ LF including cosmic variance estimates
yields the best-fit Schechter parameters of
$\phi^*=8.5_{-2.2}^{+3.0}\times10^{-4}$Mpc$^{-3}$ and
$L_{\rm Ly\alpha}^*=4.4_{-0.6}^{+0.6}\times10^{42}$erg s$^{-1}$
with a fixed $\alpha=-1.5$,
and indicates a decrease from $z=5.7$
at the $\gtrsim90$\% confidence level.
However, this decrease is not large, only $\simeq30$\% 
in Ly$\alpha$ luminosity, which is too small
to be identified in the previous studies.
A clustering signal of $z=6.6$ LAEs is detected 
for the first time. We obtain the correlation length 
of 
$r_0=2-5$ h$^{-1}_{100}$ Mpc and
bias of $b=3-6$, 
and find no significant boost
of clustering amplitude by reionization at $z=6.6$.
The average hosting dark halo mass inferred from clustering 
is 
$10^{10}-10^{11}M_\odot$, 
and duty cycle of
LAE population 
is roughly 
$\sim1$\%
albeit with large uncertainties.
The average of our 
high-quality Keck/DEIMOS spectra 
shows an FWHM velocity width of 
$251\pm 16$km s$^{-1}$.
We find no large evolution of Ly$\alpha$ line profile
from $z=5.7$ to $6.6$, and no anti-correlation 
between Ly$\alpha$ luminosity and line width
at $z=6.6$.
The combination of various reionization models
and our observational results
about the LF, clustering, and line profile
indicates that 
there would exist a small decrease
of IGM's Ly$\alpha$ transmission
owing to reionization, 
but that the hydrogen IGM is not highly neutral
at $z=6.6$.
Our neutral-hydrogen fraction constraint
implies that the major reionization process took place
at $z\gtrsim7$.
\end{abstract}


\section{Introduction}
\label{sec:introduction}

Understanding physical process of cosmic reionization is 
one of the major goals in astronomy today.
Although the increase of Gunn-Peterson (GP) 
optical depths may be contiguous from
low $z$ to $z>6$ \citep{becker2007}, 
the evolution of GP optical depths clearly show 
the steep rise at $z\sim 6$ towards high-$z$
\citep{fan2006}. On the other hand,
\citet{dunkley2009} find that
the polarization data of WMAP place the constraints that
instantaneous reionization at the late epoch below $z=8.2$ ($6.7$) is 
rejected at the $2\sigma$ ($3\sigma$) level,
and claim that the reionization process would be 
extended at $z\sim 6-11$ (see also \citealt{larson2010} for the
latest WMAP7 results). However, physical models
would not easily reproduce such a long extended reionization
due to the rapid recombination of hydrogen 
(e.g. \citealt{fukugita1994,cen2003}).
Observational measurements on neutral hydrogen fraction
of inter-galactic medium (IGM) at $z\sim 6-11$ 
are the missing pieces in this cosmological puzzle.

\begin{deluxetable*}{ccccccl}
\tablecolumns{7}
\tabletypesize{\scriptsize}
\tablecaption{Imaging Observations and Data 
\label{tab:obs}}
\tablewidth{0pt}
\setlength{\tabcolsep}{0.03in}
\tablehead{
\colhead{Band} & 
\colhead{Field Name(s)} & 
\colhead{Exposure Time} &
\colhead{PSF size\tablenotemark{a}} &
\colhead{Area} &
\colhead{$m_{\rm lim}$\tablenotemark{b}}  &
\colhead{Date of Observations} \\
\colhead{} & 
\colhead{} & 
\colhead{(sec)} &
\colhead{(arcsec)} &
\colhead{(arcmin$^2$)} &
\colhead{(3$\sigma$ AB mag)}  &
\colhead{}
}
\startdata
$NB921$ & SXDS-C & 30000 & 0.73 (0.81) & 590 & 26.2 & 2005 Oct 29, Nov 1, 2007 Oct 11-12\\
$NB921$ & SXDS-N & 37800 & 0.75 (0.85) & 734 & 26.4 & 2005 Oct 30-31, Nov 1, 2006 Nov18, 2007 Oct11-12\\
$NB921$ & SXDS-S & 37138 & 0.79 (0.83) & 783 & 26.2 & 2005 Aug29, Oct29-30, Nov1,2006 Nov18,2007 Oct12\\
$NB921$ & SXDS-E & 29400 & 0.65 (0.83) & 610 & 26.2 & 2005 Oct 31, Nov 1, 2006 Nov 18, 2007 Oct 11-12\\
$NB921$ & SXDS-W & 28101 & 0.75 (0.83) & 521 & 26.2 & 2006 Nov 18, 2007 Oct 11-12\phn \\
\cutinhead{Archival broad-band data\tablenotemark{c}.}
$B$ & SXDS-C,N,S,E,W & $19800-20700$ & $0.78-0.84$ & $915-979$ & $28.1-28.4$ & \nodata\phn \\
$V$ & SXDS-C,N,S,E,W & $17460-19260$ & $0.72-0.82$ & $915-979$ & $27.7-27.8$ & \nodata\phn \\
$R$ & SXDS-C,N,S,E,W & $13920-14880$ & $0.74-0.82$ & $915-979$ & $27.5-27.7$ & \nodata\phn \\
$i'$ & SXDS-C,N,S,E,W & $18540-38820$ & $0.68-0.82$ & $915-979$ & $27.5-27.7$ & \nodata\phn \\
$z'$ & SXDS-C,N,S,E,W & $11040-18660$ & $0.70-0.76$ & $915-979$ & $26.4-26.6$ & \nodata\phn
\enddata
\tablenotetext{a}{
FWHM of PSFs in the reduced image. Values in parenthesis
indicate the FWHMs of PSF that are matched with the broad-band images 
in each field.
}
\tablenotetext{b}{
\vspace{0mm}
Limiting magnitude defined by 
a $3\sigma$ sky noise 
in a $2''$-diameter circular aperture.
}
\tablenotetext{c}{
The archival broad-band data of SXDS presented in \citet{furusawa2008}.
We show the properties of the 5-subfield images on a single line.  
Note that \\ 
the exposure time is {\it not} a total of the 5 subfields, 
but 1 subfield, i.e. integration per pixel.
More details are presented in Table 2 of \citet{furusawa2008}.
}
\end{deluxetable*}

Studies of galaxies near the epoch of reionization (EoR) 
at $z\gtrsim 6$ are essential not only for understanding 
cosmic reionization process but also galaxy formation history.
The combination of Subaru and VLT wide-field cameras
and the newly-installed HST/WFC3
has identified a definitive decrease of 
UV-continuum luminosity function (LF)
from $z=6$ to $7-8$ (\citealt{ouchi2009b,oesch2010,bouwens2010a,
mclure2010,castellano2010,wilkins2010a,bunker2009,yan2009,wilkins2010b};
see also \citealt{hickey2009}),
which are also reproduced by recent hydrodynamic simulations
\citep{finlator2010}.
Accordingly the star-formation rates drop roughly 
by an order of magnitude from its peak of $z=2-3$ to $z=7$,
and this decrease implies that observations are
touching initial formation epoch of galaxies.
Because galaxies are thought to be sources of reionization,
the decrease of UV LF would indicate that
galaxies produce less
UV ionizing photons towards high redshifts.
The production rate of 
UV ionizing photons is close to balance with 
the recombination rate of hydrogen IGM at $z\sim 7-8$,
and the reionizing epoch may be near these redshifts.
The other interpretation of the decrease of UV LF
is that a moderately high ionizing photon escape fraction,
$f_{\rm esc}\gtrsim 0.2$, is required to keep the
universe ionized at $z\sim 7$
(e.g. \citealt{ouchi2009b,bunker2009}).
The faintest HST/WFC3 sources at $z\simeq 7-8$ show 
a very blue UV continuum slope possibly consistent 
with extremely young, metal-poor stellar populations,
which is also suggestive of large escape fraction
of $f_{\rm esc}\gtrsim 0.3$
(\citealt{bouwens2010b}; see also \citealt{finkelstein2009}),
although there are claims that their UV continuum slope measurements 
include potentially large statistical and systematic 
uncertainties \citep{schaerer2010}.

On the other hand, neutral hydrogen of IGM absorbs
Ly$\alpha$ emission line from galaxies via Ly$\alpha$ damping wing,
and dim Ly$\alpha$ luminosity, which would be identified in 
the evolution of Ly$\alpha$ emitters 
(e.g. \citealt{malhotra2004,kashikawa2006,iye2006,ota2008}).
\citet{malhotra2004} and \citet{hu2006} find that 
no significant change of Ly$\alpha$ luminosity function (LF) 
from $z=5.7$ to $6.6$,
while the study of \citet{kashikawa2006} claims that
Ly$\alpha$ LF evolves from $z=5.7$ to $6.6$. 
Similarly, Ly$\alpha$ LF evolution 
from $z=5.7$ toward higher redshifts, $z\simeq 7-7.7$,
is also under debate. Ly$\alpha$ LFs
at these redshifts are estimated with 
a reliable but only 1-2 LAEs 
at $z=7.0$ \citep{iye2006,ota2008}
and with relatively less-reliable $z\simeq 7.7$ candidates with no
spectroscopic confirmation \citep{hibon2010,tilvi2010}.
Moreover, beyond this epoch, only weak upper limits
are placed on Ly$\alpha$ LF at $z\simeq 8.8$ 
(\citealt{willis2005,cuby2007,willis2008,sobral2009}; cf. \citealt{stark2007}).
There are two conclusions from these $z=6.6-7.7$ studies, no
evolution and a decrease of Lya LF, which do not agree with
each other.
These different conclusions of Ly$\alpha$ LF evolution
may be raised by contamination, small statistics and 
systematic errors such as cosmic variance. 
Moreover, a careful argument
is needed for the interpretation of Ly$\alpha$ LF evolution,
because properties of star-formation galaxies
are changing, which are already found
in the decrease of UV LF of dropout galaxies
as discussed above.
It is also important to 
constrain reionization and galaxy formation models
with other observational quantities.
The stacked spectra of 
$z=6.6$ LAEs show no clear signal of 
Ly$\alpha$ damping wing absorption \citep{kashikawa2006,hu2006}.
Recent theoretical studies predict that 
clustering measurements of LAEs can be an independent probe
of reionization. 
Since Ly$\alpha$ lines of galaxies 
residing in ionized bubbles 
selectively escape from a partially neutral universe,
clustering amplitude of observed LAEs
would be boosted at the EoR
(\citealt{furlanetto2006,mcquinn2007,lidz2009}; cf. \citealt{iliev2008}).
Although the importance of $z>6$ LAE clustering
is claimed from the theoretical studies, no observational study has
provided a reliable measurement.
\citet{kashikawa2006} measured angular correlation function (ACF)
of 58 LAEs at $z=6.6$ which is so far the largest sample of $z=6.6$ LAEs, 
but no significant signal is detected, probably
due to the small statistics \citep{mcquinn2007}.

To address these issues in statistics of $z=6.6$ LAEs,
we are conducting an extensive survey that is the largest ever
performed in terms of area and number of objects
(cf. \citealt{malhotra2004,hu2005,kashikawa2006}).
Our survey field is a large 1 deg$^2$ area of 
Subaru/XMM-Newton Deep Survey (SXDS) field
well-separated from the Subaru Deep Field (SDF) of 
\citet{kashikawa2006}. The combination of
SXDS and SDF data will allow us to provide
a large statistical sample with cosmic variance error estimates.
The spectroscopic component of this survey is underway 
with Keck/DEIMOS and Magellan/IMACS (Ouchi et al. in prep.).
The aim of this study is to supply the reliable statistical 
measurements of $z=6.6$ LAEs based on our large sample that are useful 
for constraining the existing and forthcoming cosmological models
of cosmic reionization and galaxy formation.
Our $z=6.6$ LAE results from the large statistics will 
also be good baselines for on-going and future studies of LAEs at $z>7$
performed with the existing near-infrared imagers of VLT \citep{willis2008},
Subaru \citep{tokoku2008}, and VISTA \citep{nilsson2007} 
and the future facilities of JWST, E-ELT, GMT, and TMT.
We present our survey and sample in \S \ref{sec:observations},
luminosity functions in \S \ref{sec:luminosity_function},
clustering properties in \S \ref{sec:spatial},
Ly$\alpha$ line profiles in \S \ref{sec:line_profile},
and discuss reionization and galaxy formation
by comparisons with theoretical models in \S \ref{sec:discussion}.
Throughout this paper, magnitudes are in the AB system.
We adopt a cosmology parameter set of
$(h,\Omega_m,\Omega_\Lambda,n_s,\sigma_8)=(0.7,0.3,0.7,1.0,0.8)$
consistent with the WMAP results
\citep{komatsu2009,larson2010}.

\begin{figure}
%\epsscale{0.85}
\epsscale{1.15}
%\plotone{f1.eps}
\plotone{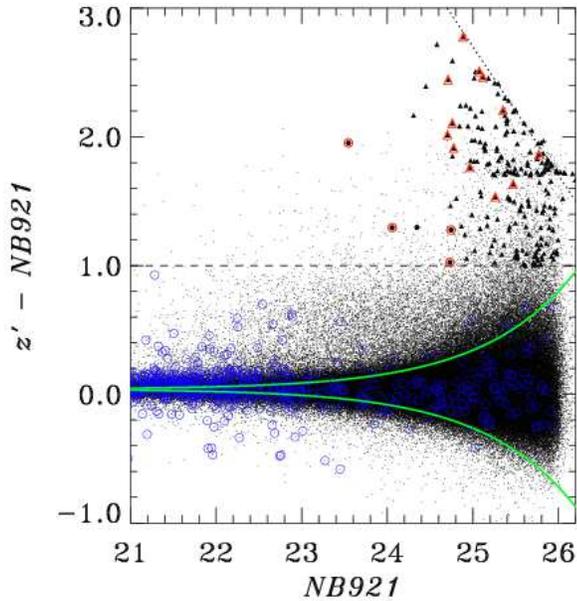}
\caption{
Color magnitude diagram of 
narrow-band excess color ($z'-NB921$)
vs. narrow-band magnitude ($NB921$). 
Black dots present colors of all the
detected objects. Black filled circles 
and triangles denote our $z=6.6$ LAEs.
Specifically, the triangles indicate 
LAEs with $z'$-band magnitudes fainter 
than that of the $3 \sigma$ level, 
and show their $1\sigma$ lower limits of 
their $z'-NB921$ colors. 
Red and blue open symbols
mark spectroscopically-identified objects
in the redshift range of LAEs and interlopers, respectively.
We define the redshift ranges of LAEs as
$6.45-6.65$
(see Figure \ref{fig:nz_zlae_NB921} for the redshift range).
The green lines indicate $2\sigma$ errors of
the color of $z'-NB921$ 
for a source with a color of $z'-NB921=0.04$, 
which corresponds to the median color of all objects.
Dashed and dotted lines represent 
our color cut for narrow-band excess
and the $1\sigma$ limit of $z'$ data,
respectively.
Note that $NB921$ magnitudes
are total magnitudes, while the colors of 
$z'-NB921$ are defined with a $2''$-diameter 
aperture.
\label{fig:cm}}
\end{figure}

\begin{figure}
%\epsscale{0.7}
\epsscale{1.15}
%\plotone{f2.eps}
\plotone{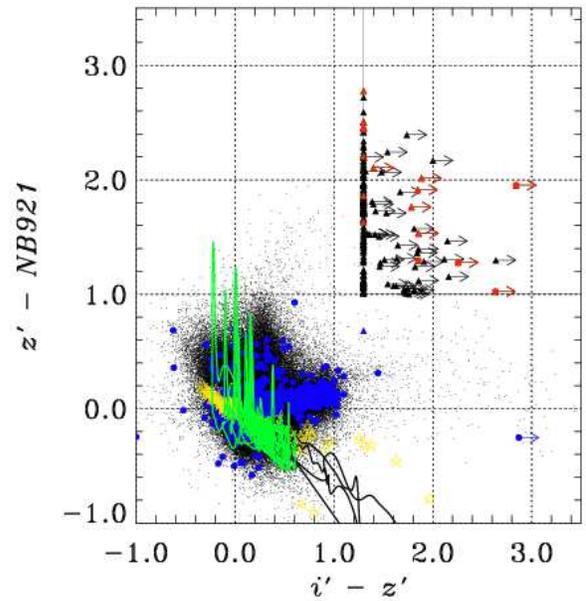}
\caption{
Two color diagram of 
narrow-band excess color ($z'-NB921$) vs.
continuum color ($i'-z'$).
Black dots present colors of all the detected objects.
Our $z=6.6$ LAEs are shown by black and red 
filled circles+triangles. The triangles have a $z'$-band 
magnitude fainter than $3 \sigma$ level,
and present the $1\sigma$ lower limit of $z'-NB921$. 
Red and blue circles+triangles
indicate spectroscopically identified objects
in and out of the LAE redshift range, respectively.
We take the same LAE redshift range as
that of Figure \ref{fig:cm}.
For display purpose, 
objects with no color measurements
(no $z'$-band detection at the $3\sigma$ level) 
are placed at a color of $i'-z'=1.3$.
The vertical and horizontal solid lines show 
our color criteria (eq. \ref{eq:selectionNB921})
for sources with a detection in $z'$ band.
Colors are defined with a $2''$-diameter aperture.
Curves present tracks of model interloper galaxies 
at different redshifts.
Green lines show 6 templates of
starburst galaxies \citep{kinney1996} up to $z=2$,
which are 6 classes of
starburst galaxies with different dust extinction
($E(B-V)=0.0-0.7$). The narrow-band excess
peaks in the green lines
correspond to the emission lines of
H$\alpha$($z=0.4$),
[{\sc Oiii}]($z=0.8$), 
H$\beta$($z=0.9$),
or [{\sc Oii}]($z=1.5$).
Black lines represent colors of typical
elliptical, spiral, and irregular galaxies \citep{coleman1980}
which are redshifted from $z=0$ to $z=2$. Yellow star marks are
175 Galactic stars given by \citet{gunn1983}.
\label{fig:cc}}
\end{figure}

\section{Observations and Data Reduction}
\label{sec:observations}

\subsection{Imaging Observations}
\label{sec:imaging_observations}

We carried out extensive deep narrow-band imaging
with Subaru/Suprime-Cam \citep{miyazaki2002}
for SXDS 
in 2005-2007.
We summarize details of our observations 
as well as image qualities
in Table \ref{tab:obs}.
We used the narrow-band filter, $NB921$,
with a central wavelength of
$\lambda_c=9196$\AA\ and a FWHM of 132\AA.
The response curves of the $NB921$-band and broad-band filters 
are presented in Figure 1 of \citet{ouchi2008}.
Five pointings of Suprime-Cam covered 5 subfields of 
SXDS, SXDS-C, N, S, E, and W \citep{furusawa2008}, 
by 8.3, 10.5, 10.3, 8.2, and 7.8 hour 
on-source integration, respectively.
In addition to these narrow-band images,
we use archival data of deep broad-band ($B$, $V$, $R$,
$i'$ and $z'$) images of the SXDS project \citep{furusawa2008}
as summarized at the second half of Table \ref{tab:obs}.
The narrow-band data are reduced with 
the Suprime-Cam Deep field REDuction package (SDFRED;
\citealt{yagi2002,ouchi2004a}). With the standard
parameter sets of SDFRED, we perform bias subtraction,
flat-fielding, distortion+atmospheric-dispersion 
corrections, sky subtraction, image alignments, and stacking. 
Before stacking, we mask out bad data areas
such as dead pixels and satellite trails. Cosmic rays
are removed in the process of stacking with the rejected-mean 
algorithm. 
The final $NB921$ images have the seeing size of $0''.7-0''.8$ in FWHM,
and reach the $3\sigma$ limiting magnitudes of 
$26.2-26.4$ mag in a $2''.0$-diameter aperture.

We use neither contaminated areas with halos of bright 
stars and CCD blooming nor low signal-to-noise (S/N) ratio 
regions located around the edge of 
the FoV, which are caused by dithering.
After we remove these low-quality regions from our catalog 
(\S \ref{sec:photometricsample}),
the effective total areas are 3238 arcmin$^2$.
This effective area corresponds to the survey volume of 
$8.0\times 10^{5}$ Mpc$^{3}$ at $z=6.565\pm 0.054$,
if we assume a simple top-hat selection function of LAEs
whose redshift distribution is defined by the FWHM of 
our narrow-band filter.

During the observations, we took images of spectrophotometric 
standard star of GD71 with $NB921$-band filter 
\citep{bohlin1995}. The standard star was observed
under photometric condition in 2006 November 18
and 2007 October 11-12.
We calculate photometric zero-points 
from the standard star data.
We check these photometric zero points 
based on colors of stellar objects in our field
and 175 Galactic stars calculated from
spectra given in \citet{gunn1983}. 
We find that colors of stellar objects in our data
are consistent with those of \citeauthor{gunn1983}'s 
(\citeyear{gunn1983}) stars
within $\simeq 0.05$ magnitude.

Our narrow-band images are aligned 
with the deep optical $BVRiz$ images from the SXDS survey
(Table \ref{tab:obs}; \citealt{furusawa2008})
based on hundreds of stellar objects 
commonly detected in both narrow-
and broad-band images. 
The astrometry of our objects is the same as 
those of SXDS version 1.0 catalog \citep{furusawa2008}.
The errors in the {\it relative} positions of objects are $\sim 0''.04$
in r.m.s. The r.m.s accuracy of the {\it absolute} positions is 
estimated in \citet{furusawa2008} to be $\sim 0''.2$.
After the image registration, we homogenize the PSF sizes of broad and
narrow-band images, referring to these stellar objects. The PSF sizes
of our narrow-band images match to those of broad-band images with
an accuracy of $\Delta{\rm FWHM} \simeq 0''.01$.

\subsection{Photometric Sample of $z=6.6$ LAEs }
\label{sec:photometricsample}

We identify 286,510 objects in the $NB921$ images 
down to $NB921=26.0$ with SExtractor \citep{bertin1996}.
We measure both MAG\_AUTO of SExtractor and
$2''.0$-diameter aperture magnitudes.
We adopt MAG\_AUTO as total magnitudes, while 
we use a $2''.0$-diameter aperture magnitude 
to measure colors of objects in order to obtain
colors of faint objects with a good signal-to-noise ratio.
We correct the magnitudes of objects
for Galactic extinction of $E(B-V)=0.020$ \citep{schlegel1998}.

We plot a color-magnitude diagram in Figure \ref{fig:cm}
for our objects.
Figure \ref{fig:cm} shows
narrow-band excess color, $z'-NB921$, and 
narrow-band magnitudes, $NB921$.
Figure \ref{fig:cc} presents a two-color diagram 
based on the $NB921$-detection catalog, together with 
colors of model galaxies and Galactic stars.
We plot colors of 3,249
spectroscopically-identified objects
which include our 16 $z=6.6$ LAEs 
(see \S \ref{sec:spectroscopic_confirmation}) and 
3,233 foreground/background interlopers \citep{ouchi2008}.
Spectroscopically-identified LAEs
are located in the upper-right part of the
two-color diagram, having 
a narrow-band excess of Ly$\alpha$ emission ($z'-NB921$) 
and red continuum colors of GP trough ($i'-z'$).

Based on the color diagram,
we select candidate LAEs with the narrow-band excess, 
no detection of blue continuum flux,
and the existence of GP trough,
by the color criteria,
{\footnotesize
\begin{eqnarray}
\label{eq:selectionNB921}
z'-NB921>1.0\ {\rm and}\ 
B>B_{2\sigma}\ {\rm and}\
V>V_{2\sigma}\ {\rm and}\ \nonumber\\
\left[(z'<z'_{3\sigma}\ {\rm and}\ i'-z'>1.3)\ {\rm or}\ (z'\ge z'_{3\sigma})\right],
\end{eqnarray}
}
which are similar to those in the study of SDF 
\citep{taniguchi2005,kashikawa2006}.
$B_{2\sigma}$ and $V_{2\sigma}$ are defined as $2\sigma$ limiting 
magnitudes of $B$ and $V$ bands, respectively ($B_{2\sigma}=28.7$ and 
$V_{2\sigma}=28.2$), which ensures 
no detection of continuum bluer than 
Lyman break ($\simeq 6900$\AA) for objects at $z=6.6$.
$z'_{3\sigma}$ is the $3\sigma$ detection 
limit ($z'_{3\sigma}=26.5$). 

We obtain a photometric sample of 
207 LAEs at $z\simeq 6.56\pm0.05$ down to $NB921=26.0$ 
in a comoving survey volume of $8\times 10^5$ Mpc$^3$ \citep{ouchi2009a}.
Our selection criteria correspond to LAEs
with the rest-frame equivalent width, $EW_0$, of $\gtrsim 36$\AA,
if a flat continuum spectrum ($f_\nu=$const)
is assumed. If the realistic spectrum of LAEs with a Gunn-Peterson trough
is assumed, the limit of rest-frame equivalent width 
is $EW_0 \gtrsim 14$ \AA\ and a limiting line flux of
$f \gtrsim 5 \times 10^{-18}$ erg s$^{-1}$ cm$^{-2}$
corresponding to $L \gtrsim 2.5 \times 10^{42}$ erg s$^{-1}$
at $z\simeq 6.56$.

\begin{deluxetable*}{ccccccccccc}
\tablecolumns{11}
\tabletypesize{\scriptsize}
\tablecaption{Ly$\alpha$ Emitters with spectroscopic redshifts
\label{tab:laes_with_redshifts}}
\setlength{\tabcolsep}{0.0in}
\tablewidth{0pt}
\tablehead{
\colhead{Object Name} & 
\colhead{$\alpha$(J2000)} &
\colhead{$\delta$(J2000)} & 
\colhead{$z$} &
\colhead{$m_{\rm NB}$} &
\colhead{$m_{\rm z'}$} &
\colhead{$L({\rm Ly \alpha})$} &
\colhead{$\Delta V_{\rm FWHM}$} &
\colhead{$EW_0$} &
\colhead{$M_{\rm UV}$} &
\colhead{SFR} \\
\colhead{} & 
\colhead{} &
\colhead{} & 
\colhead{} &
\colhead{(mag)} &
\colhead{(mag)} &
\colhead{($10^{43}$erg s$^{-1}$)} &
\colhead{(km s$^{-1}$)} &
\colhead{(\AA)} &
\colhead{(mag)} &
\colhead{($M_\odot$yr$^{-1}$)} \\
\colhead{(1)} & 
\colhead{(2)} &
\colhead{(3)} & 
\colhead{(4)} &
\colhead{(5)} &
\colhead{(6)} &
\colhead{(7)} &
\colhead{(8)} &
\colhead{(9)} &
\colhead{(10)} &
\colhead{(11)}
}
\startdata
NB921-C-106098 & 2:17:03.4908 & -4:56:19.158 & 6.589 & $24.7$ & $26.8$ & $1.2\pm 0.1$ & $269 \pm 37$ & $105.9^{+\infty}_{-65.4}$ & $\simeq -20.2$ & $\simeq 6.6$ \\
NB921-C-22057 & 2:18:20.6670 & -5:11:09.664 & 6.575 & $24.9$ & $>27.7$ & $1.0\pm 0.1$ & $304 \pm 58$ & $\gtrsim 100$ & $\lesssim -19.9$ & $\lesssim 5$ \\ 
NB921-C-34609 & 2:18:19.3797 & -5:09:00.550 & 6.563 & $24.8$ & $27.3$ & $0.9\pm 0.2$ & $247 \pm 98$ & $76.6^{+\infty}_{-57.1}$ & $\lesssim -19.9$ & $\lesssim 5$ \\ 
NB921-C-36215$^b$ & 2:17:57.5630 & -5:08:44.446 & 6.595 & $23.5$ & $25.9$ & $3.9\pm 0.2$ & $267 \pm 32$ & $116.5^{+179.5}_{-47.3}$ & $-21.1 \pm 0.7$ & $15.0 \pm 9.7$ \\
NB921-C-43803 & 2:18:26.9734 & -5:07:26.828 & 6.554 & $25.0$ & $26.9$ & $0.7\pm 0.2$ & $199 \pm 190$ & $32.2^{+828.2}_{-21.3}$ & $\simeq -20.7$ & $\simeq 10.3$ \\
NB921-C-50823$^{a}$ & 2:17:02.5772 & -5:06:04.480 & 6.545 & $24.8$ & $26.9$ & $0.9\pm 0.3$ & $257 \pm 69$ & $52.9^{+\infty}_{-35.8}$ & $\simeq -20.5$ & $\simeq 9.0$ \\
NB921-C-65683 & 2:17:01.4777 & -5:03:09.291 & 6.493 & $24.7$ & $26.1$ & $3.5\pm 12.7$ & $329 \pm 70$ & $\gtrsim 100$ & $\lesssim -19.9$ & $\lesssim 5$ \\ 
NB921-N-71598 & 2:18:44.6534 & -4:36:36.501 & 6.621 & $24.7$ & $27.4$ & $2.6\pm 0.3$ & $367 \pm 46$ & $\gtrsim 100$ & $\lesssim -19.9$ & $\lesssim 5$ \\ 
NB921-N-77765 & 2:18:23.5317 & -4:35:24.193 & 6.519 & $24.7$ & $26.4$ & $0.4\pm 0.8$ & $266 \pm 80$ & $45.2^{+\infty}_{-39.4}$ & $-21.6 \pm 0.7$ & $22.9 \pm 14.8$ \\
NB921-N-79144 & 2:18:27.0289 & -4:35:08.216 & 6.511 & $24.1$ & $25.7$ & $0.9\pm 1.2$ & $317 \pm 49$ & $24.8^{+\infty}_{-19.5}$ & $-22.2 \pm 0.4$ & $42.7 \pm 15.7$ \\
NB921-W-25755 & 2:16:58.2805 & -4:55:56.703 & 6.573 & $25.1$ & $>27.7$ & $0.7\pm 0.2$ & $134 \pm 42$ & $145.1^{+\infty}_{-115.8}$ & $\lesssim -19.9$ & $\lesssim 5$ \\ 
NB921-W-30717 & 2:16:54.5645 & -4:55:56.788 & 6.617 & $25.3$ & $26.8$ & $1.3\pm 0.2$ & $246 \pm 44$ & $153.0^{+\infty}_{-101.4}$ & $\simeq -20.1$ & $\simeq 5.9$ \\
NB921-W-30999$^{a}$ & 2:16:54.3945 & -5:00:04.162 & 6.512 & $25.1$ & $>27.7$ & $1.3\pm 0.8$ & $185 \pm 25$ & $\gtrsim 100$ & $\lesssim -19.9$ & $\lesssim 5$ \\ 
NB921-W-31790$^{a}$ & 2:16:53.9169 & -5:06:01.379 & 6.590 & $25.8$ & $>27.7$ & $0.4\pm 0.1$ & $142 \pm 154$ & $45.9^{+\infty}_{-32.9}$ & $\lesssim -19.9$ & $\lesssim 5$ \\ 
NB921-W-35578 & 2:16:50.9037 & -5:10:16.221 & 6.535 & $25.4$ & $>27.7$ & $0.6\pm 0.3$ & $279 \pm 210$ & $104.4^{+\infty}_{-89.0}$ & $\lesssim -19.9$ & $\lesssim 5$ \\ 
NB921-W-36217 & 2:16:50.4202 & -5:05:45.130 & 6.586 & $25.5$ & $27.3$ & $0.5\pm 0.2$ & $128 \pm 145$ & $34.7^{+1045.6}_{-23.3}$ & $\lesssim -19.9$ & $\lesssim 5$ \\ 
\enddata
\tablecomments{
(1): Object ID. (2)-(3): RA and Dec. (4): Redshift measured with a Ly$\alpha$ line. 
(5)-(6): Magnitudes in NB921 ($m_{\rm NB}$) and $z'$ ($m_{\rm z'}$).
$NB921$ is a total magnitude, while $z'$ is a $2''$-aperture magnitude.
The lower limit of $m_{\rm z'}$ corresponds to a $1\sigma$ limit.
(7): Ly$\alpha$ luminosity in $10^{43}$ erg s$^{-1}$.
(8): FWHM velocity width in km s$^{-1}$.
(9): Rest-frame apparent equivalent width of Ly$\alpha$ emission line. 
(10): UV total magnitude.
(11): Star-formation rate estimated from the UV magnitude
based on \citet{madau1998}. 
The quantities of (7)-(10) are estimated 
from $NB921$ and $z'$ photometry.
}
\tablenotetext{a}{
\vspace{0mm}
Ly$\alpha$ emission lines are doubly confirmed 
with our two independent spectra
taken by our redundant observations 
(see \S \ref{sec:spectroscopic_confirmation}).
}
\tablenotetext{b}{
This object is a giant LAE, {\it Himiko} \citep{ouchi2009a}.
The $EW_0$, $M_{UV}$, and SFR values listed here are
slightly different from those shown in \citet{ouchi2009a}.
In this table, we present values based on measurements of 
$2''$-aperture magnitudes to maximize a signal-to-noise ratio
for the other sources which are mostly fainter than {\it Himiko}.
For consistency, the values of
{\it Himiko} is based on $2''$-aperture magnitudes in this table.
On the other hand, \citet{ouchi2009a} estimate 
these values from total magnitudes for the bright object of {\it Himiko}
to minimize systematic errors. 
}
\end{deluxetable*}

\begin{figure}
%\epsscale{0.75}
\epsscale{1.35}
\plotone{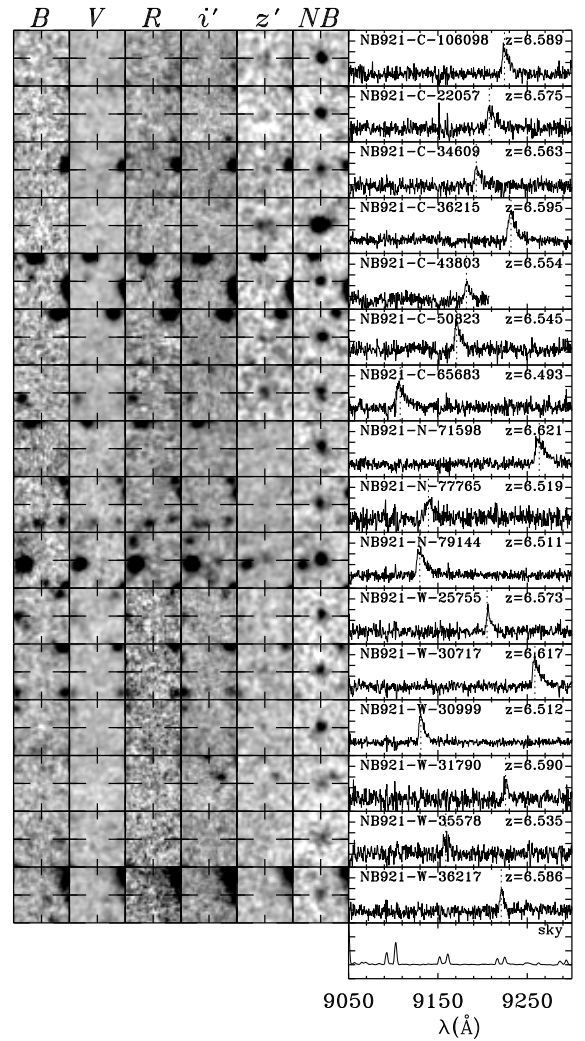}
\caption{
Spectra and snapshots of our $z=6.6$ LAEs 
confirmed with Keck/DEIMOS.
Each object has a spectrum in the right panel and
snapshots of $B$, $V$, $R$, $i'$, $z'$,
and $NB921$ images in the left panels. Each snapshot
is presented in a $6''\times 6''$ box. 
The object name and redshift
are presented in the left and right corners of each spectrum panel,
respectively.
The right bottom panel shows a typical DEIMOS spectrum of 
the sky background that is obtained in the process of 
sky subtraction.
\label{fig:image_spec_all_nb921}}
\end{figure}

\subsection{Spectroscopic Observations}
\label{sec:spectroscopic_confirmation}

We conducted deep spectroscopic follow-up for our $z=6.6$ LAEs
with Keck/DEIMOS \citep{faber2003} on 2007 November 6,
2008 August 1, and 2008 October 3. 
We took spectra of 3.0 hour on-source integration
for 3 masks and 2.3-hour integration
for 1 mask with a $1''.0$ slit and 
the 830G grating,
which covered $\simeq 6000-10000$\AA\ and
gave a medium high spectral resolution of 
$R\simeq 3600$ at $9200$\AA.
We observed 30 out of 207 LAEs
including faint LAEs whose expected flux
is below our observational limit,
and obtained 19 spectra with significant signals.
All of the 19 spectra
show a single line at around 
$9200$\AA\ with no detectable continuum. 
Because we allowed redundant observations 
for 6 LAEs that can be included in two different masks.
Three out of 19 detected objects
were taken for the same targets. 
Thus, totals of the observed and identified LAEs 
are 24 ($=30-6$) and 16 ($=19-3$), respectively.
We hereafter refer to the 24 galaxies 
as the spectroscopic sample.
The 3 duplicate spectra with the signals
present an emission line at the same wavelength
as the original spectra,
and confirm that the emission lines are real signals.
We use these 3 duplicate spectra
for the stacking analysis in \S \ref{sec:line_profile}.

\begin{figure}
%\epsscale{1.0}
\epsscale{1.15}
\plotone{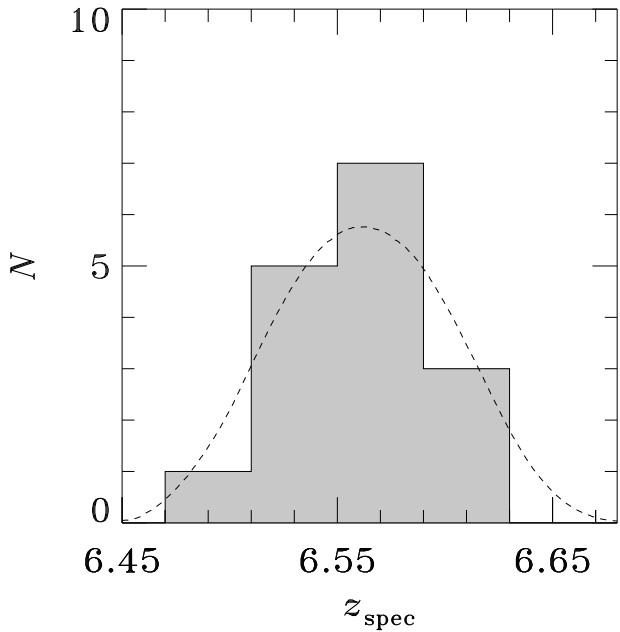}
\caption{
Redshift distribution of our LAEs
with a spectroscopic identification. 
Histogram presents LAEs confirmed
by our Keck/DEIMOS observations.
Dashed line represents the 
selection function of LAEs that is 
simply calculated from 
the response curves of $NB921$ filter.
The selection function is normalized 
by the number of the identified LAEs.\\
\label{fig:nz_zlae_NB921}}
\end{figure}

We present our spectra in Figure \ref{fig:image_spec_all_nb921}.
We have confirmed that no spectra show signatures of 
an {\sc [Oiii]} 5007 emission line (at $\sim 7000$\AA) and 
an {\sc [Oii]} 3727 emission line (at $\sim 5200$\AA) 
from a $z=0.40$ H$\alpha$ emitter or 
an {\sc [Oii]} emission line (at $\sim 6800$\AA) from a 
$z=0.84$ {\sc [Oiii]} emitter,
and found that these objects are neither a foreground H$\alpha$ nor 
{\sc [Oiii]} emitters. 
If there is a $z\simeq 1.47$ {\sc [Oii]} emitter, it should
show a $\lambda\lambda$3726,3729 doublet whose separation is 7\AA.
Our DEIMOS spectra with a FWHM spectral resolution of 
2.6\AA\ can distinguish the doublet 
from a Ly$\alpha$ line with a characteristic asymmetric
profile. Our DEIMOS spectra confirm no such signature of
{\sc [Oii]} doublet, but 
mostly clear asymmetric line profiles with an extended red wing 
that is typical for a high-$z$ Ly$\alpha$ line.
One exception is the spectrum of NB921-W-35578 that 
has three spikes in the range of $\simeq 9156-9162$\AA. 
The two-dimensional spectrum
of NB921-W-35578 presents a significant residual of sky subtraction
around $\simeq 9159-9164$\AA, although a signal of emission line is clearly 
seen in the wavelength range of free of residuals. The complicated
spectral shape of NB921-W-35578 is probably made by errors of sky subtraction.
Table \ref{tab:laes_with_redshifts} summarizes 
properties of the 16 LAEs in our spectroscopic sample.
The Ly$\alpha$ emission lines of these LAEs
are strong, $EW_0\gtrsim 20$, which
are consistent with the expected $EW_0$ from 
our color criteria ($EW_0\gtrsim 14$\AA; \S \ref{sec:photometricsample}).
Figure \ref{fig:nz_zlae_NB921} plots 
the redshift distribution of our spectroscopic sample, 
together with a selection function
simply estimated from the filter response curve of $NB921$ band.
The estimated selection function is 
similar to the redshift distribution of 
our spectroscopically identified LAEs.

\section{Luminosity Function}
\label{sec:luminosity_function}

\subsection{Completeness and Contamination of our Sample}
\label{sec:photometric_lae_completeness}

\begin{figure}
%\epsscale{0.50}
\epsscale{1.15}
\plotone{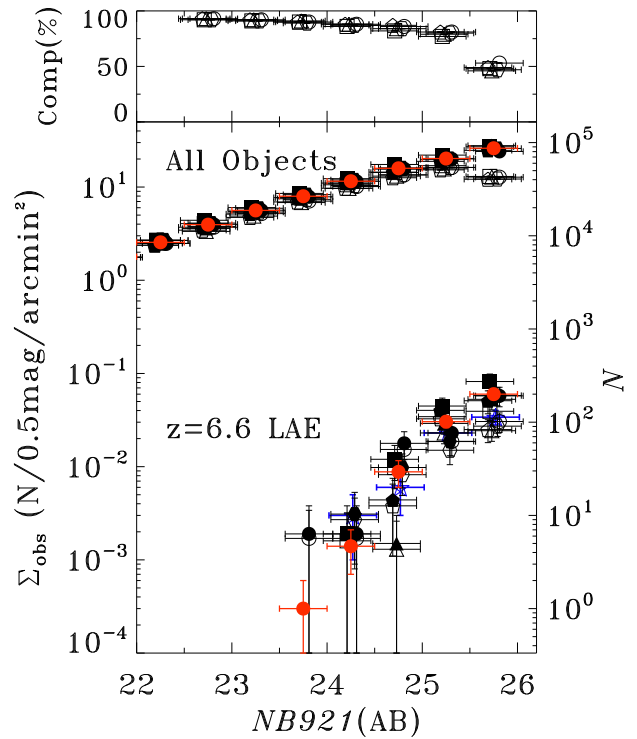}
\caption{
Top panel: Detection completeness of 
our $NB921$ images in percentage. 
Circles, hexagons, triangles,
squares, and pentagons represent the completeness
of a magnitude bin ($\Delta m=0.5$ mag) 
in the 5 subfields, SXDS-C, SXDS-N,
SXDS-S, SXDS-E, and SXDS-W, respectively.
For presentation purpose,
we slightly shift all the points along the abscissa.
Bottom panel: Surface densities of objects detected in
the $NB921$ data. The lower and upper points indicate
surface densities of our $z=6.6$ LAEs and
all the objects detected in the narrow band, respectively.
Black circles, hexagons, triangles,
squares, and pentagons plot the surface densities
in the 5 subfields of SXDS-C, SXDS-N,
SXDS-S, SXDS-E, and SXDS-W, respectively.
We distinguish between 
the raw and completeness-corrected
surface densities with the open and filled symbols,
respectively. The red filled circles represent the
surface density averaged over our entire
survey field. The errors are given by Poisson statistics
for black symbols, while the errors of red symbols
are the geometric mean of Poisson errors
and cosmic variances calculated from eq. (\ref{eq:field_variance}).
To avoid overlaps of points, 
we slightly shift all the black points along the abscissa
with respect to the corresponding red filled circles.
Right vertical axis
indicates the numbers of objects, i.e. $N/(0.5 {\rm mag})$,
identified in our entire survey area.
Blue stars denote the surface density of 
$z\simeq 6.6$ LAEs obtained by
\citet{taniguchi2005}.
\label{fig:number_density_nb921_pluscomp}}
\end{figure}

 We estimate completeness and contamination of our sample
in the same manner as \citet{ouchi2008}.
 First, we estimate the detection completeness of narrow-band
images, $f_{\rm det}$, as a function of narrow-band magnitude. 
We distribute 7400 artificial objects with a PSF profile 
that mimic LAEs on our original 3238 arcmin$^2$ images 
after adding photon noise,
and detect them in the same manner as
for the detection of our LAE catalogs with SExtractor. 
We repeat this process 10 times, and compute the detection completeness.
We plot the detection completeness as a function of
magnitude in the top panel of 
Figure \ref{fig:number_density_nb921_pluscomp}.
The top panel of Figure \ref{fig:number_density_nb921_pluscomp} presents 
that the detection completeness is typically 
$\simeq 80-90$\% for relatively luminous sources 
with $NB921<25.5$.
The detection completeness is $\simeq 50$\%
in the faintest magnitude bin of our sample, $NB921=25.5-26.0$.

 Second, we estimate the contamination 
of our LAE samples.
There are no interlopers in our spectroscopic sample.
We also use a spectroscopic catalog of 3,233 SXDS 
objects at low redshifts of $z=0-6$ \citep{ouchi2008}, 
and find none of these 3,233 low-$z$ objects are included in
our $z=6.6$ LAE sample.
We define the contamination fraction, $f_{\rm cont}$, 
with
\begin{eqnarray}
f_{\rm cont} & = & N_{\rm lowz} / N_{\rm all}
\label{eq:contamination_completeness}
\end{eqnarray}
where 
$N_{\rm lowz}$ and $N_{\rm all}$
are the numbers of low-$z$ interlopers
and all objects, respectively, 
in our spectroscopy sample.

The totals of the spectroscopically 
observed and identified LAEs are 24 and 16,
respectively (\S \ref{sec:spectroscopic_confirmation}).
Since there are 8 objects with no identification 
in our spectroscopic sample
(see Section \ref{sec:spectroscopic_confirmation}),
we calculate $f_{\rm cont}$ 
for the following two extreme cases.
If all of these unidentified objects are real
LAEs whose Ly$\alpha$ lines are simply too faint
or extended to be detected in our spectroscopy, 
we find $N_{\rm lowz} / N_{\rm all} =  0/24$
because of no interlopers
in our spectroscopic sample.
If all the unidentified objects are interlopers,
$N_{\rm lowz} / N_{\rm all} =  [24-16]/24$.
Thus, the contamination rate is taken within the range of 
$f_{\rm cont}\simeq 0-30$\% for our LAE samples.
Note that there are no obvious contaminants
in our follow-up spectroscopy and the SXDS catalogs.
Moreover, the on-going DEIMOS and IMACS spectroscopy
have already confirmed $70$\% of our LAEs 
at $\log L_{\rm Ly\alpha}>43.0$ erg s$^{-1}$,
and found no contaminants in our sample
(Ouchi et al. in preparation).
Although the effect of contaminants may be negligible,
we use this upper limit of contamination 
($f_{\rm cont}\simeq 30$\%)
to derive the upper limits of our measurements
in \S \ref{sec:spatial}.

\subsection{Surface Number Density and Cosmic Variance}
\label{sec:number_density}

The bottom panel of Figure \ref{fig:number_density_nb921_pluscomp}
shows surface densities of LAEs
and all narrow-band detected objects (designated as 'All Objects').
Red circles are the average surface densities.
Black points with 5 different symbols
indicate the surface densities in the 5 subfields
($\simeq 0.2$ deg$^2$) of Suprime-Cam, i.e., SXDS-C, -N, -S, -E, -W.
The detection completeness correction is applied
based on the simulation results described 
in Section \ref{sec:photometric_lae_completeness}.
After the completeness correction, surface densities of
$z=6.6$ LAEs among the 5 subfields differ up to by a factor of
$\sim 2$ in faint magnitude bins ($NB921 = 25-26$) and 
a factor of $\sim 10$ in bright magnitude bins ($NB921<25$), 
while the difference for 'All Objects' is negligibly small. 
Accordingly, the large variance of $z=6.6$ LAEs is not 
artifacts, but real.
These large differences of LAE surface densities 
probably come from the combination of 
cosmic variance and Poisson errors.
Following the procedures used in \citet{ouchi2008},
we evaluate the cosmic variance in our survey area, $\sigma_{\rm g}$, 
with

\begin{eqnarray}
\label{eq:field_variance}
\sigma_{\rm g} & = & 
\sigma_{\rm g:1sf} (\sigma_{\rm DM}/\sigma_{\rm DM:1sf})\\
\sigma_{\rm g:1sf}^2 & = &
[\left<(\Sigma_{\rm g:1sf}-\bar{\Sigma}_{\rm g})^2\right>-\bar{\Sigma}_{\rm g}]/
\bar{\Sigma}_{\rm g}^2
\end{eqnarray}
where $\sigma_{\rm DM}$ and $\sigma_{\rm DM:1sf}$ are
the rms fluctuation of dark matter in all the survey volume
and the volume of 5 subfields ($\simeq 0.2$ deg$^2$), respectively.
We calculate the fluctuations of dark matter with the power spectrum,
adopting the transfer function given by \citet{bardeen1986}.
$\sigma_{\rm g:1sf}$ is the fluctuation of number density of LAEs for one
subfield.
$\Sigma_{\rm g:1sf}$ and $\bar{\Sigma}_{\rm g}$ are LAE's surface densities
in a subfield and the entire survey areas, respectively.

Since these estimates of cosmic variance are based on
a large but single contiguous field, it is important to check
whether our field is located at the sky of
an overdense or underdense region.
Moreover, a large-scale overdensity or underdensity of 
Ly$\alpha$ sources could also be produced by 
an inhomogeneous distribution of Ly$\alpha$ absorbers
(i.e. neutral hydrogen) along the line of sight.
In Figure \ref{fig:number_density_nb921_pluscomp} 
we compare the surface densities of our LAEs with those 
selected with the same $NB921$ filter in 
a completely independent sky of
$\simeq 0.2$ deg$^2$ SDF
(\citealt{taniguchi2005}; see also \citealt{kashikawa2006}).
We find that the surface densities of our $z=6.6$ LAEs are 
consistent with those of \citet{taniguchi2005}
within the scatters and Poisson errors
of our 5 subfields. Moreover, Figure \ref{fig:number_density_nb921_pluscomp} 
shows that \citeauthor{taniguchi2005}'s measurements scatter around
our average surface-density curve.
Thus, we conclude that our 
field has neither signature
of overdensity nor underdensity, and that our 5 subfields
appear to represent the average cosmic variance.

\begin{figure}
%\epsscale{0.85}
\epsscale{1.15}
\plotone{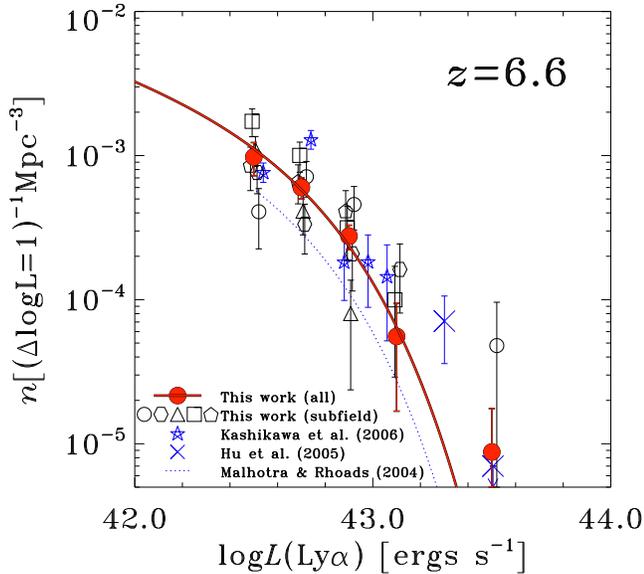}
\caption{
Luminosity functions (LFs) of LAEs at $z=6.6$,
together with those of previous measurements for $z=6.5-6.6$ LAEs.
Filled red circles denote the best-estimate LF 
at $z=6.6$ from our SXDS data. The error bars 
include uncertainties from statistics
and cosmic variance (eq. \ref{eq:field_variance}).
Black open circles, hexagons, triangles,
squares, and pentagons represent
the LFs of five $\sim 0.2$ deg$^2$ subfields, 
SXDS-C, SXDS-N, SXDS-S, SXDS-E, and SXDS-W, 
respectively.
In order to avoid the overlaps of points, we
slightly shift all the open symbols along the abscissa
with respect to the corresponding red filled circles.
Blue star marks present
$z\simeq 6.6$ LF derived in the $\sim 0.2$ deg$^2$ SDF by
\citet{kashikawa2006} (photometric sample), while blue crosses
are the measurement and upper limit of $z=6.5$ LF
given by \citet{hu2005}.
Blue dotted line shows the LF estimated by \citet{malhotra2004}.
Red solid line is the best-fit Schechter function 
to the combination of SXDS and SDF LFs at $z=6.6$.
\label{fig:lumifun_full_diff_nb921_WithComparison}}
\end{figure}

\subsection{Ly$\alpha$ Luminosity Function}
\label{sec:ly_alpha_luminosity_function}

 We derive Ly$\alpha$ LF of LAEs at $z=6.6$ 
in the same manner as \citet{ouchi2008}.
We obtain the number densities of LAEs in each magnitude bin
by dividing the observed number counts of LAEs
by the effective survey volume
defined as the FWHM of the $NB921$ bandpass times the area of the
survey. Here, we calculate the Ly$\alpha$ luminosity of each object 
with the response curves of narrow and broad bands
by subtracting the continuum emission measured from the continuum
magnitude from the total luminosity in the narrow band.
In this calculation, we use total magnitude of narrow-band images.
The continuum emission is estimated by the narrow-band excess
color
defined with a $2''$ aperture, so as to keep high
signal-to-noise ratios and to avoid the effects of source confusion
on broad-band images with high source density.
Ly$\alpha$ luminosity estimates of this photometry technique
is tested for $z=3-6$ LAEs in Figure 15 of \citet{ouchi2008}
which verifies that the ratios of these luminosity 
estimates to independent spectroscopy measurements
are about unity on average.

Although this procedure of Ly$\alpha$ LF derivation
is a simple classical method that was taken by most of previous
studies (e.g. \citealt{ouchi2003,ajiki2003,hu2004,malhotra2004}),
there are two possible sources of uncertainties
in Ly$\alpha$ LFs derived with this classical method.
(1) The narrow-band magnitude of LAEs of 
a fixed Ly$\alpha$ luminosity varies largely as a function 
of redshift. Thus Ly$\alpha$ luminosity may be over- or under-estimated 
for some LAEs. (2) The selection function of LAEs in terms of EW
also changes with redshift; the minimum EW value corresponding 
to a given (fixed) narrow-band excess, $z'- NB921$,
becomes larger when the redshift of the object is far 
from the redshift corresponding to the center of $NB921$ bandpass.
To derive Ly$\alpha$ LF
with no bias originated from (1)-(2), 
\citet{ouchi2008} have carried out 
Monte-Carlo simulations with a mock catalog of their $z=3-6$ 
LAEs uniformly distributed in comoving space
that are produced by a set of the Schechter parameters 
($\alpha, \phi^\star, L^\star$) 
and a Gaussian sigma for a probability distribution of $EW_0$.
Figures 16-18 of \citet{ouchi2008} compare
LFs from the unbiased simulations
with those from the classical method,
and indicate that
the LFs from the classical method 
are consistent with those from the simulations.
This is because 
the uncertainties (1) and (2) in the classical
method are negligible and/or cancel out \citep{ouchi2008}.
Thus, our simple classical method gives 
good estimates of Ly$\alpha$ LF.

\begin{figure}
%\epsscale{0.75}
\epsscale{1.15}
\plotone{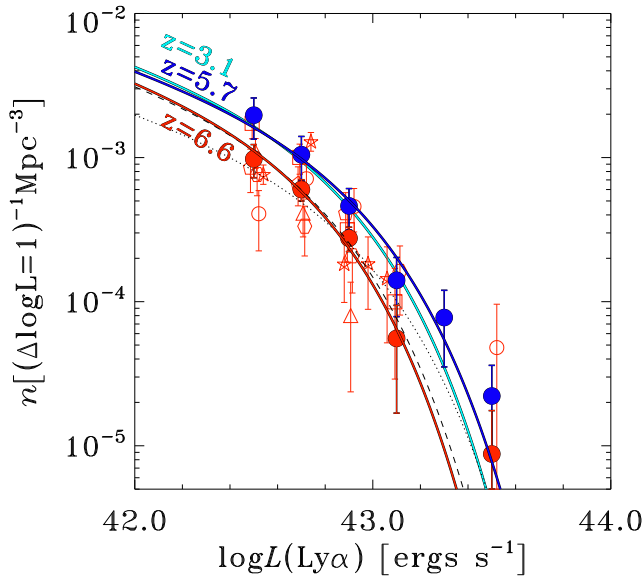}
\caption{
Evolution of Ly$\alpha$ LFs up to $z=6.6$.
Red filled circles are the best estimates
of $z=6.6$ LAEs from the entire SXDS sample,
and red solid line is the best-fit Schechter function
of $z=6.6$ LAEs.
Blue filled circles and solid line are data points
and the best-fit Schechter function, respectively, 
of $z=5.7$ LAEs given by \citet{ouchi2008}.
Note that the error bars of $z=6.6$ and $5.7$
data points (red and blue filled circles)
represent uncertainties from statistics
and cosmic variance. Cyan solid line is the
best-fit Schechter function of $z=3.1$ LAEs 
\citep{ouchi2008}. The LF decreases from
$z=5.7$ to $6.6$ significantly, while
no significant evolution can be 
found between $z=3.1$ and $5.7$.
For comparison, we plot LF estimates from 
each of the five $\sim 0.2$ deg$^2$ subfields 
with the same open symbols 
as found in Figure \ref{fig:lumifun_full_diff_nb921_WithComparison}.
These open symbols illustrate that with the data of a single
$\sim 0.2$ deg$^2$ field alone (e.g. open circles down to 
$\log L_{\rm Ly\alpha}\simeq 42.6$) which is a typical survey size of 
previous studies it is difficult to distinguish whether or not 
$z=6.6$ LFs show evolution (decrease) with respect to $z=5.7$.
Dashed and dotted lines represent the best-fit Schechter
functions to our $z=6.6$ LF
with a $\phi^*$ and $L^*$ fixed to that of $z=5.7$,
respectively.
\label{fig:lumifun_full_diff_nb921_evolution}}
\end{figure}

\begin{figure}
%\epsscale{0.85}
\epsscale{1.15}
\plotone{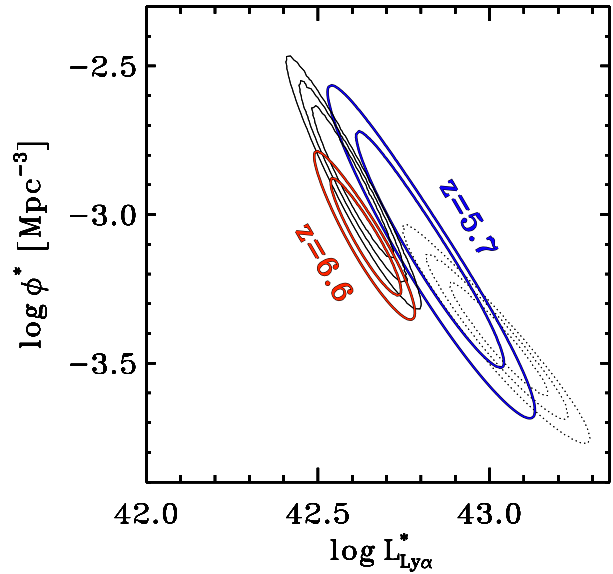}
\caption{
Error ellipses of our Schechter parameters,
$L^*_{\rm Ly\alpha}$ and $\phi^*$.
Red contours represent the fit of $z=6.6$
LF with the fixed slope of
$\alpha=-1.5$ based on SXDS and SDF data.
The inner and outer contours indicate
68\% and 90\% confidence levels, respectively,
which include cosmic variance errors.
Blue contours denote $z=5.7$ LFs given by
\citet{ouchi2008}, which are similarly derived 
with cosmic variance errors. 
For a fair comparison with our
$z=6.6$ LF, we show error ellipses of the $z=5.7$ LF
derived by the classical method (see more details
in \citealt{ouchi2008}).
The error ellipses of the $z=5.7$ LF are larger
than those of our $z=6.6$ LF. This is 
because the data of $z=5.7$ LF have
more uncertainties of cosmic variance.
Black solid and dotted lines
indicate $1$, $2$, and $3$ sigma confidence levels
of $z=6.6$ and $z=5.7$ LFs with no cosmic variance
errors previously derived solely
with the smaller data of SDF \citep{kashikawa2006}.
\label{fig:contour_schechter_LAE}}
\end{figure}

Figure \ref{fig:lumifun_full_diff_nb921_WithComparison}
presents the Ly$\alpha$ LF of our LAEs at $z=6.6$. 
Note that we have corrected the 
detection-completeness by weighting with $f_{\rm det}$
measured in Section \ref{sec:photometric_lae_completeness}.
To check the cosmic variance and the accuracy of our results, 
we plot the estimates of LFs from the entire 3238 arcmin$^2$ 
(red filled circles), together with those from 
the five subfields (black open symbols).
In Figure \ref{fig:lumifun_full_diff_nb921_WithComparison},
we calculate cosmic variances with eq. (\ref{eq:field_variance}),
and include these uncertainties in the error bars of LFs of
the entire field. We find that the scatters
of the measurements are large
among the 5 subfields
with differences by up to a factor of $\simeq 10$,
although the typical scatters of the subfield results 
are {\it not} far beyond the errors of Poisson statistics.

We plot previous measurements of Ly$\alpha$ LFs
at $z\simeq 6.6$ in Figure \ref{fig:lumifun_full_diff_nb921_WithComparison}.
The previous measurements of Ly$\alpha$ LFs show large scatters.
If one compares the LFs of
\citet{malhotra2004} (dotted line), \citet{hu2005} (crosses), and 
\citet{kashikawa2006} (stars)
which are derived from a moderate number of LAEs
and 
a moderately-wide
field,
their measurements do not appear to agree.
This large difference 
causes a long-standing argument 
of Ly$\alpha$ LF evolution
between $z=5.7$ and $6.6$.
However, all of these previous measurements
fall about within the scatters and errors
of LFs derived from our 5 subfields
each with a $\sim 0.2$ deg$^2$ area.
Thus, the discrepancies of LF measurements 
between these previous studies
are probably
due to the combination of cosmic variances
and small statistics.
Because our $z=6.6$ LFs from the 5 subfields cover
the range of measurements of \citet{malhotra2004}, \citet{hu2005}, 
and \citet{kashikawa2006}, we conclude that our
LF measurements agree well with those derived 
in previous studies.

\begin{deluxetable*}{cccccccc}
\tablecolumns{8}
\tabletypesize{\scriptsize}
\tablecaption{Ly$\alpha$ luminosity function.
\label{tab:lya_lumifun_schechter}}
\tablewidth{0pt}
\setlength{\tabcolsep}{0.0in}
\tablehead{
\colhead{$z$} &
\colhead{$\phi^*$} &
\colhead{$L_{\rm Ly\alpha}^*$\tablenotemark{a}} &
\colhead{$\alpha$} &
\colhead{$\chi_{\rm r}^2$} &
\colhead{$n^{\rm obs}$} &
\colhead{$\rho_{\rm Ly \alpha}^{\rm obs}$} &
\colhead{$\rho_{\rm Ly \alpha}^{\rm tot}$} \\
\colhead{} &
\colhead{($10^{-4}$Mpc$^{-3}$)} &
\colhead{($10^{42}$erg s$^{-1}$)} &
\colhead{} &
\colhead{} &
\colhead{($10^{-4}$Mpc$^{-3}$)} &
\colhead{($10^{39}$erg s$^{-1}$Mpc$^{-3}$)} &
\colhead{($10^{39}$erg s$^{-1}$Mpc$^{-3}$)} \\
\colhead{(1)} &
\colhead{(2)} &
\colhead{(3)} &
\colhead{(4)} &
\colhead{(5)} &
\colhead{(6)} &
\colhead{(7)} &
\colhead{(8)} 
}
\startdata
6.6 &  $8.5_{-2.2}^{+3.0}$ & $4.4_{-0.6}^{+0.6}$ & $-1.5$(fix) & $1.60$ & $4.1_{-0.8}^{+0.9}$ & $1.9_{-0.4}^{+0.5}$ & $6.6_{-0.8}^{+1.0}$ \\ 
\enddata
\tablecomments{
(1): Redshift.
(2)-(4): Best-fit Schechter parameters 
for $\phi^*$ and $L^*_{\rm Ly\alpha}$, respectively. 
$\alpha$ is fixed to $-1.5$.
(5): Reduced $\chi^2$ of the fitting.
(6)-(7): Number densities
and Ly$\alpha$ luminosity densities
calculated with the best-fit Schechter parameters down to
the observed limit of Ly$\alpha$ luminosity, i.e. 
$\log L_{\rm Ly\alpha}=42.4$ erg s$^{-1}$.
(8): Inferred total Ly$\alpha$ luminosity densities 
integrated down to $L_{\rm Ly\alpha}=0$ 
with the best-fit Schechter parameters.
}
\tablenotetext{a}{
$L^*_{\rm Ly\alpha}$ is the apparent value, i.e.
observed Ly$\alpha$ luminosity with no correction for IGM absorption.
}
\end{deluxetable*}

Schechter function \citep{schechter1976}
is fit to the Ly$\alpha$ LFs composed of 
our large area data with cosmic variance errors
and the previous LF estimate 
in the independent 876 arcmin$^2$ area of SDF \citep{kashikawa2006}.
\footnote{
Twenty six out of 29 LAEs used in \citet{malhotra2004}
are provided from a subsample of the SDF data
\citep{taniguchi2005,kashikawa2006}. To avoid
using the same LAE data,
we do not include \citet{malhotra2004} data points
for our Schechter fitting.
We also do not use data points in papers
that are not published yet in a refereed journal.
}
The Schechter function is defined by
\begin{equation}
\phi(L)dL=\phi^*(L/L^*)^\alpha \exp(-L/L^*)d(L/L^*).
\label{eq:schechter_l}
\end{equation}
With a total of 265 ($=207+58$) LAEs at $z\simeq 6.6$ in a
total area of 4114 ($=3238+876$) arcmin$^2$ data
in the SXDS and SDF, we obtain 
the best-fit Schechter parameters of
$\phi^*=8.5_{-2.2}^{+3.0}\times10^{-4}$Mpc$^{-3}$ and
$L_{\rm Ly\alpha}^*=4.4_{-0.6}^{+0.6}\times10^{42}$ erg s$^{-1}$
with a fixed $\alpha=-1.5$, which are summarized
in Table \ref{tab:lya_lumifun_schechter}.
Because the difference in $\chi^2$ for $\alpha$ values 
is insignificant, we fix $\alpha$ to $-1.5$,
which is a fiducial value used for low-$z$ Ly$\alpha$ LFs 
(e.g. \citealt{malhotra2004,kashikawa2006,ouchi2008}).
The best-fit Schechter function is
shown in Figure \ref{fig:lumifun_full_diff_nb921_WithComparison}
with the red solid line.
We also estimate the best-fit Schechter parameters
with a fixed $\alpha=-1.7$, because the value of $\alpha\simeq -1.7$ 
is recently reported for $z=2-6$ LAEs (\citealt{cassata2010};
see also \citealt{rauch2008}) 
as well as $z\sim 7$ dropouts 
(e.g. \citealt{ouchi2009b,oesch2010,mclure2010}).
The best-fit parameters for $\alpha=-1.7$
are $\phi^*=6.9_{-1.9}^{+2.6}\times10^{-4}$Mpc$^{-3}$ and
$L_{\rm Ly\alpha}^*=4.9_{-0.7}^{+0.9}\times10^{42}$ erg s$^{-1}$.
Note that our data point at the bright end
($\log L({\rm Ly\alpha})=43.5$ erg s$^{-1}$ )
appears to exceed the best-fit Schechter function
in Figure \ref{fig:lumifun_full_diff_nb921_WithComparison},
although the data point is consistent with 
the best-fit Schechter function within the error bar
that extends down to 0.
This excess data point is solely made by one exceptional LAE 
that is an extended giant LAE, {\it Himiko},
reported in \citet{ouchi2009a}.

\subsection{Evolution of Ly$\alpha$ Luminosity Function}
\label{sec:evolution_of_lya_LF}

Figure \ref{fig:lumifun_full_diff_nb921_evolution} compares 
our Ly$\alpha$ LF of LAEs at $z=6.6$ with those
at $z=3.1$ and $5.7$. Note that these low-$z$ LFs
are derived from large LAE samples of wide-field SXDS \citep{ouchi2008}
with the same procedures (including cosmic variance errors) 
and similar data sets taken with the same instrument.
In this sense, there are little systematics
among different redshift results 
caused by the sample selection and measurement technique.
While the LFs do not change within error bars from $z=3.1$ to $5.7$
\citep{ouchi2008}, the LF decreases
from $z=5.7$ to $6.6$ beyond sizes of errors, i.e. uncertainties
of statistics and cosmic variance. To quantify this decrease,
we present in Figure \ref{fig:contour_schechter_LAE}
error ellipses of Schechter parameters of 
LFs at $z=6.6$, together with those at $z=5.7$ from \citet{ouchi2008}.
Figure \ref{fig:contour_schechter_LAE} shows
that error contours of $z=6.6$ and $5.7$ 
differ at the $>90$\% significance level.
Even if we consider the maximum effect of contamination
(\S \ref{sec:photometric_lae_completeness}),
there exist a significant evolution of Ly$\alpha$ LF.
Because a contamination correction pushes down 
our $z=6.6$ LAE LF towards low number density 
in Figure \ref{fig:lumifun_full_diff_nb921_evolution},
an inclusion of contamination correction 
even strengthens the evolutionary tendency.
The $EW_0$ limit of our $z=6.6$ LAE sample
is $\gtrsim 14$\AA, which is different from and 
smaller than that of the $z=5.7$ LAE sample 
($EW_0 \gtrsim 27$\AA; \citealt{ouchi2008})
by $\Delta EW_0\simeq 10$\AA.
\citet{ouchi2008} have investigated
a possible false evolution given by 
a choice of $EW_0$ limit
for LAE samples 
at different redshifts.
They use LAE samples
at $z=3-6$ whose $EW_0$ limits
differ by 30\AA ($EW_0\gtrsim 30-60$\AA),
and fit a number-$EW_0$ distribution 
of the samples with a Gaussian function
to obtain inferred Schechter parameters
for all $z=3-6$ LAE samples with the same
$EW_0$ limit of $EW_0=0$.
\citet{ouchi2008} find that the inferred
$\phi^*$ is different from
the original $\phi^*$ only by $\lesssim 10$\%, and claim that
a choice of $EW_0$ limits provides 
a negligible impact on the results of
Ly$\alpha$ LF evolution. This is true if
a difference of $EW_0$ limit is
much smaller than the best-fit Gaussian sigma of 
number-$EW_0$ distribution ($\gtrsim 130$\AA\ for
$z=3-6$; \citealt{ouchi2008}).
Although we cannot carry out the similar
investigation for our $z=6.6$ sample due to
poorly constrained $EW_0$ values of our LAEs
(\S \ref{sec:evolution_of_lyalf} for more details),
it is very likely that the bias introduced
by the different $EW_0$ limits ($\Delta EW_0\simeq 10$\AA)
is smaller than that found in \citet{ouchi2008},
$\lesssim 10$\% corresponding to $\lesssim 0.04$ dex
in $\phi^*$. This small bias of $\lesssim 0.04$ dex
does not affect to the conclusion of LF decrease
in Figure \ref{fig:contour_schechter_LAE}. Moreover,
the conclusion of LF decrease is probably, again, 
even strengthened, because corrections for 
the incompleteness of the $EW_0$ limits
raise $\phi^*$ of the $z=5.7$ sample
($EW_0 \gtrsim 27$\AA) more than 
that of our $z=6.6$ sample
($EW_0 \gtrsim 14$\AA) in the
case of the same number-$EW_0$ distribution
at $z=5.7$ and $6.6$.
We plot error contours obtained by \citet{kashikawa2006}
in Figure \ref{fig:contour_schechter_LAE} for comparison.
Because the measurements of \citet{kashikawa2006} do not
include cosmic variance errors, their contours are
relatively small.

\begin{figure*}
%\epsscale{1.0}
\epsscale{1.2}
\plotone{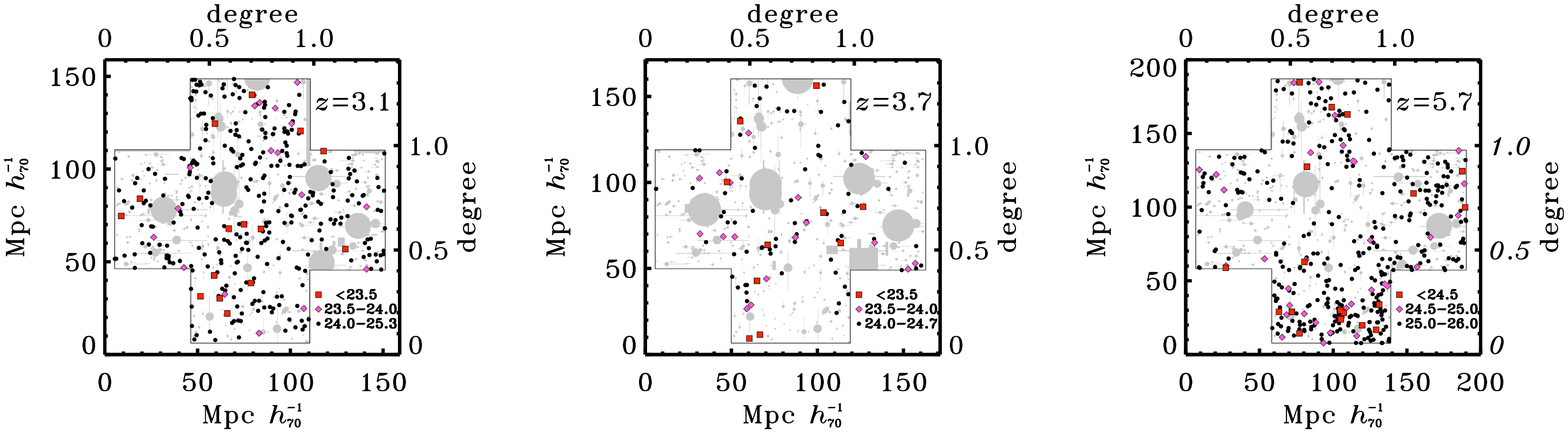}
\caption{
Sky distribution of the SXDS LAEs at $z=3.1$ (left),
$3.7$ (center), and $5.7$ (right) obtained by 
\citet{ouchi2008}.
Red squares, magenta diamonds, and black circles
present positions of narrow-band 
(bright, medium-bright, and faint) LAEs, respectively,
in narrow-band magnitudes of
$(NB503<23.5, 23.5\le NB503<24.0, 24.0\le NB503<25.3)$ (left panel),
$(NB570<23.5, 23.5\le NB570<24.0, 24.0\le NB570<24.7)$ (center panel), and
$(NB816<24.5, 24.5\le NB816<25.0, 25.0\le NB503<26.0)$ (right panel).
The gray shades represent masked areas
that are not used for sample selection.
The scale on the map is marked in both degrees and 
the projected distance in comoving megaparsecs at each redshift.
\label{fig:dist_combz3_6LAE}}
\end{figure*}

\begin{figure}
%\epsscale{1.0}
\epsscale{1.15}
\plotone{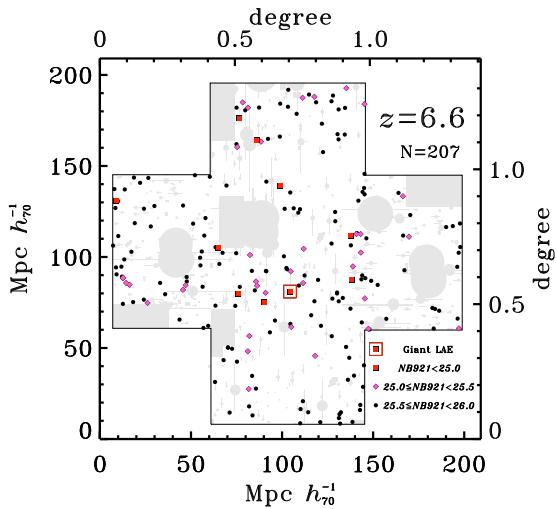}
\caption{
Same as Figure \ref{fig:dist_combz3_6LAE},
but for our 207 LAEs at $z=6.565\pm 0.054$.
Red squares, magenta diamonds, and black circles
show positions of bright ($NB921<25.0$),
medium-bright ($25.0\le NB921\le 25.5$),
and faint ($25.5\le NB921\le 26.0$) LAEs,
respectively.
The red square highlighted with 
a red open square indicates 
the giant LAE, {\it Himiko}, 
with a bright and extended Ly$\alpha$
nebular at $z=6.595$ reported by
\citet{ouchi2009a}.
\label{fig:dist_NB921LAE}}
\end{figure}

\begin{figure*}
%\epsscale{1.0}
\epsscale{1.15}
\plotone{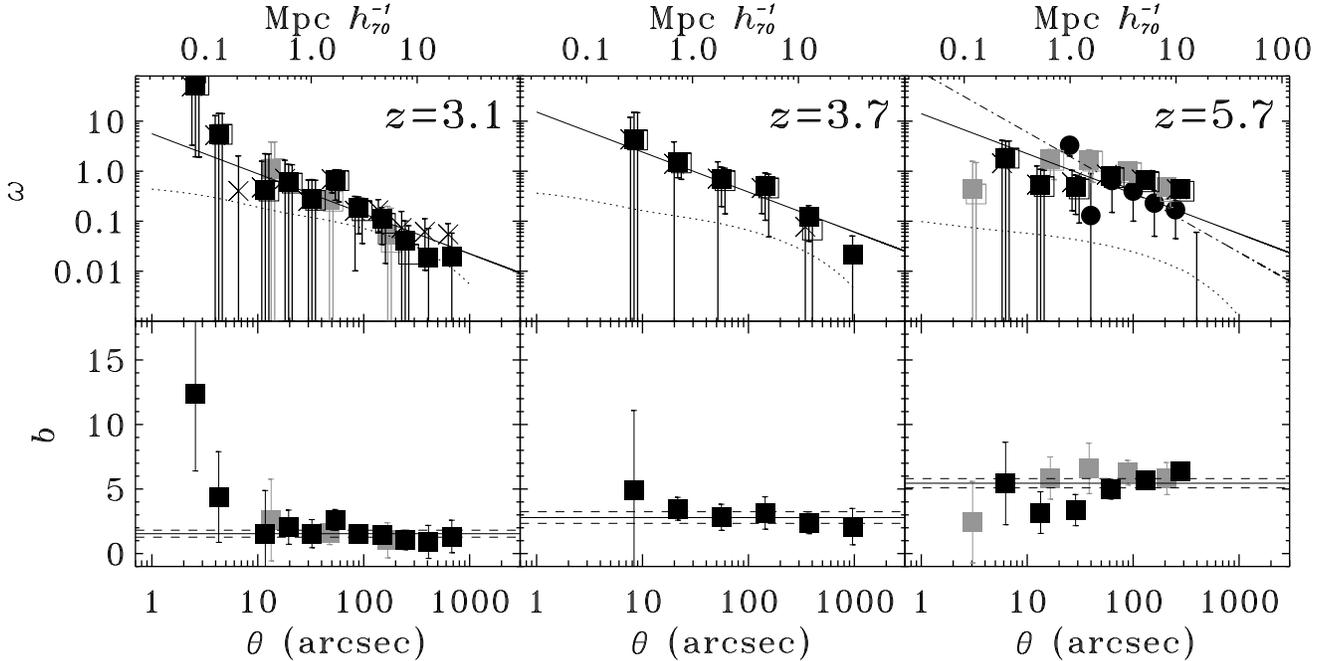}
\caption{
Angular correlation function (ACF) and bias
of the SXDS LAEs at $z=3.1$ (left), $3.7$ (center), and
$5.7$ (right). Top panels present ACFs, and bottom panels
show bias as a function of angular distance.
Black filled and open squares indicate
the best estimates for all LAEs with and without
the corrections of limited area (i.e. integral constraints) 
and number (eq. \ref{eq:acfcorrection}).
Gray filled and open squares are the same,
but for the bright subsamples, $NB503<24.7$ in
the left panels and $NB816<25.5$ in the right panels.
Crosses in the top panels represent ACFs
calculated with the Monte-Carlo simulation-based
random sources that are artificial LAEs 
distributed and detected in the real images (see the text).
In the top panels, solid lines are 
the best-fit power law functions for all LAEs,
while the dotted lines denote ACFs of 
underlying dark matter predicted 
by the CDM model \citep{peacock1996}.
In the bottom panels, solid and dashed lines
present the best estimates of bias
and the associated $1 \sigma$ errors.
In the top right panel, filled circles
and dot-dashed line are ACF and the best-fit
power law obtained by \citet{murayama2007}.
The scale on the top axis denotes
the projected distance in comoving 
megaparsecs at each redshift. 
\label{fig:acorr_z3_6comb_all_final}}
\end{figure*}

\begin{figure}
%\epsscale{1.0}
\epsscale{1.15}
\plotone{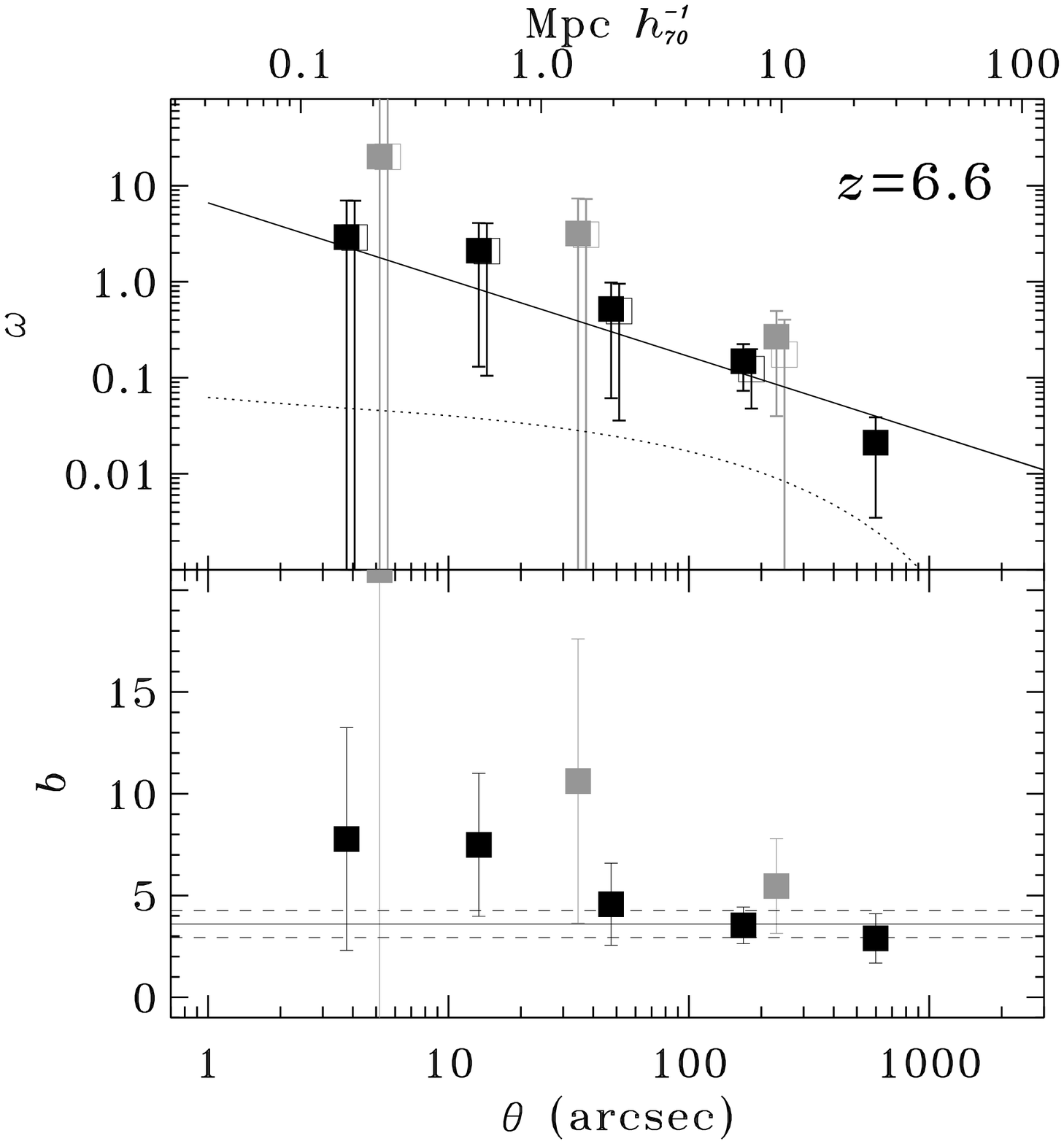}
\caption{
Same as Figure \ref{fig:acorr_z3_6comb_all_final},
but for our LAEs at $z=6.6$.
Gray filled and open squares are the estimates
for the bright ($NB921<25.5$) subsample.
\label{fig:acorr_NB921_all_final}}
\end{figure}

Although the errors of our measurements are not small,
Figure \ref{fig:contour_schechter_LAE} implies
that a decrease in $L^*$ would be the dominant factor
of the LF evolution from $z=5.7$ to $6.6$.
We investigate whether 
pure number or pure luminosity evolutions 
are dominant. We fix either $L^*$ or $\phi^*$ 
of the $z=5.7$ LF, and carry out 
Schechter function fitting to our $z=6.6$ LF.
Figure \ref{fig:lumifun_full_diff_nb921_evolution} plots
the best-fit Schechter functions for the 
two scenarios of the pure number (dotted line) and 
pure luminosity (dashed line) evolutions
from $z=5.7$ to $6.6$. 
We obtain $\phi^*(z=6.6)=0.5 \phi^*(z=5.7)$
for the pure number evolution,
and $L^*(z=6.6)=0.7 L^*(z=5.7)$ 
for the pure luminosity evolution.
The fitting to
the pure luminosity evolution is better
than that to the pure number evolution,
and the $\chi^2$ value of the pure luminosity evolution
is smaller than that of 
the pure number evolution 
by 7.40.
We confirm that, irrespective of $\alpha$, 
the pure luminosity evolution is more 
favorable than the pure number evolution
based on our Schechter function fitting
with the fixed best-fit $L^*$ or $\phi^*$ 
of the $z=5.7$ LF in the cases of $\alpha= -1.0$
and $-2.0$ (Table 5 of \citealt{ouchi2008})
that bracket a reasonable range of $\alpha$.
In the case of pure luminosity
evolution, the decrease of luminosity is 30\%
from $z=5.7$ to $6.6$. This decrease is smaller than
that obtained by \citet{kashikawa2006} who claim
a luminosity decrease by $0.75$ magnitude.
Moreover, this 30\% decrease is too small
to be distinguished by any previous studies 
that include large statistical and cosmic variance errors. 
We plot LF estimates from each of the five $\sim 0.2$ deg$^2$ subfields
in Figure \ref{fig:lumifun_full_diff_nb921_evolution}              
with the same open symbols as 
in Figure \ref{fig:lumifun_full_diff_nb921_WithComparison}.
These open symbols illustrate that with the data of a single
$\sim 0.2$ deg$^2$ field alone which is a typical survey size of 
previous studies it is difficult to distinguish whether or not $z=6.6$
LFs show evolution (decrease) with respect to $z=5.7$.
We find the decrease of Ly$\alpha$ LF between $z=5.7$ and $6.6$
at the $>90$\% significance level.
On the other hand,
\citet{malhotra2004} claim no evolution 
between $z=5.7$ and $6.6$, 
which appears to contradict with our finding.
However,
\citet{malhotra2004} carefully state that
a Ly$\alpha$ flux of $z=6.5$ LAEs 
is not attenuated by a factor of 3, i.e. 300\%,
compared with $z=5.7$ LAEs. Because 
\citet{malhotra2004} do not conclude
evolutionary tendency in the change by a few 10\% level,
but by a few 100\% level. In this sense,
the conclusion of \citet{malhotra2004}
quantitatively agrees with ours. 
Another issue is the baseline LF at $z=5.7$ used in
these comparisons.
The baseline $z=5.7$ LFs are different between
our study, \citet{malhotra2004}, and \citet{hu2005}.
We use the up-to-date $z=5.7$ LF of \citet{ouchi2008} shown 
in Figure \ref{fig:lumifun_full_diff_nb921_evolution},
while \citet{malhotra2004} and \citet{hu2005} use
their own LFs. Figure 19 of \citet{ouchi2008} compares
all of these three $z=5.7$ LFs, and indicates
that our baseline $z=5.7$ LF of \citet{ouchi2008} 
comes around the
middle of previous results and falls between
those of \citet{malhotra2004} and \citet{hu2004};
the one used by \citet{malhotra2004} is placed below ours,
while the one used by \citet{hu2005} is above ours.
Because we see exactly the same tendency
in $z=6.6$ LF measurements in Figure 
\ref{fig:lumifun_full_diff_nb921_WithComparison}
(i.e. \citeauthor{malhotra2004}'s LF below ours,
and \citeauthor{hu2005}'s LF above ours),
the 'no evolution' is more clearly
found in Figure 2 of \citet{malhotra2004} and 
Figure 3 of \citet{hu2005},
even without considering their large uncertainties
of statistics and cosmic variance.

In summary, we conclude that Ly$\alpha$ LF
decreases from $z=5.7$ to $6.6$, and
that no evolution of Ly$\alpha$ LF is ruled out 
at the $>90$\% significance level
with errors including uncertainties of statistics
and cosmic variance. The pure luminosity evolution
is more preferable to the pure number density evolution.
If the pure luminosity evolution is
assumed, $L_{\rm Ly\alpha}^*$ dims by 30\% 
from $z=5.7$ to $6.6$.

\section{Spatial Distribution}
\label{sec:spatial}

In this section, we discuss clustering of our $z=6.6$ LAEs.
Since we need to know an evolutionary trend of clustering at $z=3-6$
to distinguish cosmic reionization and galaxy evolution effects
at $z=6.6$ (\S \ref{sec:evolution_of_clustering}),
we derive correlation functions of not only
our 207 LAEs at $z=6.6$, but also 
those at lower redshifts.
We use 356, 101, and 401 LAEs at $z=3.1$, $3.7$, and $5.7$, respectively,
given by \citet{ouchi2008}. Since we have a moderately
large number ($\gtrsim 200$) of LAEs at $z=3.1$, $5.7$, and 
$6.6$ down to narrow-band magnitude limits of $25.3$, $26.0$, 
and $26.0$, respectively, we make bright subsamples
with narrow-band limiting magnitudes brighter by 0.5 magnitude;
$24.8$ ($z=3.1$), $25.5$ ($z=5.7$), and $25.5$ ($z=6.6$).
Readers should refer \citet{ouchi2008} 
for the details of the SXDS  $z=3.1-5.7$ LAE
samples.

\subsection{Angular Correlation Function}
\label{sec:spatial_angular}

Figures \ref{fig:dist_combz3_6LAE}-\ref{fig:dist_NB921LAE} 
present sky distributions of the $z=3.1-5.7$ LAEs
and our $z=6.6$ LAEs, respectively.
In order to quantitatively 
measure the inhomogeneity of spatial distribution,
we derive the angular two-point correlation function
(ACF) in the same manner as \citet{ouchi2003},
\citet{ouchi2004b} and \citet{ouchi2005b}.
According to \citet{landy1993},
the ACF is calculated by:

\begin{equation}
\omega_{\rm obs}(\theta)
  = [DD(\theta)-2DR(\theta)+RR(\theta)]/RR(\theta),
\label{eq:landyszalay}
\end{equation}
where $DD(\theta)$, $DR(\theta)$, and $RR(\theta)$ are numbers of
galaxy-galaxy, galaxy-random, and random-random pairs normalized by
the total number of pairs in each of the three samples.
We first create a pure random sample composed of 100,000 sources
with the same geometrical constraints as of the data sample,
and estimate errors with the bootstrap technique \citep{ling1986}.
Figures \ref{fig:acorr_z3_6comb_all_final}-\ref{fig:acorr_NB921_all_final} 
show the ACFs, $\omega_{\rm obs}(\theta)$, of LAEs from the observations
at $z=3.1-5.7$ and $6.6$.
We find significant clustering signals for our $z=6.6$ LAEs
as well as the $z=3.1-5.7$ LAEs.

\begin{deluxetable*}{crccccc}
\tablecolumns{7}
\tabletypesize{\scriptsize}
\tablecaption{Clustering Measurements
\label{tab:clustering_measurements}}
\tablewidth{0pt}
\tablehead{
\colhead{$z$} &
\colhead{Sample} &
\colhead{$C+1/N$} &
\colhead{$A_\omega$} &
\colhead{$\beta$} &
\colhead{$r_0$} &
\colhead{$r_0^{\rm max}$} \\
\colhead{} &
\colhead{} &
\colhead{} &
\colhead{(arcsec$^\beta$)} &
\colhead{} &
\colhead{($h^{-1}_{100}$ Mpc)} &
\colhead{($h^{-1}_{100}$ Mpc)} \\
\colhead{(1)} &
\colhead{(2)} &
\colhead{(3)} &
\colhead{(4)} &
\colhead{(5)} &
\colhead{(6)} &
\colhead{(7)}
}
\startdata
$3.1$ & All ($NB503<25.3$) & $1.99\times 10^{-2}$ & $5.61\pm2.46$ & $0.80$ (fix) & $1.70^{+0.39}_{-0.46}$ & $1.99^{+0.45}_{-0.55}$ \\ 
$3.1$ & $NB503<24.8$ & $1.66\times 10^{-2}$ & $3.13\pm3.12$ & $0.80$ (fix) & $1.23^{+0.58}_{-1.18}$ & $1.44^{+0.67}_{-1.37}$ \\ 
$3.7$ & All ($NB570<24.7$) & $5.59\times 10^{-2}$ & $15.24\pm6.38$ & $0.80$ (fix) & $2.74^{+0.58}_{-0.72}$ & $3.24^{+0.69}_{-0.85}$ \\ 
$5.7$ & All ($NB816<26.0$) & $4.52\times 10^{-2}$ & $14.13\pm2.80$ & $0.80$ (fix) & $3.12^{+0.33}_{-0.36}$ & $4.30^{+0.45}_{-0.50}$ \\ 
$5.7$ & $NB816<25.5$ & $8.76\times 10^{-2}$ & $26.37\pm6.59$ & $0.80$ (fix) & $4.41^{+0.59}_{-0.65}$ & $6.08^{+0.80}_{-0.90}$ \\ 
$6.6$ & All ($NB921<26.0$) & $2.53\times 10^{-2}$ & $6.63\pm3.73$ & $0.80$ (fix) & $2.31^{+0.65}_{-0.85}$ & $3.60^{+1.02}_{-1.32}$ \\ 
$6.6$ & $NB921<25.5$ & $9.35\times 10^{-2}$ & $23.25\pm22.31$ & $0.80$ (fix) & $4.64^{+2.10}_{-3.86}$ & $7.23^{+3.28}_{-6.01}$ \\ 
\enddata
\tablecomments{
(1): Redshift,
(2): LAE sample,
(3): offset value for $A_{\omega}$ 
that corrects for limited survey area and number (eq. \ref{eq:acfcorrection}),
(4): best-fit amplitude of power-law function 
for the angular correlation function in arcsec$^\beta$.
(5): the fixed $\beta$ for the power-law fitting,
(6)-(7): correlation lengths in $h^{-1}_{100}$ Mpc. 
The best-estimate and the maximally-contamination corrected $r_0$
in (6) and (7), respectively.
To facilitate comparison with previous results,
we express $r_0$ using $h_{100}$, the Hubble constant 
in units of 100 km s$^{-1}$ Mpc$^{-1}$,
instead of 70 km s$^{-1}$ Mpc$^{-1}$.
}
\end{deluxetable*}

We then confirm that these clustering signals are not
artifacts produced by
the slight inhomogeneous quality over 
the images or occultation by foreground objects
on the basis of our Monte Carlo simulations. 
We use mock catalogs of LAEs 
obtained by simulations of \citet{ouchi2008}, 
which have number counts and color distribution 
that agree with observational measurements.
We generate 50,000 artificial LAEs based on the mock catalog,
and distribute them randomly on the original 1 deg$^2$ 
narrow- and broad-band images after adding Poisson noise 
according to their brightness. Since most of LAEs are nearly 
point sources,
we assume profiles of point-spread functions that are the same
as the original images.
Then, we detect these simulated LAEs and measure their
brightness in the same manner as the real LAEs.
We iterate this process 10 times and select
LAEs with the same color criteria as the real LAEs.
We thus obtain $\sim 200,000$ simulation-based 
random sources at each redshift
whose positions are affected by 
the inhomogeneity of LAE detectability
and the occultation of foreground objects,
and therefore slightly different from the pure random sample.
We use these simulation-based random sources for our ACF calculation.
Crosses in Figure \ref{fig:acorr_z3_6comb_all_final}
present estimates of ACFs with these random sources.
The ACFs estimated with these simulation-based random sources (crosses)
and the pure random-distribution sources (squares) are consistent.
Accordingly, we conclude that the clustering signals are
not artifacts
given by the slight inhomogeneity of image qualities
or the occultation of foreground objects.

To evaluate observational offsets 
included in $\omega_{\rm obs}(\theta)$
due to the limited area and object number,
we assume the real ACF, $\omega(\theta)$, 
is approximated by the power law,
\begin{equation}
\omega(\theta) = A_\omega \theta^{-\beta}.
\label{eq:acorr_powerlaw}
\end{equation}
Then, the offset from the observed ACF, $\omega_{obs}(\theta)$,
is given by the integral constraint, $C$, \citep{groth1977},
and the number of objects in the sample, $N$,
\begin{eqnarray}
\omega(\theta) & = & \omega_{obs}(\theta)+C+\frac{1}{N} 
\label{eq:acfcorrection} \\
C & = & \frac{\Sigma RR(\theta) A_\omega \theta^{-\beta}}{\Sigma RR(\theta)}.
\label{eq:integralconstant}
\end{eqnarray}
The term, $1/N$, in eq. (\ref{eq:acfcorrection}) corrects 
for the difference between the number of object pairs,
$N(N-1)/2$, and its approximation, $N^2/2$ \citep{peebles1980}.
Note that most of the previous clustering studies for high-$z$
galaxies neglect this $1/N$ term 
(e.g. \citealt{roche1999,daddi2000}) probably
because of their large samples. However,
it should be applied 
for samples with a small number of objects
such as LAE samples to obtain more accurate
ACF at a large scale. The ACF corrected with
eq. (\ref{eq:acfcorrection}) are also presented
in Figures \ref{fig:acorr_z3_6comb_all_final}-\ref{fig:acorr_NB921_all_final} .

We fit the power law (eq. \ref{eq:acorr_powerlaw})
over $10''<\theta<1000''$
with the corrections. The lower limit of the fitting range, 
$\theta=10''$, is placed, because 1-halo term of high-$z$ galaxies 
is dominant at this small scale 
\citep{ouchi2005b,lee2006,hildebrandt2007,hildebrandt2009}.
The upper limit of the range is nearly the limit
of our ACF measurements with significant signals. We apply the
upper limit of $400''$ for the $z=5.7$ LAEs, since
a large-scale clustering ($\gtrsim 20$ Mpc) 
with proto-clusters is reported in this survey volume
\citep{ouchi2005a}.
Because we obtain no meaningful constraints on $\beta$,
we fix $\beta$ with the fiducial value of $0.8$
following the previous clustering analyses
(e.g. \citealt{ouchi2003,gawiser2007,kovac2007}).
We summarize the best-fit values 
in Table \ref{tab:clustering_measurements}.

Foreground contamination to a galaxy sample 
dilutes the apparent clustering amplitude of galaxies. 
If the fraction of contaminants is $f_{\rm c}$,
the apparent $A_\omega$ value  
can be reduced by a factor of up to $(1 - f_{\rm c})^2$.
The correlation amplitude with the contamination correction, 
$A_{\omega}^{\rm max}$, is 
given by
\begin{equation}
A_\omega^{\rm max} = \frac{A_\omega}{(1-f_{\rm c})^2}.
\label{eq:clustering_dilution}
\end{equation}
This is the maximum reduction of the correlation
amplitude that occurs when the contaminants 
are not at all clustered.
In reality, contaminants, if any, are the sum of 
foreground galaxies mostly at some specific redshifts, 
and thus would be clustered to some extent on the sky.
We use the maximum values of $f_{\rm c}$,
and obtain maximal $A_\omega^{\rm max}$ values inferred 
from the LAE data. The maximum values of $f_{\rm c}$ are
$(0.13, 0.14, 0.25)$ for redshifts of 
$(3.1, 3.7, 5.7)$ \citep{ouchi2008}
and 
$0.33$ for $z=6.6$ 
(\S \ref{sec:photometric_lae_completeness}).
In the next section, we will place conservative
upper limits on correlation lengths and bias measurements
with the maximal values of $A_\omega^{\rm max}$.

\begin{deluxetable*}{cccccccc}
\tablecolumns{8}
\tabletypesize{\scriptsize}
\tablecaption{Bias and Hosting Dark Halos
\label{tab:bias_darkhalo}}
\tablewidth{0pt}
\tablehead{
\colhead{$z$} &
\colhead{$\log(L_{\rm Lim})$} &
\colhead{$n_{\rm g}$} &
\colhead{$b_{\rm g}$} &
\colhead{$b_{\rm g}^{\rm max}$} &
\colhead{$M_{\rm h}$} &
\colhead{$M_{\rm h}^{\rm min}$} &
\colhead{$M_{\rm h}^{\rm max}$} \\
\colhead{} &
\colhead{(erg s$^{-1}$)} &
\colhead{(Mpc$^{-3}$)} &
\colhead{} &
\colhead{} &
\colhead{($M_\odot$)} &
\colhead{($M_\odot$)} &
\colhead{($M_\odot$)} \\
\colhead{(1)} &
\colhead{(2)} &
\colhead{(3)} &
\colhead{(4)} &
\colhead{(5)} &
\colhead{(6)} &
\colhead{(7)} &
\colhead{(8)}
}
\startdata
$3.1$ & $42.3$ & $8.9^{+1.9}_{-1.2} \times 10^{-4}$ & $1.5\pm0.7$ & $1.7\pm0.8$ & $2.9^{+24.0}_{-2.9} \times 10^{10}$ & $5.2^{+73.0}_{-5.2} \times 10^{9}$ & $6.7^{+42.0}_{-6.7} \times 10^{10}$ \\ 
$3.1$ & $42.1$ & $1.5^{+0.3}_{-0.2} \times 10^{-3}$ & $1.5\pm0.3$ & $1.8\pm0.3$ & $3.6^{+5.6}_{-2.7} \times 10^{10}$ & $6.8^{+15.0}_{-5.5} \times 10^{9}$ & $8.2^{+11.0}_{-5.7} \times 10^{10}$ \\ 
$3.7$ & $42.6$ & $2.8^{+0.6}_{-0.6} \times 10^{-4}$ & $2.8\pm0.5$ & $3.2\pm0.5$ & $3.0^{+2.7}_{-1.7} \times 10^{11}$ & $1.1^{+1.2}_{-0.7} \times 10^{11}$ & $5.7^{+4.6}_{-3.0} \times 10^{11}$ \\ 
$5.7$ & $42.6$ & $3.5^{+2.5}_{-1.7} \times 10^{-4}$ & $6.1\pm0.7$ & $8.2\pm0.9$ & $6.1^{+2.8}_{-2.2} \times 10^{11}$ & $3.5^{+1.9}_{-1.4} \times 10^{11}$ & $1.7^{+0.7}_{-0.5} \times 10^{12}$ \\ 
$5.7$ & $42.4$ & $6.8^{+3.8}_{-2.7} \times 10^{-4}$ & $5.5\pm0.4$ & $7.3\pm0.5$ & $3.8^{+1.2}_{-0.9} \times 10^{11}$ & $2.1^{+0.7}_{-0.6} \times 10^{11}$ & $1.2^{+0.3}_{-0.3} \times 10^{12}$ \\ 
$6.6$ & $42.6$ & $1.8^{+0.5}_{-0.4} \times 10^{-4}$ & $6.0\pm2.2$ & $8.9\pm3.3$ & $2.4^{+5.6}_{-2.1} \times 10^{11}$ & $1.3^{+3.9}_{-1.2} \times 10^{11}$ & $1.1^{+2.1}_{-0.9} \times 10^{12}$ \\ 
$6.6$ & $42.4$ & $4.1^{+0.9}_{-0.8} \times 10^{-4}$ & $3.6\pm0.7$ & $5.4\pm1.0$ & $2.4^{+3.1}_{-1.6} \times 10^{10}$ & $9.0^{+15.0}_{-6.5} \times 10^{9}$ & $1.5^{+1.5}_{-0.9} \times 10^{11}$ \\ 
\enddata
\tablecomments{
(1): Redshift,
(2): approximated limiting Ly$\alpha$ luminosity estimated from narrow-band magnitudes,
(3): number density of the observed LAEs
that is derived with the best-fit Schechter function (Table \ref{tab:lya_lumifun_schechter}),
(4): best estimate of bias,
(5): upper limit of bias with the maximal contamination correction, 
(6): average hosting dark halo mass inferred from the best estimate of bias, 
(7): minimum hosting dark halo mass, and  
(8): upper limit of the average hosting dark halo mass corresponding to the upper limit of bias.
}
\end{deluxetable*}

\subsection{Correlation Length and Bias}
\label{sec:spatial_correlation_bias}

The spatial correlation function of galaxies is 
well-approximated by a power law as
\begin{equation}
\xi=(r/r_0)^{-\gamma}
\label{eq:3dcorrelation}
\end{equation}
only with a subtle departure
from the real spatial correlation function
\citep{zehavi2004},
where $r$ is the spatial separation between two objects,
$r_0$ is the correlation length, and $\gamma$ is the slope
of the power law.
The correlation length, $r_0$, 
is related to the correlation amplitude, $A_\omega$, 
with the integral equation
called Limber equation \citep{peebles1980,efstathiou1991},
\begin{equation}
      A_{\omega}= k r_0^\gamma
      \int_{0}^{\infty} F(z) D_\theta^{1-\gamma}(z)N(z)^2g(z) dz 
      \left[ \int_{0}^{\infty} N(z) dz \right]^{-2},
 \label{eq:limber}
\end{equation}
where $F(z)$ 
\footnote{
Assuming that the clustering pattern 
is fixed in comoving coordinates in the
redshift range of our sample, 
we take the functional form,
$F(z)={(1+z)/(1+z_c)}^{-(3+\epsilon)}$,
where $z_c$ is the central redshift of the sample
and $\epsilon=-1.2$. 
The effect of the change in $\epsilon$ over
$0<\epsilon<-3$ on $r_0$ is, however, very small.
}
describes the redshift dependence of $\xi(r)$, 
$D_\theta(z)$ is the angular diameter distance, 
$N(z)$ is the redshift distribution of objects,
$
g(z)=H_0/c[(1+z)^2(1+\Omega_m z 
    + \Omega_\Lambda((1+z)^{-2}-1))^{1/2}],
$
and $k$ is a numerical constant, 
$k=\sqrt{\pi} \Gamma[(\gamma -1)/2]/\Gamma(\gamma/2)$.
The slope, $\beta$, of the angular correlation function
is related to $\gamma$ by 
\begin{equation}
\gamma=\beta+1.
\label{eq:gamma}
\end{equation}
We adopt the redshift distribution of the LAEs
calculated with the simulation output sources 
(i.e. referred to as simulation-based random sources 
in our Monte Carlo simulations; \S \ref{sec:spatial_angular}).
The correlation lengths thus obtained are summarized in
Table \ref{tab:clustering_measurements}. These correlation
lengths are expressed with $h_{100}$, the Hubble constant in
units of 100 km s$^{-1}$, instead of 70 km s$^{-1}$, to facilitate
comparison with previous results.
In Table \ref{tab:clustering_measurements},
we present $r_0$ and its upper limit, $r_0^{\rm max}$, that
are the correlation lengths obtained from
$A_\omega$ and its maximal value, $A_\omega^{\rm max}$, 
respectively. The correlation lengths are typically
$1-2$ $h_{100}^{-1}$ Mpc at $z\sim 3$ and
$2-6$ $h_{100}^{-1}$ Mpc at $z=4-6$.
Our estimates of $r_0$ are 
consistent with the previous measurements.
\citet{ouchi2003} and \citet{kovac2007}
obtain ($r_0$, $r_0^{\rm max}$) of ($3.5\pm 0.3$, $6.2\pm 0.5$) $h_{100}^{-1}$ Mpc
at $z=4.9$ and ($3.20\pm 0.42$, $4.61\pm 0.60$) $h_{100}^{-1}$ Mpc
at $z=4.5$, respectively. 
\citet{gawiser2007} and \citet{guaita2010} estimate $r_0$ to be 
$2.5^{+0.6}_{-0.7}$ $h_{100}^{-1}$ Mpc at $z=3.1$ 
and $3.4\pm 0.6$ $h_{100}^{-1}$ Mpc at $z=2.1$, respectively,
where $r_0$ is converted to values in units of $h_{100}^{-1}$ Mpc.
At $z=3.1$, the value of \citet{gawiser2007}
is slightly higher than that of our $z=3.1$ LAEs,
but these are still consistent within $1\sigma$ errors.
The correlation length of our $z=6.6$ LAEs
is $r_0=2-5$ $h_{100}^{-1}$ Mpc, while that for $r_0^{\rm max}$
is $3-7$ $h_{100}^{-1}$ Mpc, as presented in
Table \ref{tab:clustering_measurements}.

Note that these correlation lengths
are estimated by the Limber equation of 
eq. (\ref{eq:limber}). \citet{simon2007}
claims that the Limber equation
is accurate for small galaxy separation
but breaks down beyond a
certain separation that mainly depends on
a width of galaxy distribution and
distance. In this sense, eq. (\ref{eq:limber}) is 
regard as Limber approximation that is valid only 
in a small angular separation.
\citet{simon2007} presents the exact equation
connecting the spatial and angular correlation functions, 
which is hereafter referred as
\citeauthor{simon2007}'s equation.
On the other hand, \citet{sobral2010} have demonstrated that 
there is only a small, $\simeq 4$\%-level, difference
between the correlation lengths calculated
with Limber's and \citeauthor{simon2007}'s equations
based on their NB-selected sample of H$\alpha$ emitters at $z=0.8$.
Although their survey area is similar to that of our study, 
it is not clear whether the Limber approximation 
gives only a negligible impact
on our results at the different redshift of $z=6.6$.
We have compared ACFs calculated with Limber's and
\citeauthor{simon2007}'s equations 
in the same manner as shown in Figure 4 of \citet{sobral2010}. 
We have found that 
there are no apparent differences between two results up to $\theta \sim 200"$.
We have used our data of $10"-1000"$ 
for our power-law fitting that is mostly not affected by
the bias of Limber approximation. 
Moreover, the ACFs of Limber's and 
\citeauthor{simon2007}'s equations 
differ only by a factor of $\sim 2$ even at $\theta$ near $1000"$,
the largest angular separation of our fitting
where the error of our data point is
comparably large. Thus, the $r_0$ values change 
negligibly small for our galaxies at higher redshifts,
compared with their relatively large statistical errors.

We define galaxy-dark matter bias 
in the framework of $\Lambda$CDM model by
\begin{equation}
b_{\rm g}(\theta)^2
 \equiv
  \omega(\theta)/\omega_{\rm dm}(\theta),
\label{eq:bias}
\end{equation}
where $\omega_{\rm dm}(\theta)$ is the ACF predicted 
by the non-linear model \citep{peacock1996}
at the cosmic volume defined with $N(z)$
for each sample.
Note that $\omega_{\rm dm}(\theta)$ is defined
in the same survey volume as our LAE sample,
and that this bias value is equivalent to bias given by
the original definition, $b_{\rm g}^2=\xi(r)/\xi_{\rm dm}(r)$,
where $\xi_{\rm dm}(r)$ is the spatial correlation function
of dark matter.
Top panels of Figures 
\ref{fig:acorr_z3_6comb_all_final}-\ref{fig:acorr_NB921_all_final}
plot $\omega_{\rm dm}(\theta)$, and 
the bottom panels of Figures 
\ref{fig:acorr_z3_6comb_all_final}-\ref{fig:acorr_NB921_all_final}
show $b_{\rm g}(\theta)$.
We obtain an average bias, $b_{\rm g}$, that
is a value of $b_{\rm g}(\theta)$ averaged
over the same angular range 
as the one for the $A_{\omega}$ fitting 
(\S \ref{sec:spatial_angular})
with error weights.
The upper limits of bias values, $b_{\rm g}^{\rm max}$, 
are also calculated with $A_{\omega}^{\rm max}$
for the maximal contamination
correction case of eq. (\ref{eq:clustering_dilution}).
The estimates of $b_{\rm g}$ and $b_{\rm g}^{\rm max}$ are presented in 
Table \ref{tab:bias_darkhalo}.
For our $z=6.6$ LAEs, 
we obtain $b_{\rm g}=3-6$ and $b_{\rm g}^{\rm max}=5-9$.

\begin{figure}
%\epsscale{0.45}
\epsscale{1.25}
\plotone{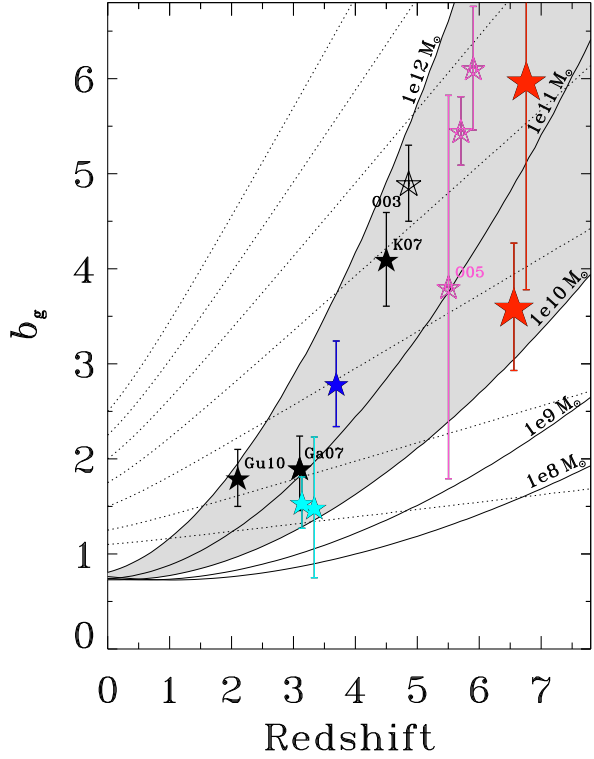}
\caption{
Bias of LAEs at $z=2-7$.
Red stars present our LAEs at $z=6.6$
with the bright (top) and
all (bottom) samples. Cyan, blue,
magenta stars plot bias of SXDS LAEs 
at $z=3.1$, $3.7$, and $5.7$, respectively,
estimated with the samples of \citet{ouchi2008}.
Small offsets are given to these points
in the direction of redshift to avoid overlapping 
symbols.
Right and Left cyan stars denote the bright
and all samples at $z=3.1$, respectively. 
For $z=5.7$ LAEs, we show the bias of
the bright LAE sample at right, all LAE sample
in the middle, and the similar sample but 
estimated by \citet{ouchi2005a} (O05)
with the count-in-cell technique
in a large scale of 20-Mpc radius
at left.
Black stars represent estimates
given by 
\citet{guaita2010} (Gu10; $z=2.1$),
\citet{gawiser2007} (Ga07; $z=3.1$),
\citet{kovac2007} (K07; $z=4.5$), and
\citet{ouchi2003} (O03; $z=4.9$).
Because the fields of the SXDS LAEs at $z=5.7$
and \citet{ouchi2003} have a strong large-scale clustering
with proto-cluster candidates 
made of LAE overdensity, we use open-star symbols
to indicate the existence of these structures
that may boost the bias measurements. Solid lines
denote bias as a function of redshift for dark halos
with a mass of $10^{8}$, $10^{9}$, $10^{10}$, $10^{11}$,
and $10^{12} M_\odot$ predicted by the \citet{sheth1999} model
in the case of one-to-one correspondence between galaxies and dark halos.
Gray area denotes the dark halo mass range of $10^{10}-10^{12} M_\odot$
where LAEs at $z=2-7$ falls.
Dotted lines represent evolutionary tracks of bias under
the galaxy-conserving model (eq. \ref{eq:galaxy_conserving}).
This plot assumes $\sigma_8=0.8$.
We apply a correction to
all of the previous estimates
that assume $\sigma_8=0.9$,
except for the one of \citet{guaita2010}.
\label{fig:redshift_bias_LBGLAE}}
\end{figure}

\begin{figure*}
%\epsscale{0.9}
\epsscale{1.15}
\plotone{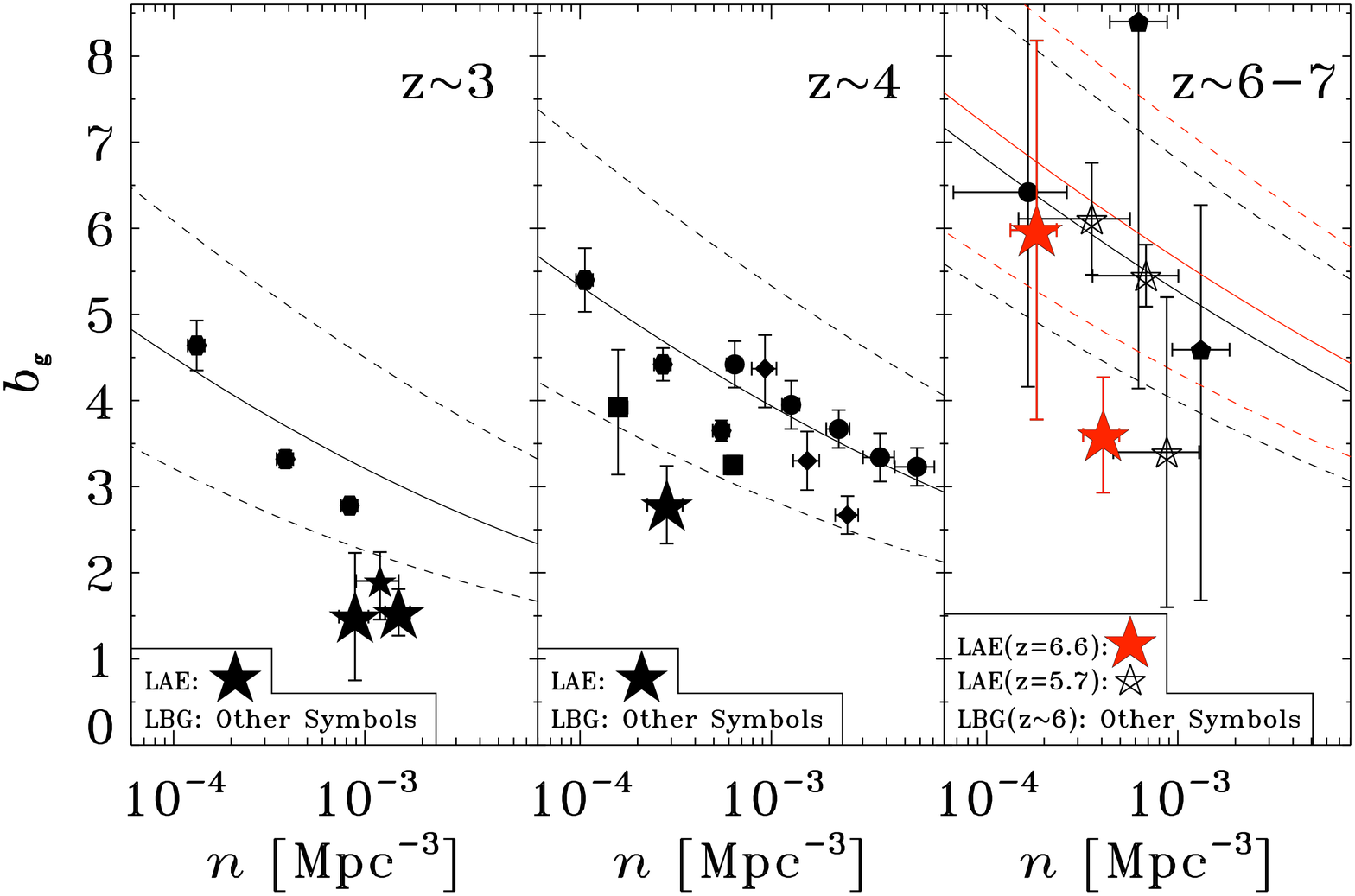}
\caption{
Bias as a function of number density
at $z\sim 3$ (left), $z\sim 4$ (center),
and $z\sim 6-7$ (right). Star marks represent
LAEs, while the other symbols denote LBGs
(including dropout galaxies). Specifically, the red stars indicate
our LAEs at $z=6.6$. The open star marks in the right panel
are $z=5.7$ SXDS LAEs whose bias values might be boosted
by the large-scale clustering with proto-clusters \citep{ouchi2005a}. 
All of
the star marks are our estimates from samples
of this study and \citeauthor{ouchi2008}'s (\citeyear{ouchi2008}) studies, 
except the small star mark in the left panel that
is given by \citet{gawiser2007}. 
Because our estimates of LAEs overlap with that of
\citet{gawiser2007}, we slightly
shift their point in the number density direction
for clarification.
We plot the bias and number densities of LBGs
with squares ($z\simeq 4$; \citealt{ouchi2004b}),
diamonds ($z\simeq 4$; \citealt{lee2006}),
pentagons ($z\simeq 6$; \citealt{overzier2006}),
hexagons ($z\simeq 3.3$ and $3.8$; \citealt{hildebrandt2009}),
and circles ($z\simeq 4$ \citealt{ouchi2005b} and $z\simeq 6$; 
\citet{ouchi2005c}). Black solid lines
indicate the relation of bias and number density for
the case of one-to-one correspondence between galaxies and dark halos
\citep{sheth1999} at $z=3$, $4$, and $6$, 
in the left, center, and right panels,
respectively. Black dashed lines 
present the same relation but their number densities
are multiplied by $1/10$ and $10$ from bottom to top.
Red solid and dashed lines in the right panel 
are the same, but for $z=6.6$, the redshift of our LAE samples.
This plot assumes $\sigma_8=0.8$.
Because all of the previous bias estimates here,
are obtained with a cosmological
parameter of $\sigma_8=0.9$,
we apply the correction to the previous estimates
for the different $\sigma_8$ values.
\label{fig:numdensity_bias_LBGLAE_z3z4z67}}
\end{figure*}

\subsection{Hosting Dark Halo}
\label{sec:halo_mass}

We estimate dark halo masses of LAEs
with bias values obtained 
in Section \ref{sec:spatial_correlation_bias}
in the framework of $\Lambda$CDM model.
Because, as discussed in \S \ref{sec:evolution_of_clustering},
we find that clustering of $z=6.6$ LAEs
is negligibly affected by cosmic reionization,
compared with the size of errors,
bias of $z=6.6$ LAEs mostly depends
on properties of hosting dark halos.

Generally, the average dark halo mass, $M_{\rm h}$, and 
the mean galaxy bias, $b_{\rm g}$,
are given by
\begin{eqnarray}
M_{\rm h} & = & \frac{\int_{M_{\rm h}^{\rm min}}^{\infty}\ M\ f_{\rm duty}^{\rm LAE}(M)\ N_{\rm g}(M)\ n(M)\ dM}{\int_{M_{\rm h}^{\rm min}}^{\infty}\ f_{\rm duty}^{\rm LAE}(M)\ N_{\rm g}(M)\ n(M)\ dM}\ \ \ {\rm and}
\label{eq:average_DMmass} \\
b_{\rm g} & = & \frac{\int_{M_{\rm h}^{\rm min}}^{\infty}\ b(M)\ f_{\rm duty}^{\rm LAE}(M)\ N_{\rm g}(M)\ n(M)\ dM}{\int_{M_{\rm h}^{\rm min}}^{\infty}\ f_{\rm duty}^{\rm LAE}(M)\ N_{\rm g}(M)\ n(M)\ dM},
\label{eq:average_bias}
\end{eqnarray}
where $n(M)$ and $b(M)$ are number density and bias of dark halos, respectively,
with a mass of $M$, and $N_{\rm g}(M)$ is 
a galaxy occupation function at the 
halo mass, $M$. 
$f_{\rm duty}^{\rm LAE}(M)$ is 
a fraction of dark halos hosting LAEs
to those hosting any galaxies, 
which we refer to as duty cycle of LAEs.
$M_{\rm h}^{\rm min}$ is the minimum dark halo mass 
that can host a galaxy. To obtain
$n(M)$ and $b(M)$, we apply the model of \citet{sheth1999}
(see also \citealt{mo2002}),
\begin{equation}
n(M)dM = A(1+\frac{1}{\nu'^{2q}})\sqrt{\frac{2}{\pi}}
\frac{\bar{\rho_0}}{M} \frac{d\nu'}{dM} \exp \left(-\frac{\nu'^2}{2}\right)dM,
\label{eq:mass_func_shethtormen}
\end{equation}
where $\nu'=\sqrt{a}\nu$, $a=0.707$, $A\simeq 0.322$ and $q=0.3$.
$\bar{\rho_0}$ is the current 
mean density of the Universe.
Here, $\nu$ is a function of the growth factor, $D(z)$, 
and the rms of the density fluctuations on mass scale $M$, $\sigma (M)$,
as defined by
\begin{equation}
\nu\equiv \frac{\delta_c}{D(z)\sigma(M)},
\label{eq:nu}
\end{equation}
where $\delta_c=1.69$ represents the critical amplitude
of the perturbation for collapse. 
We calculate the growth factor, $D(z)$,
at redshift, $z$, following \citet{carroll1992}.
Again the power spectrum of the density fluctuations is
calculated with the transfer function of \citet{bardeen1986}.
The bias of dark halos is estimated by
\begin{equation}
b = 1+\frac{1}{\delta_c} \left[\nu'^2+b\nu'^{2(1-c)}-
\frac{\nu'^{2c}/\sqrt{a}}{\nu'^{2c}+b(1-c)(1-c/2)} \right],
\label{eq:bias_sheth}
\end{equation}
where $b=0.5$, $c=0.6$, and the other parameters
($\nu'$, $\delta_c$, and $a$) are the same as
those in eqs. (\ref{eq:mass_func_shethtormen}) 
and (\ref{eq:nu}).

Note that our clustering measurements do not have
accuracies high enough to determine the galaxy occupation function
which is often described with three free parameters
(see e.g. \citealt{hamana2004,ouchi2005b,lee2006,kovac2007}).
Different choices of galaxy occupation functions
change average halo masses only by a factor of a few, 
but not over an order of magnitude for LAEs \citep{hamana2004},
unless extremely unphysical parameter sets are taken.
Because we aim to accomplish an accuracy of an order of magnitude
for halo properties inferred from clustering,
we assume $N_{\rm g}(M)=1$, a one-to-one correspondence 
between galaxies and dark halos. We also assume that 
duty cycle of LAEs is mass independent for simplicity
(see more discussions about duty cycle in \S \ref{sec:duty_cycle}).
In this case, $M_{\rm h}$ and $b_{\rm g}$ are irrespective of
$f_{\rm duty}^{\rm LAE}$. We determine $M_{\rm h}^{\rm min}$
with our observational estimates of LAE bias, $b_{\rm g}$, via
eq. (\ref{eq:average_bias}), and calculate
the average dark halo mass, $M_{\rm h}$, from the $M_{\rm h}^{\rm min}$ values
with eq. (\ref{eq:average_DMmass}).
Table \ref{tab:bias_darkhalo} presents $M_{\rm h}$ and $M_{\rm h}^{\rm min}$
of our LAEs, together with the maximum of average halo mass, $M_{\rm h}^{\rm max}$,
estimated with the maximally contamination corrected bias of $b^{\rm max}$.

Figure \ref{fig:redshift_bias_LBGLAE} plots 
our $b_{\rm g}$ values as a function of redshift. 
This figure also compares $b_{\rm g}$ of LAEs obtained
in the previous studies. 
Note that in our study
we assume $\sigma_8=0.8$ that is consistent with the latest WMAP results
\citep{komatsu2009,larson2010}.
Because all of the previous
studies, except \citet{guaita2010},
assume $\sigma_8=0.9$, 
we multiply
the bias values of these previous results
by 1.1 to correct for the difference of $\sigma_8$.
This correction factor of 1.1, irrespective of
redshift and selection function, is estimated
from $\sqrt{\omega_{\rm dm}(\sigma_8=0.9)/\omega_{\rm dm}(\sigma_8=0.8)}$
with the model of \citet{peacock1996}, where $\omega_{\rm dm}(\sigma_8=0.9)$
and $\omega_{\rm dm}(\sigma_8=0.8)$ are ACFs predicted 
by the non-linear model under $\sigma_8=0.9$ and $0.8$,
respectively.
Figure \ref{fig:redshift_bias_LBGLAE} shows that
our all and bright subsample measurements
($z=3.1$ and $6.6$) agree within 
$1\sigma$ errors at each redshift. 
This indicates that a clear luminosity segregation
cannot be found beyond the size of our relatively
large error bars for samples whose luminosity
differs only by a factor of $\simeq 1.6$
(see Tables \ref{tab:clustering_measurements}-\ref{tab:bias_darkhalo}).
The LAE samples of the previous studies
and ours have similar Ly$\alpha$ luminosity limits 
($L_{\rm Ly\alpha}\simeq 2\times 10^{42}$ erg s$^{-1}$)
only with small differences by a factor of 2, and 
the effect of luminosity dependence is probably
smaller than the given relatively large error bars.
Thus, we consider that the luminosity dependence
between the LAE samples is negligibly small in 
Figure \ref{fig:redshift_bias_LBGLAE} 
(see also the arguments for the relation between bias and number
density described below).
Figure \ref{fig:redshift_bias_LBGLAE} shows
that our measurements at $z=3.1$ are consistent
with the one at the same redshift 
given by \citet{gawiser2007}.
The bias of LAEs appear to increase from $z=2-3$
to $6.6$. Because there are strong large-scale clustering
with proto-clusters or proto-cluster candidates reported in
the data of $z=4.9$ in the SDF \citep{shimasaku2003} 
and $z=5.7$ in the SXDS \citep{ouchi2005a}.
Figure \ref{fig:redshift_bias_LBGLAE} plots 
open star marks for the bias estimates
at $z=4.9$ in the SDF \citep{ouchi2003}
and $z=5.7$ in the SXDS \citep{ouchi2005a}
to indicate that these bias estimates
might be boosted by these obvious cosmic structures.
Even if we omit the points of these open star marks
in Figure \ref{fig:redshift_bias_LBGLAE}, 
we confirm the trend of increasing bias 
toward high-$z$. To understand this increase,
we plot the \citeauthor{sheth1999}'s (\citeyear{sheth1999}) model of 
a constant dark halo mass under the assumption of
one-to-one correspondence 
between galaxies and dark halos. 
The dark halo masses
of our $z=6.6$ LAEs are about $10^{10}-10^{11}M_\odot$
in Figure \ref{fig:redshift_bias_LBGLAE}, which are
the same as those listed in Table \ref{tab:bias_darkhalo}.
At $z=2-7$, the bias values fall 
in the range of $\simeq 10^{10}-10^{12}M_\odot$ for dark halo masses. 
Thus, the average dark halo mass of LAEs is roughly
$\sim 10^{11\pm 1}M_\odot$ at $z=2-7$. The dark halo masses of LAEs show
no significant evolution at $z=2-7$ 
beyond the mass-estimate scattering by an order of magnitude.
It would indicate that LAEs, observed so far, 
are galaxies at the evolutionary stage
for all or some type of galaxies 
whose dark halos have reached a mass of
$\sim 10^{11\pm 1}M_\odot$.

Figure \ref{fig:numdensity_bias_LBGLAE_z3z4z67}
plots bias and number density of LAEs,
and compares those with Lyman break galaxies (LBGs) 
including $i$-dropouts.
The solid lines present the relation
of bias and number density for all dark halos,
which correspond to the case of one-to-one correspondence
between galaxies and dark halos. The dashed lines
are the same as the solid lines but with their number densities 
multiplied by $10$ and $1/10$.
The bias values of LAEs are generally smaller
than those of LBGs at redshifts over $3-7$.
One exception is LAEs at $z=5.7$ that have a bias value 
as large as those of $z\sim 6$ LBGs. However,
bias measurements of these LAEs could be
boosted by strong large-scale clustering with proto-clusters 
as discussed above.
On the other hand, LBGs have a roughly
one-to-one correspondence
between dark halos and galaxies within 
a scatter of one-order of magnitude 
in number density. Comparing with LBGs,
we find that LAEs have significantly less
number density and/or bias. This weak bias
at the small number density for LAEs
indicates that LAEs reside in
less massive dark halos on average, and that
not all but some of less massive dark halos
can host LAEs. Because it is unlikely that
some dark halos always have LAEs and 
the other dark halos with the same mass 
never host LAEs, dark halos become 
the ones hosting LAEs by duty cycle,
stochastic processes made by
star-formation history and changes of
interstellar medium geometry and dynamics \citep{nagamine2008}.

\subsection{Duty Cycle}
\label{sec:duty_cycle}

In Figure \ref{fig:numdensity_bias_LBGLAE_z3z4z67},
the simplest estimate of duty cycle can be obtained from the
ratio of the number density (ordinate of Figure \ref{fig:numdensity_bias_LBGLAE_z3z4z67}) 
of all halos (solid line) to that of LAEs (star marks).
However, due to a mild gradient of the solid curves, especially
in the small bias range, even very small uncertainties of bias
estimates of LAEs give very different duty cycle values.
By this reason, a determination
of duty cycle requires a precision measurement of bias of LAEs.
Nevertheless, in the following subsection, we try to
constrain the duty cycle of LAEs with the given bias
measurements and errors of the present study.

The duty cycle of LAEs can be constrained with
a galaxy number density, $n_{\rm g}$,
\begin{equation}
n_{\rm g}  = \int_{M_{\rm h}^{\rm min}}^{\infty}\ f_{\rm duty}^{\rm LAE}(M)\ N_{\rm g}(M)\ n(M)\ dM.
\label{eq:galaxy_number_density}
\end{equation}
Following the assumptions, $N_{\rm g}(M)=1$ and constant $f_{\rm duty}^{\rm LAE}$,
made in \S \ref{sec:halo_mass}, 
the equation of eq. (\ref{eq:galaxy_number_density})
can be expressed as 
$f_{\rm duty}^{\rm LAE}=n_{\rm g}/\int_{M_{\rm h}^{\rm min}}^{\infty}\ n(M)\ dM$.
We estimate $f_{\rm duty}^{\rm LAE}$ with $M_{\rm h}^{\rm min}$ values
and the observed number densities of LAEs (Table \ref{tab:bias_darkhalo})
to be $f_{\rm duty}^{\rm LAE}=(0.008\pm 0.03, 0.04\pm 0.1, 1\pm 1, 0.009\pm 0.04)$
for our all LAEs at $z=(3.1,3.7,5.7,6.6)$.
Note that errors of $f_{\rm duty}^{\rm LAE}$ are raised 
by the combination of $M_{\rm h}^{\rm min}$ and $n_{\rm g}$ estimates.
The uncertainties of our $f_{\rm duty}^{\rm LAE}$ estimates 
are quite large as discussed above.
Especially, the $f_{\rm duty}^{\rm LAE}$ value of $z=5.7$ provides 
no meaningful constraints, because its error of $n_{\rm g}$ 
is larger than the others due to a relatively 
large cosmic variance observed in our survey volume
(see the caption of Figure \ref{fig:contour_schechter_LAE}).
Because of these large uncertainties, we cannot
constrain the evolution of duty cycle. 
However, these $f_{\rm duty}^{\rm LAE}$ estimates indicate that 
$f_{\rm duty}^{\rm LAE}$ is a few $0.1$ to a few percent, 
roughly $\sim 1$\%.

By definition, the duty cycle of LAEs, $f_{\rm duty}^{\rm LAE}$, 
is a fraction of galaxies that are LAEs. 
It can be expressed by the product, 
$f_{\rm duty}^{\rm LAE}=f_{\rm duty}^{\rm SF} \times f_{\rm duty}^{\rm Ly\alpha}$,
where $f_{\rm duty}^{\rm SF}$ is a duty cycle of star-forming activities
in all dark halos, and $f_{\rm duty}^{\rm Ly\alpha}$ is the one for Ly$\alpha$
emitting galaxies among star-forming galaxies.
\citet{ouchi2004b} estimate $f_{\rm duty}^{\rm SF}$
of LBGs, UV-bright star-forming galaxies, 
to be $0.1-1$
\footnote{
These numbers are given in the sixth column
in Table 6 of \citet{ouchi2004b}.
}
from their clustering analyses, 
and a fainter galaxy have a smaller
$f_{\rm duty}^{\rm SF}$ of $0.2$ at $i\simeq 25.5$.
Similarly, \citet{lee2009} estimate $f_{\rm duty}^{\rm SF}$
to be 0.15-0.6 for their LBGs with $z'<27$ \citep{lee2006}.
Because our LAEs have a UV continuum magnitude
of $\sim 27$ mag on average \citep{ono2010a,ono2010b}, 
results of \citet{ouchi2004b} and \citet{lee2009}
suggest that $f_{\rm duty}^{\rm SF}$ is $\lesssim 0.2$
for LBGs whose UV luminosity is comparable to our LAEs.
On the other hand, \citet{stark2010} obtain 
$f_{\rm duty}^{\rm Ly\alpha}\sim 0.5$ at $\sim 27$ mag,
which is a fraction of Ly$\alpha$ emitting dropouts
with $EW_0>50$\AA\ to all LBGs at $z\simeq 3-6$.
Combining these previous results, we find
$f_{\rm duty}^{\rm LAE}=f_{\rm duty}^{\rm SF} \times f_{\rm duty}^{\rm Ly\alpha}\lesssim 0.1$,
which is consistent with our estimates.
\citet{hayes2010} claim $f_{\rm duty}^{\rm Ly\alpha}=0.05$
from the comparison of Ly$\alpha$ and H$\alpha$ emitting galaxies
based on Ly$\alpha$ and H$\alpha$ luminosity functions. 
If we naively assume that $f_{\rm duty}^{\rm SF}$ of H$\alpha$ emitting galaxies
is the same as that of LBGs, i.e. $f_{\rm duty}^{\rm SF} \lesssim 0.2$,
we obtain $f_{\rm duty}^{\rm LAE} \lesssim 0.01 (=0.2\times 0.05)$,
which is roughly consistent with our result.

\begin{figure}
%\epsscale{0.95}
\epsscale{1.15}
\plotone{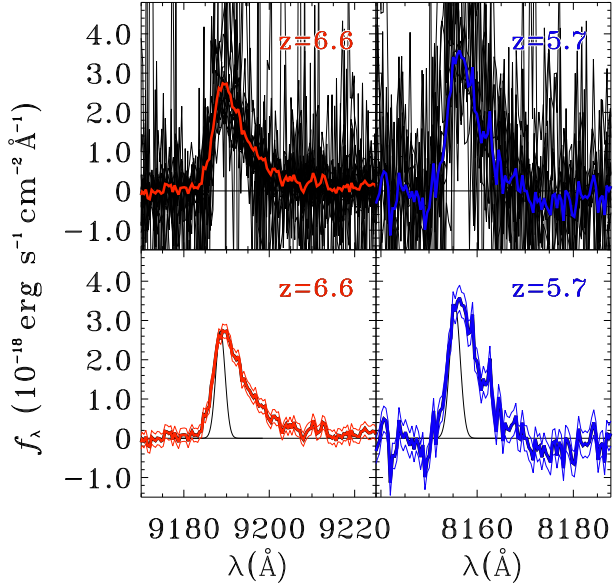}
\caption{
Ly$\alpha$ lines of LAEs.
{\it Top left:} Red line represents the
composite spectrum of our $z=6.6$ LAEs.
Black lines denote our 19 DEIMOS spectra
whose Ly$\alpha$ lines are aligned to
the average redshift and scaled to
the flux of the composite spectrum.
{\it Bottom left:} The composite spectrum
of our $z=6.6$ LAEs is shown with
a thick red line. One sigma error of 
the composite spectra are shown 
with thin red lines.
Black curve presents
the instrumental resolution of our DEIMOS
spectra whose shape is approximated with
a Gaussian profile.
{\it Top right:} Same as the top left panel,
but for the $z=5.7$ LAEs. Blue line
is the composite
spectrum of the $z=5.7$ LAEs.
{\it Bottom right:} Same as the bottom left panel,
but for the $z=5.7$ LAEs.
Blue lines are used for $z=5.7$ LAEs.
\label{fig:line_profile_all}}
\end{figure}

\section{Ly$\alpha$ Line Profile}
\label{sec:line_profile}

We investigate Ly$\alpha$ line profiles
of our $z=6.6$ LAEs, exploiting
our high quality medium-high
resolution spectra of Keck/DEIMOS.
In this analysis, we add 
the 3 duplicate spectra
to the 16 spectra (\S \ref{sec:spectroscopic_confirmation}), 
and use a total of 19 DEIMOS spectra
for $z=6.6$ LAEs.  
In order to study the evolution of Ly$\alpha$ 
line profiles between $z=5.7$
and $6.6$, 
we also perform analysis of 11 DEIMOS
spectra of $z=5.7$ LAEs.
Ten out of the 11 DEIMOS spectra are taken
in the SDF by \citet{shimasaku2006} with
exactly the same instrumental configuration as 
that of our $z=6.6$ spectra, 
and the remaining one DEIMOS spectrum is
obtained for the SXDS $z=5.7$ LAE sample \citep{ouchi2008}
as mask fillers of our DEIMOS 
observations on 2008 October 3
(\S \ref{sec:spectroscopic_confirmation}).
The top panels of Figure \ref{fig:line_profile_all}
present all of the DEIMOS spectra at $z=6.6$ and $5.7$.

\subsection{Composite Spectra and Ly$\alpha$ Velocity Widths}
\label{sec:composite_spectra}

We make composite spectra of our 19 spectra of $z=6.6$ LAEs,
and the 11 spectra of $z=5.7$ LAEs to investigate
the average Ly$\alpha$ line profiles.
The line centers of the spectra 
are aligned to the wavelengths of 
9190\AA\ and 8157\AA\ that
are average ones of the $z=6.6$ and $5.7$ samples,
respectively.
The line fluxes are normalized to
average Ly$\alpha$ fluxes,
since we calculate rejected-mean
and median spectra for our stacking.
In this process, we remove
lowest-quality five and two spectra
from the $z=6.6$ and $5.7$ LAE spectra,
respectively, so as to avoid 
possible systematic bias
given by uncertainties of 
line center determination
and/or flux normalization.
We compute rejected-mean spectra 
weighted with S/N ratios of Ly$\alpha$ lines,
in the same manner as \citet{ouchi2008}.
At each wavelength bin, the rejection is 
made for data points that fall beyond 
68 percentile among all data points,
which corresponds to the rejection of 
two (one) data points (point) above and below 
the 68 percentile for the LAEs at $z=6.6$
($5.7$). By this rejection algorithm,
our composite spectra
are secure against systematic 
residuals of sky subtraction.
We compare these composite spectra
with the ones 
computed by simple median statistics, 
and check whether these rejection procedures 
introduce a systematic bias.
We find that the two composite spectra agree
very well and that the rejected-mean composite 
spectra do not show any sign of bias in the
overall shapes and velocity width of Ly$\alpha$ lines.
Because improvements of S/N are generally
better in the rejected-mean stacking
than in the simple median stacking,
we use the rejected-mean spectra
for our composite spectra in the following
discussion.
Figure \ref{fig:line_profile_all} plots 
our composite spectra
of $z=6.6$ and $5.7$ LAEs.
The quality of the spectrum of $z=5.7$ LAE
is not as good as that of $z=6.6$ LAE.
This is because 
i) the total number of
available $z=5.7$ spectra (11) is small and
ii) the average S/N ratio of individual $z=5.7$ spectra
is relatively low due to their short ($\sim 2$ hr)
integration time \citep{shimasaku2006}.
The instrumental spectral resolution of 
the DEIMOS data are presented 
in the bottom panels of 
Figure \ref{fig:line_profile_all},
which illustrate that our composite spectra are 
well resolved beyond the instrumental resolution.
Our composite spectra show a very clear
asymmetric profile because of the
high S/N ratio.

\begin{figure}
%\epsscale{1.0}
\epsscale{1.15}
\plotone{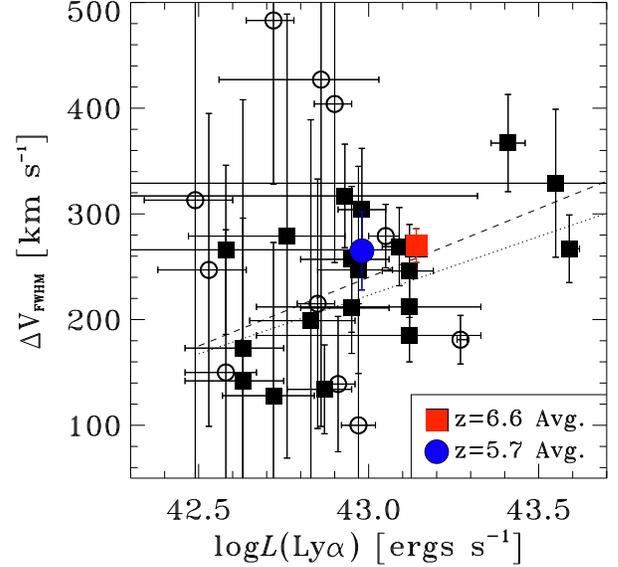}
\caption{
FWHM velocity width of our $z=6.6$ LAEs,
together with the $z=5.7$ LAEs.
Red square and blue circle denote
velocity widths of the composite
spectra. Black squares and open circles
are velocity widths measured for 
individual LAEs at $z=6.6$ and $5.7$,
respectively. Dotted and dashed lines
represent the best-fit linear functions
to the individual LAEs at $z=6.6$
with and without measurement errors
in the fitting, respectively.
These FWHM velocity widths are the ones
given by the Gaussian profile fitting, which
provides robust measurements
for relatively poor quality data
of individual spectra.
\label{fig:lya_vfhwm}}
\end{figure}

\begin{figure}
%\epsscale{1.0}
\epsscale{1.15}
\plotone{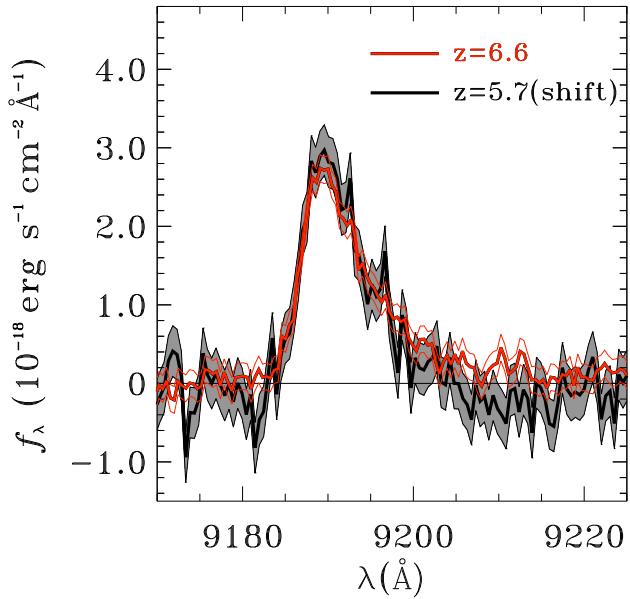}
\caption{
Evolution of LAE spectra.
Red and black thick lines represent
composite spectra of LAEs at $z=6.6$
and $5.7$, respectively. 
For comparison of line shapes, 
the composite spectrum of 
the $z=5.7$ LAEs is redshifted to
$z=6.6$, and scaled by an arbitrary factor.
One sigma errors of the composite spectra
are shown with thin red lines for
$z=6.6$ LAEs and a gray shade
for $z=5.7$ LAEs.
\label{fig:line_profile_evolution}}
\end{figure}

To quantify the Ly$\alpha$ line profiles,
we evaluate an FWHM velocity width, $\Delta V_{\rm FWHM}$.
We fit a Gaussian profile to the individual spectra
as well as the composite spectra, and obtain 
$\Delta V_{\rm FWHM}$. We correct for the instrumental
broadening of line profile, and estimate intrinsic
$\Delta V_{\rm FWHM}$ by 
$\Delta V_{\rm FWHM}=\sqrt{v_{\rm obs}^2 - v_{\rm inst}^2}$,
where $v_{\rm obs}$ and $v_{\rm inst}$ are
FWHM velocity widths for the measured Ly$\alpha$ lines
and the instrumental resolution, respectively.
We estimate errors of 
$\Delta V_{\rm FWHM}$ including uncertainties
raised by a choice of profile fitting range and 
Poisson statistics based on Monte-Carlo simulations.
Figure \ref{fig:lya_vfhwm} and Table \ref{tab:laes_with_redshifts} 
present the $\Delta V_{\rm FWHM}$ measured from the spectra
and total Ly$\alpha$ luminosity estimated
from the spectroscopic redshifts and the imaging data.
Figure \ref{fig:lya_vfhwm} indicates that
the $\Delta V_{\rm FWHM}$ values of $z=6.6$ LAEs
are distributed in the range of 100-400 km s$^{-1}$
at $\log L({\rm Ly\alpha})\simeq 42.6-43.6$.
The data points of $z=5.7$ LAEs are
similarly distributed,
but with a larger scatter given by the larger
errors than those of $z=6.6$ LAEs.
Our composite spectra have the FWHM velocity widths of
$270\pm 16$ and $265\pm 37$ km s$^{-1}$ at 
$z=6.6$ and $5.7$, respectively.
We also evaluate additional possible errors of 
$\Delta V_{\rm FWHM}$ introduced by
uncertainties of line-center alignment in 
the process of spectrum stacking. 
We find that the typical line-center determinations
are as good as 0.32\AA\ and 0.90\AA\ for
$z=6.6$ and $5.7$ LAE spectra, respectively, 
and that the additional errors of line-center alignment
contributing to $\Delta V_{\rm FWHM}$ 
are negligibly small, only a 0.1-1\% level.
We compare $\Delta V_{\rm FWHM}$ of the $z=6.6$ and $5.7$
composite spectra in Figure \ref{fig:lya_vfhwm}.
We find that $\Delta V_{\rm FWHM}$ does not evolve
from $z=6.6$ to $5.7$ beyond the $1\sigma$ level.
Because the Ly$\alpha$ luminosities of our composite
spectra are almost the same, $\log L({\rm Ly\alpha})\simeq 43.0$,
this comparison includes little bias of
luminosity dependence.

Note that these $\Delta V_{\rm FWHM}$ are estimated by
Gaussian profile fitting. The real Ly$\alpha$ profile
is not Gaussian, but asymmetric,
as clearly found in our composite spectra.
Although one cannot obtain $\Delta V_{\rm FWHM}$
without fitting of a profile, such as Gaussian,
for our relatively poor S/N data of individual spectra,
$\Delta V_{\rm FWHM}$ can be directly measured
at least for the high quality composite spectra.
The direct measurements of FWHM velocity widths
corrected for the instrumental broadening are 
$251$ and $260$ km s$^{-1}$ for the $z=6.6$ and $5.7$ composite
spectra, respectively.
Table \ref{tab:average_lya_line_profile} 
summarizes the properties of our composite spectra,
and indicates that the FWHM velocity widths of the direct measurements 
are comparable to those of the Gaussian fitting
within the errors.

Figure \ref{fig:line_profile_evolution} compares 
the composite spectra of $z=6.6$ and $5.7$ LAEs.
The composite spectrum of $z=6.6$ LAEs
is similar to the one of $z=5.7$ LAEs within the
errors. We find no large evolution 
beyond the $1\sigma$ errors of our measurements.
In fact, as shown in Table \ref{tab:average_lya_line_profile}, 
the direct (Gaussian) FWHM velocity widths 
obtained above are $251$ $(270)$ $\pm 16$ km s$^{-1}$
at $z=6.6$ and $260$ $(265)$ $\pm 37$ km s$^{-1}$
at $z=5.7$, and there are no differences between
the $z=6.6$ and $5.7$ composite spectra beyond
the $1\sigma$ error
of $\simeq 40$ km s$^{-1}$,
which is dominated 
by the error of $z=5.7$ composite spectrum.
\citet{hu2006} also find 
no evolution of line profiles,
although there are no quantitative
comparisons in their study.
Our composite spectrum of $z=5.7$ has an S/N ratio lower
than that of $z=6.6$ by a factor of 2
in the FWHM measurements, which do not
allow us to investigate a slight profile change
from $z=6.6$ to $5.7$ that may exist.
If one compares the best-estimate spectra
of our $z=6.6$ and $5.7$ LAEs (thick red and
black lines, respectively, 
in Figure \ref{fig:line_profile_evolution}), 
the slope of red wing is found to 
be slightly sharper for $z=5.7$ LAEs 
than $z=6.6$ LAEs. This would be a hint
of flattening of Ly$\alpha$ line profiles
from $z=5.7$ to $6.6$.
A future study with better spectroscopic data 
would address the issue of this slight
evolutionary effect.

Interestingly, our composite
spectra indicate that Ly$\alpha$ emission
may not be simple smooth asymmetric lines,
but with a substructure.
Comparing our composite spectra with
the curves of instrumental spectral resolution
in the bottom panels of Figure \ref{fig:line_profile_all},
we find knees in the blue tail of the composite
LAE spectra at $\simeq 9185$ and $8152$\AA\ for 
$z=6.6$ and $5.7$, respectively,
both of which correspond to the same {\it rest-frame} 
wavelength about $\simeq 0.7$\AA\ bluer 
than their line peaks.
Since similar knees are not seen in our sky line spectra (e.g. 
the bottom right panel of Figure \ref{fig:image_spec_all_nb921}),
these knees are not made by DEIMOS's 
instrumental line profile. Another possibility
may be a residual of sky subtraction, but
there are no clear reasons why the residual
is mostly positively scattered in statistical sense
to make the knees.
The peaks of the knees are detected
at the $3.5$ and $2.2$ sigma levels for
LAEs at $z=6.6$ and $5.7$, respectively,
at the same rest-frame wavelength,
which is difficult to be explained by
the random errors
(see $1\sigma$ errors associated with the composite
spectra in Figure \ref{fig:line_profile_all}).
Thus, the knees in the blue wings of $z=6.6$ and $5.7$
spectra are probably real.

\begin{deluxetable*}{ccccc}
\tablecolumns{5}
\tabletypesize{\scriptsize}
\tablecaption{Average Ly$\alpha$ Line Properties
\label{tab:average_lya_line_profile}}
\tablewidth{0pt}
\tablehead{
\colhead{$\left < z \right >$} &
\colhead{$\left < L_{\rm Ly\alpha} \right >$} &
\colhead{$\Delta V_{\rm FWHM}$(direct)} &
\colhead{$\Delta V_{\rm FWHM}$(Gaussian)} &
\colhead{Error($\Delta V_{\rm FWHM}$)} \\
\colhead{} &
\colhead{($10^{43}$erg s$^{-1}$)} &
\colhead{(km s$^{-1}$)} &
\colhead{(km s$^{-1}$)} &
\colhead{(km s$^{-1}$)} \\
\colhead{(1)} &
\colhead{(2)} &
\colhead{(3)} &
\colhead{(4)} &
\colhead{(5)} 
}
\startdata
$6.56$ & $1.39 \pm 0.06$ & $251$ & $270$ & $16$ \\ 
$5.71$ & $0.96 \pm 0.03$ & $260$ & $265$ & $37$ \\ 
\enddata
\tablecomments{
(1): Average redshift and
(2): average Ly$\alpha$ luminosity
of our spectroscopic data. 
(3): Average FWHM velocity width given by the direct measurement,
(4): average FWHM velocity width determined by the Gaussian fitting,
and
(5): error of FWHM velocity width measurement.
}
\end{deluxetable*}

\subsection{Ly$\alpha$ Velocity-Luminosity Relation}
\label{sec:lya_velocity_luminosity_relation}

Figure \ref{fig:lya_vfhwm} presents a weak,
but a positive correlation
between $\Delta V_{\rm FWHM}$ and $\log L({\rm Ly\alpha})$
in our $z=6.6$ LAEs. A linear function fit to
the $z=6.6$ $\Delta V_{\rm FWHM}$-$\log L({\rm Ly\alpha})$ data
indicates that the positive correlation is
found at the $2.5\sigma$ level.
This trend is opposite to the one found
by \citet{kashikawa2006}. \citet{kashikawa2006}
study LAEs with $\log L({\rm Ly\alpha})\simeq 42.3-43.0$
that are fainter than ours by a factor of 2-4
on average. A distribution of data points of 
\citet{kashikawa2006} is similar to ours
in the luminosity range where both studies
have measurements. Figure 11 of \citet{kashikawa2006}
presents two faint LAEs that have large FWHMs,
which apparently make the anti-correlation.
If these two faint LAEs are largely up-scattered
by statistical errors or sample variance,
the anti-correlation is not clearly found.
Moreover, 
\citet{kashikawa2006} consider 
no velocity-width measurement errors in their linear-function fitting,
while the uncertainties of velocity-width
measurements increase towards faint luminosity.
Thus, the previous conclusion of the anti-correlation
is probably not strong.
On the other hand, in Figure \ref{fig:lya_vfhwm},
our positive correlation is apparent
when our data of three bright LAEs 
in $\log L({\rm Ly\alpha})\simeq 43.3-43.6$
are included. 
When these three data points are not used for
our fitting, neither of positive nor negative correlation
is identified beyond the $1\sigma$ level.
If these three bright LAEs are not typical ones 
due to a sample variance, 
the clear positive correlation would not be found.
Although there is a possibility that these three
LAEs are not typical due to sample variance,
these bright LAEs have small errors that are difficult to
produce largely up-scattered measurements.
We conclude that
there is no anti-correlation 
between Ly$\alpha$ luminosity and line width
at $z=6.6$, and that,
if our spectroscopic sample is not a biased one, 
Ly$\alpha$ velocity width 
positively correlates with Ly$\alpha$ luminosity
at $z=6.6$ in the luminosity range of
$\log L({\rm Ly\alpha})\simeq 42.6-43.6$.

\section{Discussion}
\label{sec:discussion}

\subsection{Constraints on the Cosmic Reionization History}
\label{sec:constraints_on_reionization}

In this subsection, we discuss implications 
for cosmic reionization based on our LF,
clustering, and Ly$\alpha$ line profile results 
with the aid of theoretical models. 
Because some conclusions of 
theoretical models would depend on their
assumptions and methods such
as analytical, semi-analytical, 
and numerical techniques including radiative transfer, 
we compare our observational
measurements with as many various theoretical models 
available to date as possible.
In this way, we aim to
obtain implications for cosmic reionization 
with less model dependencies.

\subsubsection{Evolution of Ly$\alpha$ LF and Luminosity Density}
\label{sec:evolution_of_lyalf}

\begin{figure}
%\epsscale{0.9}
\epsscale{1.15}
\plotone{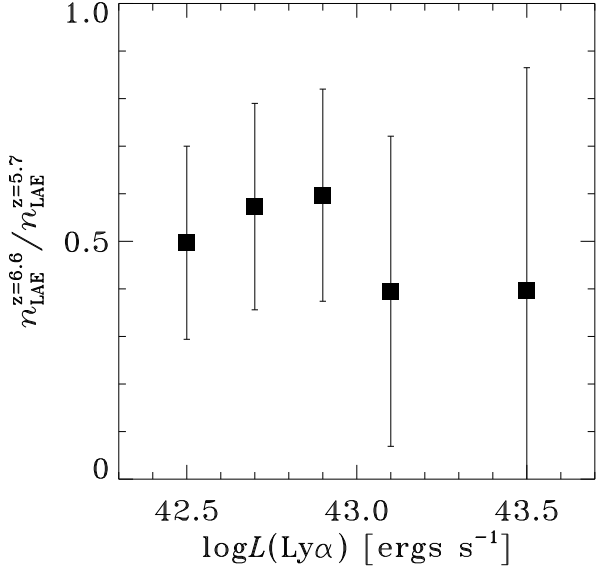}
\caption{
Ratio of number densities of $z=6.6$ LAEs to $z=5.7$ LAEs,
as a function of Ly$\alpha$ luminosity. 
Squares plot observational results from LAEs at $z=6.6$ (this study)
and $z=5.7$ \citep{ouchi2008}.
The data binning is the same as that in 
Figures \ref{fig:lumifun_full_diff_nb921_WithComparison}-\ref{fig:lumifun_full_diff_nb921_evolution}.
\label{fig:numberevolz66z57}}
\end{figure}

In Section \ref{sec:evolution_of_lya_LF},
we have found that Ly$\alpha$ LF decreases
from $z=5.7$ to $6.6$ at the $>90$\% confidence level,
and that the decrease is 30\% in luminosity
for the case of pure luminosity evolution.
Because Ly$\alpha$ LF evolution is made not only
by cosmic reionization but also by
galaxy evolution, we should interpret 
the decrease of Ly$\alpha$ LF carefully.
In fact, UV-continuum luminosity function
of dropout galaxies also decreases from 
$z=6$ to $7-8$ in observational data
(e.g. \citealt{ouchi2009b,bouwens2010a,castellano2010}).
Some fraction of Ly$\alpha$ LF decrease
may be explained by this galaxy evolution effect.
Figure \ref{fig:numberevolz66z57} presents a number density ratio
of $z=6.6$ to $5.7$ LAEs as a function of Ly$\alpha$ luminosity.
We find no significant dependence on luminosity 
for the ratio within errors,
i.e. by a factor of $\simeq 2$ at the luminosity range of $\log L=42.5-43.5$. 
However, it appears that the ratio is relatively smaller
at the bright luminosity ($\log L=43.0-43.5$) than 
the faint luminosity ($\log L=42.5-43.0$). Note that
Figure \ref{fig:numberevolz66z57} misses a measurement
at the bin of $\log L=43.3$, because no LAEs are found
at $z=6.6$ in this bin. If the trend of relatively small
number density of $z=6.6$ LAEs at the bright luminosity is true,
galaxy evolution may be more dominant on
the LF evolution between $z=5.7$ and $6.6$
than the cosmic reionization effect. Since brighter LAEs
are likely hosted by more massive dark halos
whose formation epoch is near these redshifts 
(see halo mass functions, e.g., \citealt{sheth1999}), 
evolution of number density of bright LAEs would 
be more affected by galaxy evolution. On the other hand,
simple models of LAEs suggest
that brighter LAEs have a less reduction of Ly$\alpha$ flux
due to large ionized bubbles surrounding bright LAEs \citep{haiman2002}, 
which would imply that evolution of bright LAEs
are milder than that of faint LAEs
in the reionization-effect dominant case.

We need to quantify how much decrease of Ly$\alpha$ LF
is contributed from cosmic reionization or galaxy evolution.
If we have an assumption of no galaxy formation effects
in Ly$\alpha$ LF evolution, which is made in
previous studies (e.g. \citealt{malhotra2004,kashikawa2006}),
a ratio of IGM's Ly$\alpha$ transmission at $z=6.6$
to the one at $z=5.7$ is 0.7. However, it is not clear
whether no galaxy evolution assumption is 
correct. UV LF evolution of LAEs 
between $z=5.7$ and $6.6$ 
could resolve the degeneracy
of Ly$\alpha$ LF evolution between cosmic reionization
and galaxy formation. Although \citet{kashikawa2006} claim that
there is no evolution of UV LF of LAEs between $z=5.7$
and $6.6$ based on their Ly$\alpha$-subtracted $z'$ photometry,
we have found that UV LF of LAEs cannot 
be derived from Ly$\alpha$-subtracted $z'$ photometry
with an accuracy better than a factor of 2-3,
due to large uncertainties. In fact, UV magnitudes estimated
with the Ly$\alpha$ subtracted $z'$ photometry
have large errors, $>0.4-0.7$, as shown 
in Table \ref{tab:laes_with_redshifts}. 
These large errors are raised, because fluxes 
in $z'$ band are dominated by the strong Ly$\alpha$ line, 
but not by the faint UV continuum. Only with optical $z'$
band photometry, UV LFs of LAEs can be
reliably derived up to $z\lesssim 6$ LAEs whose
Ly$\alpha$ lines do not enter the $z'$ band
(e.g. \citealt{hu2004,shimasaku2006,ouchi2008}),
and near-infrared photometry is required to
derive reliable UV LF of LAEs at $z=6.6$.
It should be noted that, by the same reason,
$EW_0$ values of UV-continuum faint LAEs
($EW_0\gtrsim 100$\AA) are very poorly constrained
(Table \ref{tab:laes_with_redshifts}), and that
a number-$EW_0$ distribution of our $z=6.6$ LAEs 
is not obtained reliably.
Since we cannot use an uncertain UV LF of 
LAEs at $z=6.6$, we investigate effects of galaxy formation
with the other methods.

\begin{figure}
%\epsscale{0.55}
\epsscale{1.15}
\plotone{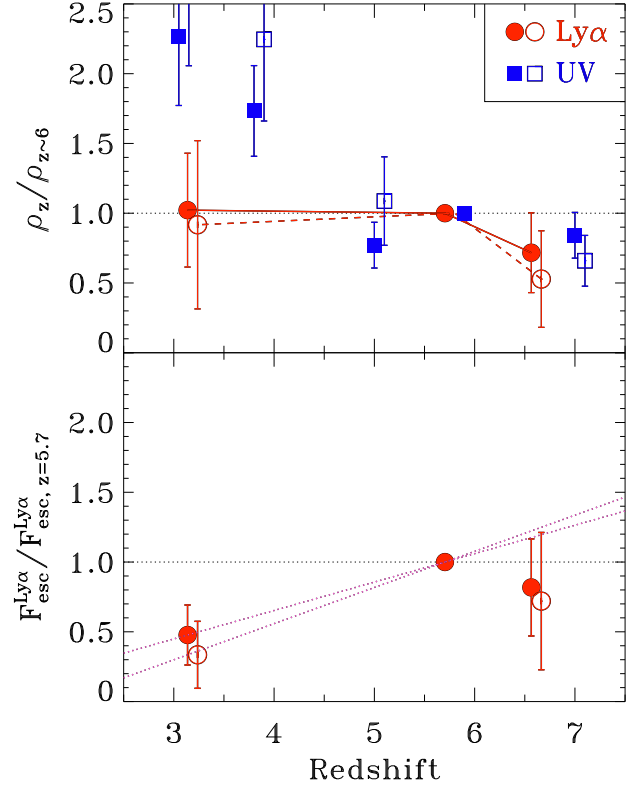}
\caption{
Top panel shows ratios of
a luminosity density at each redshift
to the one at $z\sim 6$. Red circles and
blue squares represent the ratios 
for Ly$\alpha$ and UV, respectively.
The luminosity density of $z\sim 6$
is defined with the one at $z=5.7$ for Ly$\alpha$
and $z\simeq 6$ for UV.
The open symbols are the ratios of luminosity
densities integrated down to the observed luminosity
($\log L_{\rm Ly\alpha}=42.4$ for Ly$\alpha$ or
$M_{\rm UV}=-18$ for UV), while the filled
symbols are the ratios of estimated total
luminosity densities integrated to zero luminosity.
Open symbols are shifted by +0.1 along redshift for clarity.
Red solid and dashed lines simply connect 
the Ly$\alpha$ points of red filled and open circles, respectively.
Note that the ratios at $z\simeq 6$ have no error bars,
because this is a definition of the ratios.
Instead, the errors of luminosity density measurements
at $z\simeq 6$ are included in the ratios at the other
redshifts ($z\neq 6$).
Bottom panel plots ratios of total Ly$\alpha$
escape fraction, $F_{\rm esc}^{\rm Ly\alpha}$, as a function
of redshift. The ratios are $F_{\rm esc}^{\rm Ly\alpha}$
at each redshift divided by $F_{\rm esc}^{\rm Ly\alpha}$
at $z=5.7$. Again, the open circles are those
calculated with the luminosity densities integrated down to 
the observed luminosity, while the filled circles are 
estimated from the total
luminosity densities integrated down to zero luminosity.
Magenta dotted lines present linear fits
of $F_{\rm esc}^{\rm Ly\alpha}/F_{\rm esc, z=5.7}^{\rm Ly\alpha}$ and redshift
to filled and open circles in the low-redshift range of $z=3.1-5.7$.
\label{fig:rho_Talpha}}
\end{figure}

\begin{figure*}
%\epsscale{1.0}
\epsscale{1.15}
\plotone{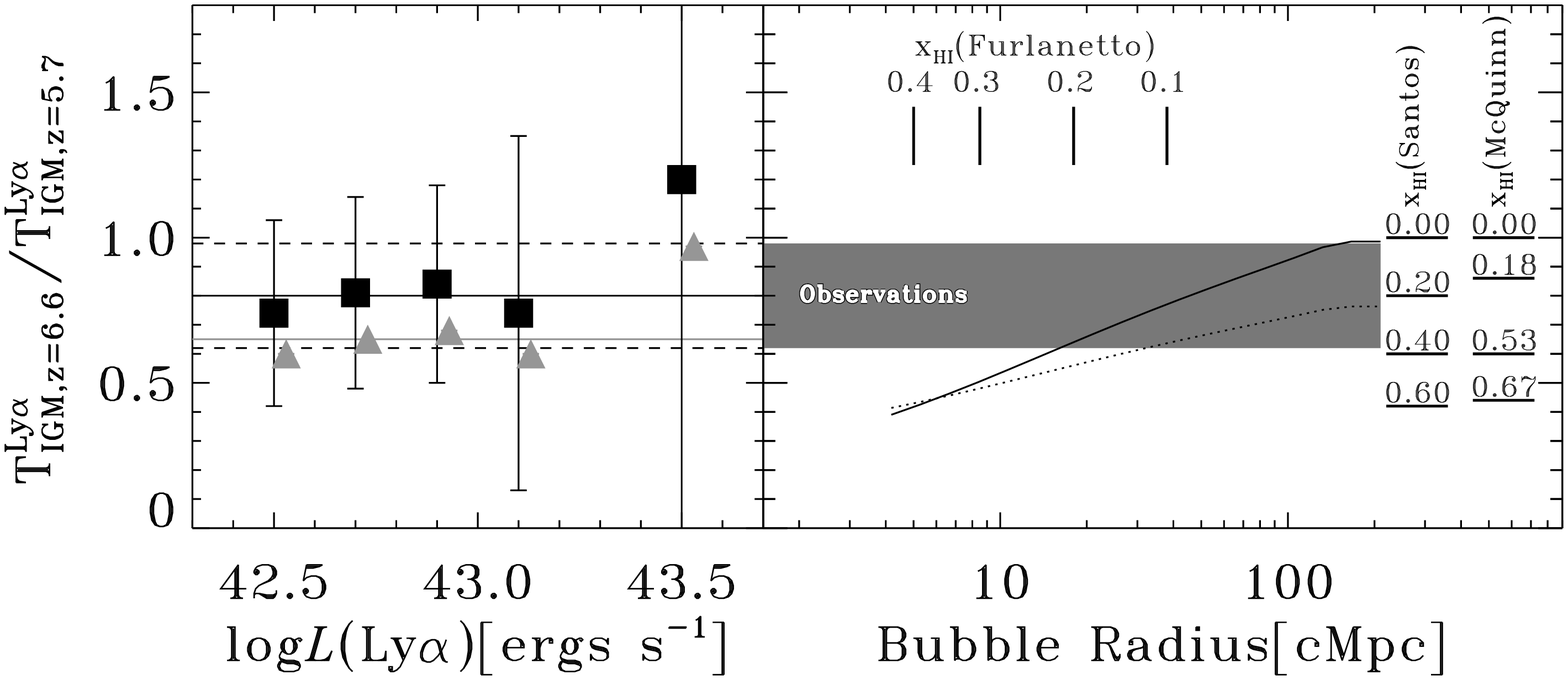}
\caption{
Ly$\alpha$ transmission through IGM
at $z=6.6$ that is normalized by the one at $z=5.7$.
Left panel shows our observational estimates
of $T_{\rm IGM,z=6.6}^{\rm Ly\alpha}/T_{\rm IGM,z=5.7}^{\rm Ly\alpha}$.
Black squares plot $T_{\rm IGM,z=6.6}^{\rm Ly\alpha}/T_{\rm IGM,z=5.7}^{\rm Ly\alpha}$
in the case of no evolution of Ly$\alpha$ escape fraction ($f_{\rm esc}^{\rm Ly\alpha}$).
Black solid and dashed lines are
the best estimate and $\pm 1\sigma$ errors 
of $T_{\rm IGM,z=6.6}^{\rm Ly\alpha}/T_{\rm IGM,z=5.7}^{\rm Ly\alpha}$, respectively,
for the no $f_{\rm esc}^{\rm Ly\alpha}$ evolution case.
Similarly, gray triangles and line represent the lower limit 
in the case of the $f_{\rm esc}^{\rm Ly\alpha}$ evolution.
Right panel compares these estimates with model predictions.
Gray region shows our observational constraints ($\pm 1\sigma$) 
on $T_{\rm IGM,z=6.6}^{\rm Ly\alpha}/T_{\rm IGM,z=5.7}^{\rm Ly\alpha}$.
Solid and dotted lines represent the model predictions
of \citet{dijkstra2007b} as a function of ionized bubble radius
in comoving Mpc for the cases with and without 
ionizing background boosts of undetected surrounding sources,
respectively. Ticks at top present the relation between
typical bubble radius and neutral hydrogen fraction of IGM
for $x_{\rm HI}\simeq 0.4$, $0.3$, $0.2$, and $0.1$,
which are predicted by the analytic model of \citet{furlanetto2006}.
Two sets of ticks in the right hand side of the panel denote
the predicted relations between $x_{\rm HI}$ and $T_{\rm IGM,z=6.6}^{\rm Ly\alpha}/T_{\rm IGM,z=5.7}^{\rm Ly\alpha}$ 
in the models of \citet{santos2004} and \citet{mcquinn2007}.
\label{fig:R_Lobs}}
\end{figure*}

We calculate luminosity densities, $\rho$, of 
Ly$\alpha$ lines from LAEs and UV continua
from dropout galaxies, and present ratios
of $\rho_z/\rho_{z\sim 6}$ in the top panel of Figure \ref{fig:rho_Talpha}, 
where $\rho_z$ and $\rho_{z\sim 6}$ are
luminosity densities of redshifts $z$ and $6$, respectively.
For these calculations, we use Ly$\alpha$ LFs at $z=6.6$
(this study) and $3.1$ \citep{ouchi2008} as well as
UV LFs at $z=2-7$ \citep{bouwens2007,reddy2009,oesch2010}.
We estimate two sets of $\rho$ from integrals down to
the observed luminosity
($\log L_{\rm Ly\alpha}=42.4$ for Ly$\alpha$ or
$M_{\rm UV}=-18$ for UV) and down to
zero luminosity; the latter is probably
near a total $\rho$. We confirm that 
the ratios of these different estimates 
agree within the error bars in Figure \ref{fig:rho_Talpha}.
We, thus, refer the latter estimate
for our fiducial results including 
no systematic bias from observations.
The top panel of Figure \ref{fig:rho_Talpha} indicates that
Ly$\alpha$ $\rho_z/\rho_{z\sim 6}$ stays constant within
the errors between $z=3.1$ and $5.7$, but
there is a drop from $z=5.7$ to $6.6$ beyond the error bar.
This drop of Ly$\alpha$ $\rho_z/\rho_{z\sim 6}$
is originated from the decrease of Ly$\alpha$ LF
in this redshift range. On the other hand, 
a ratio of UV $\rho$ monotonically decreases 
from $z\sim 3$ to $7$. The decrease
of UV $\rho_z/\rho_{z\sim 6}$ from $z\sim 6$ to $7$
suggests that the cosmic star-formation rate density (SFRD)
of galaxies decline at this redshift range,
and this cosmic SFRD decline would contribute to 
the decrease of Ly$\alpha$ $\rho_z/\rho_{z\sim 6}$
from $z=5.7$ to $6.6$. 

We evaluate the effect of cosmic SFRD decline 
on the basis of evolution of UV luminosity density.
We assume that Ly$\alpha$ luminosity
density, $\rho^{\rm Ly\alpha}$, is proportional to UV luminosity
density, $\rho^{\rm UV}$,
\begin{equation}
\rho^{\rm Ly\alpha} = \kappa\ T_{\rm IGM}^{\rm Ly\alpha}\ f_{\rm esc}^{\rm Ly\alpha}\ \rho^{\rm UV},
\label{eq:rho_Lya}
\end{equation}
where $T_{\rm IGM}^{\rm Ly\alpha}$ is a transmission fraction of 
Ly$\alpha$ through IGM and $f_{\rm esc}^{\rm Ly\alpha}$ is a Ly$\alpha$ 
escape fraction
within a galaxy through their inter-stellar medium (ISM).
$f_{\rm esc}^{\rm Ly\alpha}$ depends on gas infall+outflow \citep{santos2004,dijkstra2007a,dijkstra2010},
distribution of galactic hydrogen \citep{zzheng2009,zzheng2010}, and dust obscuration \citep{dayal2010}.
$\kappa$ is a factor converting from UV to Ly$\alpha$ luminosities,
which depend on stellar population, i.e., IMF, age, and metallicity.
Assuming that stellar population of LAEs is the same at $z=5.7$
and a given redshift, $z$, we can obtain a ratio of 
$T_{\rm IGM}^{\rm Ly\alpha}\ f_{\rm esc}^{\rm Ly\alpha}$ at $z=5.7$ to a redshift, $z$,
only with Ly$\alpha$ and UV luminosity densities;
\begin{equation}
\frac{T_{\rm IGM,z}^{\rm Ly\alpha}\ f_{\rm esc,z}^{\rm Ly\alpha}}{T_{\rm IGM,z=5.7}^{\rm Ly\alpha}\ f_{\rm esc,z=5.7}^{\rm Ly\alpha}} = \frac{\rho_{z}^{\rm Ly\alpha}/\rho_{z=5.7}^{\rm Ly\alpha}}{\rho_{z}^{\rm UV}/\rho_{z=5.7}^{\rm UV}},
\label{eq:Tf}
\end{equation}
where the indices, $z$ and $z=5.7$, show redshifts.
Because the product, $T_{\rm IGM}^{\rm Ly\alpha}\ f_{\rm esc}^{\rm Ly\alpha}$,
is a total of Ly$\alpha$ escape fractions, i.e. a fraction of 
Ly$\alpha$ photons escaping from IGM and ISM,
we refer to this product as a total Ly$\alpha$ escape fraction.
We write
\begin{equation}
F_{\rm esc}^{\rm Ly\alpha} \equiv T_{\rm IGM}^{\rm Ly\alpha}\ f_{\rm esc}^{\rm Ly\alpha}
\label{eq:Fesc}
\end{equation}
for simplicity.
With this definition, the left hand side of eq. (\ref{eq:Tf})
is rewritten as $F_{\rm esc,z}^{\rm Ly\alpha}/F_{\rm esc,z=5.7}^{\rm Ly\alpha}$
showing redshifts with the indices.
The eq. (\ref{eq:Tf}) means that
$F_{\rm esc,z}^{\rm Ly\alpha}/F_{\rm esc,z=5.7}^{\rm Ly\alpha}$ 
is determined by the ratio of Ly$\alpha$ luminosity densities
divided by the ratio of UV luminosity densities. Since the
ratio of UV luminosity densities, $\rho_{z}^{\rm UV}/\rho_{z=5.7}^{\rm UV}$,
plays a role of a correction factor of cosmic SFRD evolution,
we refer this to a cosmic SFRD correction factor.
The bottom panel of Figure \ref{fig:rho_Talpha} plots
$F_{\rm esc,z}^{\rm Ly\alpha}/F_{\rm esc,z=5.7}^{\rm Ly\alpha}$ as a function of
redshift. 
We find that the total Ly$\alpha$ escape fraction
might show a possible decrease by $\sim 20$\% from $z=5.7$ to $6.6$,
albeit with a large error.
Thus, the evolution of $F_{\rm esc,z}^{\rm Ly\alpha}$ is
small and less significant than that of 
the  ratio of Ly$\alpha$ $\rho_{\rm z}$ shown
in the top panel of Figure \ref{fig:rho_Talpha}.
On the other hand, $F_{\rm esc,z}^{\rm Ly\alpha}$ increases from
$z=3.1$ to $5.7$ in the bottom panel of Figure \ref{fig:rho_Talpha}.
This is probably because a fraction of Ly$\alpha$ escaping from ISM 
(i.e. $f_{\rm esc}^{\rm Ly\alpha}$) 
increases from low-$z$ towards $z=5.7$. If one naively extrapolates
the linear fit of the relation at $z=3.1-5.7$ between
$F_{\rm esc,z}^{\rm Ly\alpha}/F_{\rm esc,z=5.7}^{\rm Ly\alpha}$ and redshift towards high-$z$
in the bottom panel of Figure \ref{fig:rho_Talpha}, 
$F_{\rm esc}^{\rm Ly\alpha}$ at $z=6.6$ is
smaller than that of the extrapolation 
(magenta dotted lines in Figure \ref{fig:rho_Talpha})
just beyond the $1\sigma$ error. 
However, the evolution of
IGM ($T_{\rm IGM}^{\rm Ly\alpha}$) and galaxies ($f_{\rm esc}^{\rm Ly\alpha}$)
cannot be clearly distinguished.

\begin{figure*}
%\epsscale{1.0}
\epsscale{1.15}
\plotone{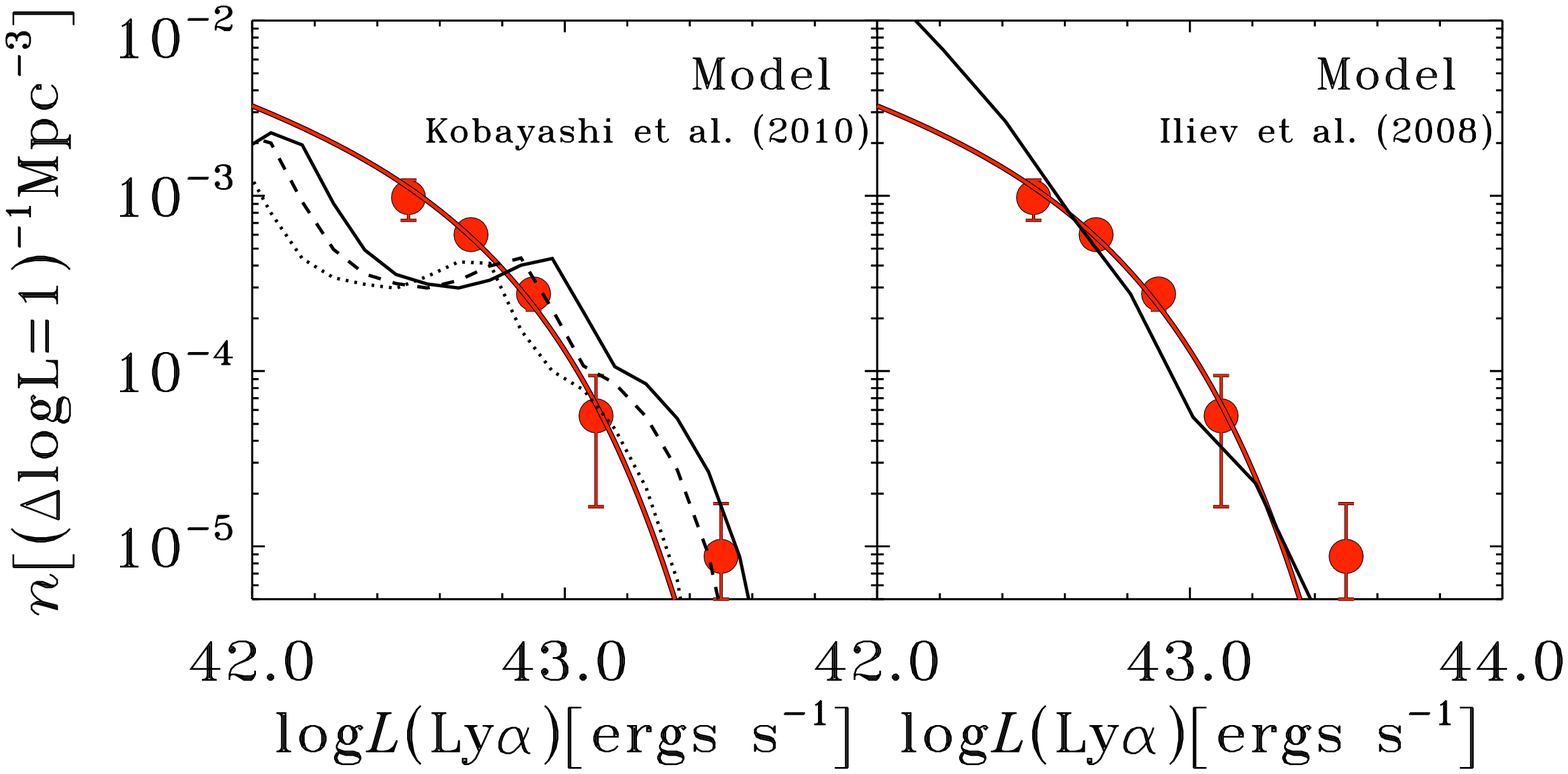}
\caption{
Comparisons of our Ly$\alpha$ LF at $z=6.6$ 
with those given by theoretical models.
Red circles and line are our best-estimate LF
as shown in Figure \ref{fig:lumifun_full_diff_nb921_evolution}.
Left panel shows LFs predicted with semi-analytic
models of \citet{kobayashi2010}. Solid, dashed, and
dotted lines are predicted LFs with 
$T_{\rm IGM,z=6.6}^{\rm Ly\alpha}/T_{\rm IGM,z=5.7}^{\rm Ly\alpha} = 1.0$, $0.8$,
and $0.6$, respectively.
Right panel plots a LF reproduced by the radiative
transfer model of \citet{iliev2008}
with the applied shift of $\Delta \log L({\rm Ly\alpha}) = +31.6$.
\label{fig:lumifun_full_diff_nb921_modelcomparison}}
\end{figure*}

Thus, we evaluate $f_{\rm esc}^{\rm Ly\alpha}$,
and estimate evolution of Ly$\alpha$ transmission of IGM
from $z=5.7$ to $6.6$ with a quantity of
$T_{\rm IGM,z=6.6}^{\rm Ly\alpha}/T_{\rm IGM,z=5.7}^{\rm Ly\alpha}$.
Because the estimates based on $\rho_{z}^{\rm Ly\alpha}/\rho_{z=5.7}^{\rm Ly\alpha}$
shown above would provide additional uncertainties raised by
the integration ranges for Ly$\alpha$ luminosity densities, 
we perform calculations with ratios of Ly$\alpha$ luminosity
by replacing $\rho_{z}^{\rm Ly\alpha}/\rho_{z=5.7}^{\rm Ly\alpha}$
with the ratio of Ly$\alpha$ luminosity,
$L_{z}^{\rm Ly\alpha}/L_{z=5.7}^{\rm Ly\alpha}$, in eq. (\ref{eq:Tf})
at given data points of observed number density, although
this improvement of calculations provide 
only negligible changes in our results. 
Note that the quantity of $T_{\rm IGM,z=6.6}^{\rm Ly\alpha}/T_{\rm IGM,z=5.7}^{\rm Ly\alpha}$ 
can be written as
$T_{\rm IGM,z=6.6}^{\rm Ly\alpha}/T_{\rm IGM,z=5.7}^{\rm Ly\alpha}=
(F_{\rm esc,z=6.6}^{\rm Ly\alpha}/F_{\rm esc,z=5.7}^{\rm Ly\alpha})
(f_{\rm esc,z=5.7}^{\rm Ly\alpha}/f_{\rm esc,z=6.6}^{\rm Ly\alpha})$
by the definition (eq. \ref{eq:Fesc}),
Here, the unknown physical quantity is the evolution
of the ratio of Ly$\alpha$ escaping fractions of ISM, i.e.,
$(f_{\rm esc,z=5.7}^{\rm Ly\alpha}/f_{\rm esc,z=6.6}^{\rm Ly\alpha})$.
We consider the two cases of $f_{\rm esc}^{\rm Ly\alpha}$ evolution
under an assumption of
no luminosity dependence of $f_{\rm esc}^{\rm Ly\alpha}$ for
simplicity.
The first case is no evolution of $f_{\rm esc}^{\rm Ly\alpha}$ between $z=5.7$ and $6.6$, 
which gives $(f_{\rm esc,z=5.7}^{\rm Ly\alpha}/f_{\rm esc,z=6.6}^{\rm Ly\alpha})=1$.
We find that the ratio is 
$T_{\rm IGM,z=6.6}^{\rm Ly\alpha}/T_{\rm IGM,z=5.7}^{\rm Ly\alpha} = 0.80 \pm 0.18$,
and that IGM transmission at $z=6.6$ 
decreases at the $1\sigma$ level.
For the second case, we estimate it with
a $f_{\rm esc}^{\rm Ly\alpha}$ evolution between $z=5.7$ and $6.6$.
We assume that the redshift evolution of $f_{\rm esc}^{\rm Ly\alpha}$ 
between $z=5.7$ and $6.6$ is the same as that of linear increase per redshift 
found in the low redshift regime of $z=3.1-5.7$. 
To quantify the evolution of $f_{\rm esc}^{\rm Ly\alpha}$
at the low-$z$ ($z=3.1-5.7$), we can rewrite the eq. (\ref{eq:Fesc});
$f_{\rm esc,z=3.1}^{\rm Ly\alpha}/f_{\rm esc,z=5.7}^{\rm Ly\alpha}=
(F_{\rm esc,z=3.1}^{\rm Ly\alpha}/F_{\rm esc,z=5.7}^{\rm Ly\alpha})(T_{\rm IGM,z=5.7}^{\rm Ly\alpha}/T_{\rm IGM,z=3.1}^{\rm Ly\alpha})$.
Figure \ref{fig:rho_Talpha} shows that
$(F_{\rm esc,z=3.1}^{\rm Ly\alpha}/F_{\rm esc,z=5.7}^{\rm Ly\alpha})=0.48$.
We estimate the ratio of $(T_{\rm IGM,z=5.7}^{\rm Ly\alpha}/T_{\rm IGM,z=3.1}^{\rm Ly\alpha})$
with the average GP optical depths. Assuming that IGM absorbs a blue half of symmetric
Ly$\alpha$ emission line, we obtain
$(T_{\rm IGM,z=5.7}^{\rm Ly\alpha}/T_{\rm IGM,z=3.1}^{\rm Ly\alpha})=(0.53/0.85)$
based on the estimates of \citet{fan2006}.
\footnote{
The estimate of $T_{\rm IGM}^{\rm Ly\alpha}$
differs only by $3-5$\% from those with \citet{madau1995} and \citet{meiksin2006} 
(see \citealt{ouchi2008}). 
}
Note that this number is a lower limit,
because Ly$\alpha$ lines are redshifted from
the systemic velocity by a few hundred km s$^{-1}$
and less absorbed by IGM \citep{pettini2001,mclinden2010,steidel2010}. 
The real number of
$(T_{\rm IGM,z=5.7}^{\rm Ly\alpha}/T_{\rm IGM,z=3.1}^{\rm Ly\alpha})$
is $>(0.53/0.85)$, closer to unity.
Hence the value of 
$f_{\rm esc,z=3.1}^{\rm Ly\alpha}/f_{\rm esc,z=5.7}^{\rm Ly\alpha}$ is $>0.30$ ($=0.48\times 0.53/0.85$).
Extrapolating the redshift evolution of $f_{\rm esc}^{\rm Ly\alpha}$
to $6.6$, we obtain the ratio of Ly$\alpha$ escape fraction
at $z=5.7$ to $6.6$,
$(f_{\rm esc,z=5.7}^{\rm Ly\alpha}/f_{\rm esc,z=6.6}^{\rm Ly\alpha})> 0.81$.
Thus, $T_{\rm IGM,z=6.6}^{\rm Ly\alpha}/T_{\rm IGM,z=5.7}^{\rm Ly\alpha} > 0.65$ in this second case
with the redshift evolution of $f_{\rm esc}^{\rm Ly\alpha}$.

The left panel of Figure \ref{fig:R_Lobs} presents 
$T_{\rm IGM,z=6.6}^{\rm Ly\alpha}/T_{\rm IGM,z=5.7}^{\rm Ly\alpha}$
for the first and second cases with black and gray lines,
respectively.
Figure \ref{fig:R_Lobs} also plots the ratio
as a function of Ly$\alpha$ luminosity
for each observational data point with squares and triangles.
Note that the second case only gives the lower limit,
and that the gray triangles in the left panel of
Figure \ref{fig:R_Lobs} represent the lower limits.
Figure \ref{fig:R_Lobs} shows that 
these lower limits are consistent with the results of 
the first case within the $-1\sigma$ error,
and, thus, the second-case result is included
in the first-case result of $T_{\rm IGM,z=6.6}^{\rm Ly\alpha}/T_{\rm IGM,z=5.7}^{\rm Ly\alpha} = 0.80 \pm 0.18$.
It indicates that the IGM transmission at $z=6.6$,
$T_{\rm IGM,z=6.6}^{\rm Ly\alpha}$,
is smaller than the one at $z=5.7$ by 20\% but just beyond the $1\sigma$ error.
We conclude that there would exist a small decrease
of IGM transmission at the $1\sigma$ level probably
contributed by cosmic reionization, and that
the ratio of $T_{\rm IGM,z=6.6}^{\rm Ly\alpha}/T_{\rm IGM,z=5.7}^{\rm Ly\alpha}$
is $\simeq 0.8$ and no smaller than $0.6$.

We obtain another independent constraint
on 
$T_{\rm IGM,z=6.6}^{\rm Ly\alpha}/T_{\rm IGM,z=5.7}^{\rm Ly\alpha}$ by comparison with
physical models including galaxy evolution and cosmic reionization effects.
Left panel of Figure \ref{fig:lumifun_full_diff_nb921_modelcomparison}
compares Ly$\alpha$ LFs predicted with semi-analytic models
of \citet{kobayashi2010}. 
Because the model of \citet{kobayashi2010} reproduces
observed Ly$\alpha$ LFs between $z=3.1$ and $5.7$,
the galaxy evolution component of this model at $z\simeq 3-6$
is probably reliable.
We plot their Ly$\alpha$ LFs at $z=6.56$
in three cases of $T_{\rm IGM,z=6.6}^{\rm Ly\alpha}/T_{\rm IGM,z=5.7}^{\rm Ly\alpha}=1.0$, $0.8$,
and $0.6$
\footnote{
Because \citet{kobayashi2010} define $T_{\rm IGM,z=5.7}^{\rm Ly\alpha}=1.0$,
these three models correspond to their models of
$T_{\rm Ly\alpha}=1.0$, $0.8$, and $0.6$.
}
. Although the models under-predict number densities of LAEs at the
faint end ($\log L\lesssim 42.8$) for all the cases, shapes of the predicted LFs
agree at the bright luminosity ($\log L\gtrsim 42.8$).
Because their faint-end LF may be strongly affected by
complicated star-formation and supernova feedback processes 
depending on model assumptions (\citealt{kobayashi2010}; see
\citealt{nagashima2004} for their models of galaxy formation components),
we compare only their bright-end LFs. 
The left panel of Figure \ref{fig:lumifun_full_diff_nb921_modelcomparison}
shows that their bright-end LF of
the $T_{\rm IGM,z=6.6}^{\rm Ly\alpha}/T_{\rm IGM,z=5.7}^{\rm Ly\alpha}=0.8$ model
reproduces our observational results.
This constraint on $T_{\rm IGM,z=6.6}^{\rm Ly\alpha}/T_{\rm IGM,z=5.7}$
is consistent with our estimate of
$T_{\rm IGM,z=6.6}^{\rm Ly\alpha}/T_{\rm IGM,z=5.7}=0.80\pm 0.18$.
Right panel of Figure \ref{fig:lumifun_full_diff_nb921_modelcomparison}
presents Ly$\alpha$ LF predicted with the radiative
transfer model of \citet{iliev2008}.
Because their luminosity is arbitrary,
we have applied a shift of $\Delta \log L = +31.6$ to their model
that is roughly matched to our observational data points.
The LF shape of the model agrees with that of our observations
in the luminosity range of $\log L\simeq 42.5-43.5$,
although there is a hint of a steeper LF slope
in the model than our observational measurements.

Next, we place a constraint
on a neutral hydrogen fraction of IGM, $x_{\rm HI}$,
and a typical ionized bubble radius at $z=6.6$
based on our estimates of 
$T_{\rm IGM,z=6.6}^{\rm Ly\alpha}/T_{\rm IGM,z=5.7}\simeq 0.8$.
Right panel of Figure \ref{fig:R_Lobs} ticks
$x_{\rm HI}$ values at the corresponding
$T_{\rm IGM,z=6.6}^{\rm Ly\alpha}/T_{\rm IGM,z=5.7}$ 
obtained by analytic models of \citet{santos2004}
and radiative transfer models of \citet{mcquinn2007}.
In these comparisons, we assume that 
$T_{\rm IGM}$ is contributed by 
scattering of Ly$\alpha$ damping wing 
of hydrogen IGM alone.
In the models of \citet{santos2004},
we have applied the realistic models with
a Ly$\alpha$ line redshifted by $360$ km s$^{-1}$ from a systemic redshift,
since redshifted Ly$\alpha$ lines by a few hundred km s$^{-1}$ 
are observationally found not only in LBGs \citep{pettini2001,steidel2010},
but also in LAEs \citep{mclinden2010}.
In the \citet{mcquinn2007} models,
we use ratios of intrinsic to observed 
Ly$\alpha$ luminosities at the cumulative number density of 
$\simeq 10^{-5}$ Mpc$^{-3}$, which corresponds to
$\log L\simeq 43$ near to the luminosity range 
of our observations.
Both of the models indicate that 
$T_{\rm IGM,z=6.6}^{\rm Ly\alpha}/T_{\rm IGM,z=5.7}\simeq 0.8$
corresponds to $x_{\rm HI}\simeq 0.2$. Even with
the errors of $T_{\rm IGM,z=6.6}^{\rm Ly\alpha}/T_{\rm IGM,z=5.7}$,
we find that $x_{\rm HI}$ is smaller than $\lesssim 0.4-0.5$.

The right panel of Figure \ref{fig:R_Lobs} also
presents cosmic reionization models of 
\citet{dijkstra2007b}
and \citet{furlanetto2006}. The models of 
\citet{dijkstra2007a} imply that
our observational estimate of 
$T_{\rm IGM,z=6.6}^{\rm Ly\alpha}/T_{\rm IGM,z=5.7}^{\rm Ly\alpha} \simeq 0.8$
corresponds to a typical ionized bubble size
of $\gtrsim 40$ comoving Mpc. 
Following the discussion of \citet{dijkstra2007b},
we apply the analytic model of \citet{furlanetto2006}
to this typical ionized bubble size.
At the top of the right panel of Figure \ref{fig:R_Lobs},
we tick the volume averaged neutral fraction predicated
by \citet{furlanetto2006}. We find that 
the model of \citet{furlanetto2006}
infers that our lower limit of $\gtrsim 40$ comoving Mpc
corresponds to the neutral fraction of $x_{\rm H_I} \lesssim 0.1$.

In summary, all of these various theoretical models, i.e. 
analytic, semi-analytic, and numerical models,
indicate that our observational estimate of 
$T_{\rm IGM,z=6.6}^{\rm Ly\alpha}/T_{\rm IGM,z=5.7}^{\rm Ly\alpha} = 0.80 \pm 0.18$
corresponds to $x_{\rm H_I} \lesssim 0.2\pm 0.2$.
In the previous studies of Ly$\alpha$ LFs,
\citet{malhotra2004} and \citet{kashikawa2006} place
upper limits of neutral fraction for $x_{\rm HI} \lesssim 0.3$ 
and $\lesssim 0.45$, respectively,
with the combination of their measurements and one analytic model
of \citet{santos2004}. Our result is consistent with
these previous studies, but our constraint is stronger
and more robust than these previous results 
because of our better LF evolution determination
with cosmic variance errors
and the inclusion of model dependent errors
based on various independent reionization models.
If all of our deep fields are not strongly 
biased by patchy IGM distribution by chance, our conclusion
of IGM neutral fraction, $x_{\rm H_I} \lesssim 0.2\pm 0.2$, 
is not significantly changed.

\begin{figure}
%\epsscale{0.8}
\epsscale{1.15}
\plotone{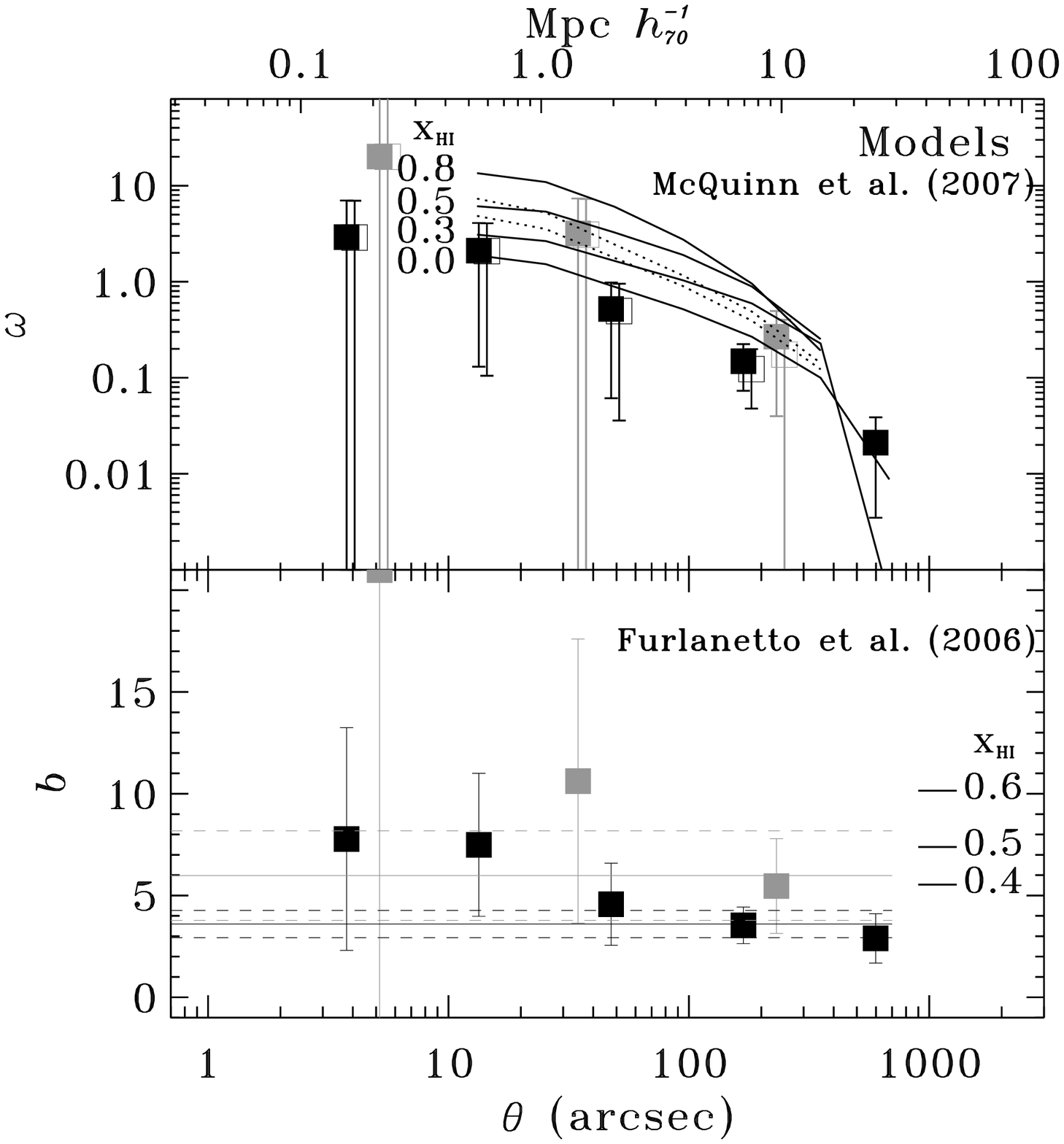}
\caption{
Comparison with our clustering measurements and 
reionization models. This plot is the
same as Figure \ref{fig:acorr_NB921_all_final},
but with the model predictions of angular correlation function
in the top panel \citep{mcquinn2007}
and bias in the bottom panel \citep{furlanetto2006}.
$Top:$ Solid lines are the model predictions of \citet{mcquinn2007}
for $z=6.6$ LAEs whose hosting dark halo mass is 
$3\times 10^{10}M_\odot$.
Neutral fractions of the universe are $x_{\rm HI}=0.0$,
$0.3$, $0.5$, and $0.8$ from bottom to top lines.
Dotted lines indicate $z=6.6$ LAEs with a dark halo mass
of $7\times 10^{10}M_\odot$ (bottom) and $1\times 10^{11}M_\odot$ (top)
in the case of $x_{\rm HI}=0.0$.
$Bottom:$ At the right side of this plot,
we present \citeauthor{furlanetto2006}'s (\citeyear{furlanetto2006})
predicted bias values under the assumptions
described in the text. For clarity, we also plot
the best-bias estimate and $1\sigma$ ranges
for our bright ($NB921<25.5$) LAE subsample 
with gray solid and dashed lines, respectively.
\label{fig:acorr_NB921_all_model_McQuinnFurlanetto_final}}
\end{figure}

\subsubsection{Evolution of Clustering}
\label{sec:evolution_of_clustering}

Theoretical models predict that
a clustering amplitude of observed LAEs 
is boosted, due to
the additional clustering of LAEs
whose Ly$\alpha$ photons can escape
in the case that these LAEs reside
in an ionized bubble at the reionization epoch 
(\citealt{furlanetto2006,mcquinn2007,lidz2009}; 
cf. \citealt{iliev2008}).
In Figure \ref{fig:redshift_bias_LBGLAE},
we find no sudden rise of bias from $z=5.7$ to $6.6$
beyond the error bars.
This indicates that clustering of $z=6.6$ LAEs
is weakly affected by cosmic reionization.
With this result on bias evolution,
we constrain cosmic reionization models
of LAE clustering predictions.

Figure \ref{fig:acorr_NB921_all_model_McQuinnFurlanetto_final}
compares the ACFs and bias of our LAEs at $z=6.6$
with theoretical predictions of \citet{mcquinn2007}
and \citet{furlanetto2006}.
\citet{mcquinn2007} models are based on radiative transfer simulations.
In their models, Ly$\alpha$ emission fluxes are assigned 
to dark halos with a mass above a minimum halo mass.
We find that the average hosting dark halo masses of LAEs
are about $10^{11\pm 1}M_\odot$ over $z=2-7$ in Section
\ref{sec:halo_mass}. This result is also correct at only $z=2-6$, 
even if we omit the results of $z=6.6$ LAEs whose
clustering might be contaminated by the cosmic reionization effect.
If we assume that hosting halo masses of LAEs do not change significantly
from $z\sim 6$ to $6.6$ so as true for $z=3-6$ LAEs,
the possible minimum halo mass of $z=6.6$ LAEs 
is about $10^{10}M_\odot$. The top panel of
Figure \ref{fig:acorr_NB921_all_model_McQuinnFurlanetto_final}
plots the models of 
\citet{mcquinn2007} with the minimum halo mass of 
$3\times 10^{10}M_\odot$ which is the closest 
to $10^{10}M_\odot$ and the smallest halo mass 
in their models. The comparison of our $z=6.6$
LAEs and these models indicates that 
the neutral hydrogen fraction of IGM is less than
$x_{\rm HI}\lesssim 0.5$ at $z=6.6$.
We also plot the rest of models
with masses of $7\times 10^{10}M_\odot$ and 
$1\times 10^{11}M_\odot$ provided by
\citet{mcquinn2007}, and confirm that
these massive halo models generally give
a strong clustering, and that
their constraints on $x_{\rm HI}$
are all consistent 
with this relatively weak upper limit of 
$x_{\rm HI}\lesssim 0.5$.
Thus, the comparison with 
\citet{mcquinn2007} models
places an upper limit of $x_{\rm HI}\lesssim 0.5$.
The comparison with \citet{furlanetto2006} models
is shown in the bottom panel of 
Figure \ref{fig:acorr_NB921_all_model_McQuinnFurlanetto_final}.
We use ratios of small-scale bias
to hosting halo bias for LAEs in \citet{furlanetto2006}. 
\citet{furlanetto2006} define the small-scale
bias as a bias at a scale smaller than a typical 
ionized bubble size at the redshift.
At $z=6.6$, near the end of reionization, a typical
size of ionized bubbles is $\gtrsim 10-100$ Mpc.
Because our bias estimates of $z=6.6$ LAEs are
mostly made at the scale of $\simeq 1-10$ Mpc 
(see Figure \ref{fig:acorr_NB921_all_final}),
we, thus, regard our bias values as a small-scale bias
that is defined in the study of \citet{furlanetto2006}.
For a hosting halo bias, we assume again
no evolution of halo mass,
and use $b=3.7$ that is a bias of $10^{10}M_\odot$
at $z=6.6$ given by eq. (\ref{eq:bias_sheth}). 
Comparing the $1\sigma$ limits of
average bias for all sample (black dashed line)
and the bright subsample (dark dashed line)
in the bottom panel of 
Figure \ref{fig:acorr_NB921_all_model_McQuinnFurlanetto_final},
we find that a neutral hydrogen fraction of IGM
should be smaller than $x_{\rm HI}\lesssim 0.5$
at $z=6.6$. Either \citet{mcquinn2007}
or \citet{furlanetto2006} models 
would imply that a neutral hydrogen fraction 
is $x_{\rm HI}\lesssim 0.5$ at $z=6.6$ 
based on our clustering estimates.

\subsubsection{Evolution of Ly$\alpha$ Line Profile}
\label{sec:evolution_of_lya_line_profile}

Ly$\alpha$ damping wing of neutral hydrogen IGM 
absorbs a Ly$\alpha$ emission line 
at $>1216$\AA\ in the partially and fully neutral universe.
Because Ly$\alpha$ damping wing absorption 
at $>1216$\AA\ monotonically weakens towards red wavelengths,
the damping wing absorption changes a shape
of Ly$\alpha$ line profile. 
Since a Ly$\alpha$ flux absorbed by Ly$\alpha$ forest
has an asymmetric profile with its flux peak near 1216\AA, i.e. 
in the bluest wavelength of high-$z$ Ly$\alpha$ line, 
the absorption component of Ly$\alpha$ damping wing (having a stronger
attenuation in bluer wavelengths)
should broaden the Ly$\alpha$ line.
We use models of Ly$\alpha$ lines
made by \citet{dijkstra2007a},
and evaluate the Ly$\alpha$ line broadening.
We find that an FWHM of Ly$\alpha$ line broadens
by $7-10$\% in a very neutral universe
where a LAE resides in an ionized bubble
with a radius of $0.7$ physical Mpc
($5$ comoving Mpc at $z=6.6$),
compared with a LAE in more ionized universe
with an ionized bubble radius of 
$2-10$ physical Mpc
($15-80$ comoving Mpc at $z=6.6$).
Because the curvature of Ly$\alpha$ damping
wing absorption is not steep, 
the Ly$\alpha$ line broadening is not large,
only by a few-10 percent level.
Thus, high quality spectra are needed to
identify the Ly$\alpha$ line broadening, 
if any, at an observed redshift.
In \S \ref{sec:composite_spectra}, we obtain 
the direct (Gaussian) FWHM velocity widths 
of $251$ $(270)$ $\pm 16$ km s$^{-1}$
at $z=6.6$ and $260$ $(265)$ $\pm 37$ km s$^{-1}$
at $z=5.7$. These measurements include
errors of 6\% and 14\%, which allow us
to marginally identify a few-10 percent level
broadening of Ly$\alpha$.
Our data show that there is no large evolution
of Ly$\alpha$ FWHM beyond $\simeq 14$\%, 
but it is not clear whether
a few-percent level FWHM broadening exists.
In fact, there is a possible hint of
flattening of Ly$\alpha$ profile from $z=5.7$
to $6.6$ (\S \ref{sec:composite_spectra}), 
although its significance level
is below the $1\sigma$ level.
One of the goals of future LAE studies
will be testing the broadening of Ly$\alpha$ FWHM
down to a few percent level.

\citet{haiman2005} suggest a test of reionization
with a relation between Ly$\alpha$ luminosity and
line width. They claim that
LAEs in a neutral universe have an anti-correlation
between Ly$\alpha$ luminosity and line width,
because fainter LAEs preferably residing in smaller HII
regions are affected by a stronger damping wing attenuation than
brighter LAEs. We compare our observational results with the model of
\citet{haiman2005} in the luminosity range and velocity width
similar to our observations.
The model of \citet{haiman2005} indicates
that, if the universe is fully neutral ($x_{\rm HI}=1$),
an FWHM velocity width (line width) decreases only by 30 km s$^{-1}$ (1\AA)
from $\log L({\rm Ly\alpha})=42.5$ to $\log L({\rm Ly\alpha})=43.5$
for LAEs with a velocity width of 300 km s$^{-1}$.
In their model, no anti-correlation, but rather positive-correlation, 
is found for more ionized universe with a neutral fraction
of $x_{\rm HI}=0.5$ and $0.25$ in this relatively bright
luminosity range.
Thus, in this luminosity range, 
they predict that a weak anti-correlation appears
only when the universe is nearly neutral.
Our observational results indicate that 
there is no anti-correlation between 
Ly$\alpha$ luminosity and FWHM velocity width
(\S \ref{sec:lya_velocity_luminosity_relation}). Our best-fit function
presented in Figure \ref{fig:lya_vfhwm} shows a rather positive
correlation and an increase of FWHM velocity width by $111 \pm 45$ km s$^{-1}$
from $\log L({\rm Ly\alpha})=42.5$ to $\log L({\rm Ly\alpha})=43.5$.
If the model prediction of \citet{haiman2005} is correct,
our observations rule out the fully neutral universe
at $z=6.6$ at the $\simeq 3\sigma$ level.

In \S \ref{sec:composite_spectra},
we find an interesting knee feature in Ly$\alpha$ line profiles
of LAEs at $z=6.6$ as well as $z=5.7$. 
It would be suggestive of
galaxy outflow or proximity effect, but it is not clear whether
it can be interpreted with a reasonable physical picture.
Because we find these knee features not only at $z=6.6$, 
but also $z=5.7$ when the universe is highly ionized ($x_{\rm HI}\sim 10^{-4}$),
the knee features would not be related to cosmic reionization
but galaxy formation. The knee features may be important for understanding
dynamics, UV radiation field, and structure of LAEs at high redshifts.

\subsubsection{Cosmic Reionization}
\label{sec:cosmic_reionization_history}

\begin{figure}
%\epsscale{0.5}
\epsscale{1.15}
\plotone{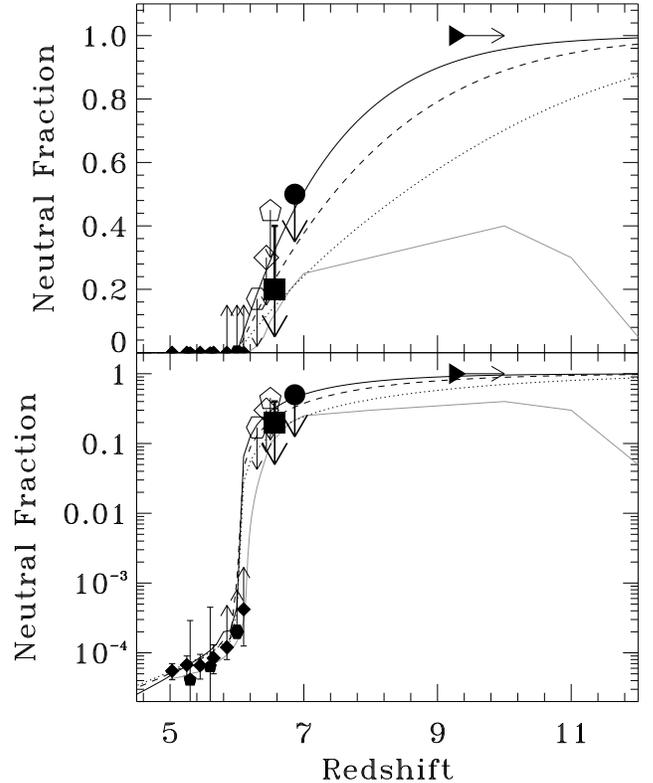}
\caption{
Neutral hydrogen fraction, $x_{\rm HI}$,
of IGM as a function of redshift.
Top and bottom panels are the same,
but with a vertical axis of
linear and log scales, respectively.
Filled square and circle are the upper
limits of $x_{\rm HI}$ that we obtain 
from the evolution of Ly$\alpha$ LF 
and clustering, 
respectively. 
Open diamond and
pentagon denote the upper limits 
from Ly$\alpha$ LF at $z=6.5$ given by
\citet{malhotra2004} and \citet{kashikawa2006}.
Open hexagon is the upper limit
estimated from the constraints of Ly$\alpha$
damping wing of GRB at $z=6.3$ \citep{totani2006}.
Filled hexagon and pentagons indicate
constraints given by GRB spectra \citep{gallerani2008b}
and QSO dark gap statistics \citep{gallerani2008a},
respectively.
Filled diamonds represent the measurements
from GP optical depth of SDSS QSOs \citep{fan2006}.
Triangle plots the $1\sigma$ lower-limit of
redshift of a neutral universe 
given by WMAP7 \citet{larson2010}
in the case of instantaneous reionization.
Avoiding overlapping symbols, 
we give a small offset
along redshift 
to the positions of 
the filled circle and the open diamond.
Dotted, dashed, and solid lines show
the evolution of $x_{\rm HI}$ 
for minihalo, small, and large halo
cases, respectively, predicted
by \citet{choudhury2008}. Gray solid 
line presents the prediction 
in the double reionization scenario 
suggested by \citet{cen2003}.
\label{fig:z_xHI}}
\end{figure}

In Sections 
\ref{sec:evolution_of_lyalf}-\ref{sec:evolution_of_lya_line_profile},
we have obtained the constraints of
neutral hydrogen fraction of $x_{\rm HI}\lesssim 0.2\pm 0.2$
and $x_{\rm HI}\lesssim 0.5$ at $z=6.6$ from
the evolution of Ly$\alpha$ LF and clustering,
respectively, and ruled out the fully neutral universe 
at $z=6.6$ at the $\simeq 3\sigma$ level
by Ly$\alpha$ line profiles.
It should be noted that these three constraints agree,
even though these results are given by three independent observational
quantities, i.e. Ly$\alpha$ LF, clustering, and 
Ly$\alpha$ line properties.
In Figure \ref{fig:z_xHI}, we plot 
our two relatively strong constraints of $x_{\rm HI}$
from Ly$\alpha$ LF and clustering, and compare
with the previous estimates and theoretical models.
Our results are consistent 
with the previous ones based on
LAEs as well as GRBs
\footnote{
The estimates of a single GRB 
would include an additional
systematic error of $\delta x_{\rm HI}\sim 0.3$
owing to the patchiness of reionization
\citep{mcquinn2008}.
Since the $z\simeq 6.7$ GRB estimate
of \citet{greiner2009}
allows $x_{\rm HI}=0.001-1$
within the $1.2\sigma$ level
due to the strong degeneracy between
absorptions of a damped Ly$\alpha$ absorber
and IGM, the result of \citet{greiner2009} 
is not included in Figure \ref{fig:z_xHI}.
}
and QSOs, and
our constraint from Ly$\alpha$ LF, $x_{\rm HI}\lesssim 0.2\pm 0.2$,
is generally stronger than the previous ones
at $z\simeq 6.5-6.6$.
This upper limit
at $z=6.6$ prefers an early ($z\gtrsim 7$) reionization 
on average, which requires more ionizing photons
at $z\gtrsim 7$.

Triangle in Figure \ref{fig:z_xHI} is
the $1\sigma$ lower limit of redshift
given by the WMAP7 result
in the instantaneous reionazation case
($z=10.5\pm 1.2$; \citealt{larson2010}).
This rules out the instantaneous reionization
at $z<9.3$, i.e. 100\%-neutral fraction at $z<9.3$ ,
at the $1\sigma$ level.
Figure \ref{fig:z_xHI} shows predictions
of semi-analytic reionization models
of \citet{choudhury2008}. For sources of reionization,
they assume three minimum halo masses for star-forming
galaxies, $\sim 5\times 10^5 M_\odot$,
$\sim 10^8M_\odot$, and $\sim 10^9M_\odot$ 
at $z=6$ for mini, small, and large halo
cases, respectively. 
\citet{choudhury2008} have concluded
that reionization only by large halos
is ruled out by the constraints of
the electron scattering optical depth 
(WMAP3; \citealt{spergel2007}) and various
neutral hydrogen fraction upper limits
available at the time of their study.
Our results support their claim, and
our relatively strong upper limit of
$x_{\rm HI}\lesssim 0.2\pm 0.2$ favor their
model including minihalos. Minihalos
can accomplish the relatively early reionization,
as suggested by various theoretical studies.
On the other hand, a change of star-forming galaxy
properties is also possible. These could be
an increase of escape fraction of ionizing photons
such suggested by high-$z$ dropout observations
\citep{ouchi2009b,bunker2009,finkelstein2009,bouwens2010b}.
It is also possible that the
changes of IMF and metallicity are
important (e.g. \citealt{stiavelli2004}).
In fact, the model of \citet{cen2003} 
shown Figure \ref{fig:z_xHI}
includes the effects of 
IMF and metallicity changes.
\citet{cen2003} argues that reionization took place
twice; the first at $z\sim 15-16$ by metal-free 
Population III stars
with a top-heavy IMF and the second at $z\sim 6$ by 
Population II stars, which give a medium large
Thomson scattering optical depth of $0.10\pm 0.03$ 
similar to WMAP7 results \citep{larson2010}. 
This scenario gives a fairly low neutral fraction 
at $z=6.6$ that is consistent with our LAE constraints.
Although it is not clear whether reionization took place twice,
metal-free star-formation would characterize
the history of cosmic reionization.

\subsection{Role of LAEs in Galaxy Formation History}
\label{sec:role_of_LAEs}

The evolution of LAE LF, clustering, and Ly$\alpha$ line profiles 
includes various hints
for understanding LAEs in galaxy formation history.
As shown in Figure \ref{fig:numdensity_bias_LBGLAE_z3z4z67},
a typical bias of LAEs is smaller than that of LBGs over $z=2-7$.
Thus, a typical halo mass of LBG is estimated to be $\sim 10^{12\pm 1}M_\odot$,
about one order of magnitude larger than that of LAEs
\citep{hamana2004,ouchi2004b,ouchi2005b,lee2006,lee2009,mclure2009,hildebrandt2009}.
\footnote{
Note that LBGs referred here are typical dropout galaxies
with a UV luminosity near $L^*$, and that, in our discussion, 
we do not include the population of the LAE analog of dropout galaxies 
with a very faint UV magnitude such those studied by \citet{stark2010}. 
Similarly, LAEs are typical LAEs so far observed in $\log L({\rm Ly\alpha})\simeq 42-44$,
and not those fainter than this luminosity range.
}
It is possible that some galaxies become LAEs with a halo mass
of $\sim 10^{11}M_\odot$ then evolve into LBGs with a halo mass of 
$\sim 10^{12}M_\odot$ via mass assembly such as mergers and accretion.
If this scenario is true, a typical LAE could be 
a progenitor of typical LBG. In this scenario, two
observational questions are consistently answered.
The first question is the different trend of LBG and LAE evolution.
The LF of LBGs decreases from $z=3$ to $6$ (e.g. \citealt{bouwens2007}), 
while the LF of LAEs does not change much in $z=3-6$ 
(e.g. \citealt{ouchi2008}). In this scenario,
LAEs would form earlier than LBGs, and the significant drop of LF
may be found only for LBGs in this intermediate
redshift of $z=3-6$. This is because the number density
of LBG's massive ($\sim 10^{12}M_\odot$) hosting halos sharply drops
from $z=3$ to $6$, while that of LAEs'
less-massive ($\sim 10^{11}M_\odot$) hosting halos does not decrease
much 
based on the halo model \citep{sheth1999}. 
The redshift of $z=3-6$ 
might be still the major formation epoch of LBGs,
while the major formation epoch of LAEs would be earlier than
LBGs, i.e. at $z>6$, near the epoch when we find a decrease of LAE LF 
(from $z=5.7$ to $6.6$) in this study. 
The second question is the deficit of 
strong Ly$\alpha$ emitting galaxies
among a UV bright population.
It is found that
UV-bright galaxies do not have
a strong Ly$\alpha$ emission line
\citep{ouchi2003,ando2006,shimasaku2006,vanzella2007,vanzella2009,ouchi2008}.
This trend is usually discussed on a plane of UV luminosity and Ly$\alpha$ EW,
and no objects are found in the UV-luminous and EW-large regime (e.g. \citealt{ando2006}).
In the case of our scenario, a typical LAE evolves into a typical LBG, i.e.
less-massive galaxies with a strong Ly$\alpha$ emission line 
become massive galaxies with a weak or no Ly$\alpha$ emission line.
When this evolution is investigated on this plane, 
objects move from UV-faint to UV-luminous regimes,
because UV luminosity positively correlates with
stellar mass at high redshifts \citep{papovich2001,sawicki2007,yabe2009}.
The UV-faint (less-massive) objects are, first, distributed widely 
in EW values on this plane before the evolution.
Then, these objects end up 
in the UV-bright and small-EW regime on this plane after the evolution.
The tendency of Ly$\alpha$ deficit among UV-bright galaxies
would be consistent with the scenario of this LAE-LBG evolution
sequence.
In this way, this evolutionary scenario 
would provide the answers to these two observational questions.
The recent study of \citet{vanzella2009} has found that
LAEs have a more compact UV-continuum morphology than
LBGs with no Ly$\alpha$ emission at $z\sim 4$. 
If our scenario of LAE-LBG evolution is correct,
star-formation activities start at the center
then extend to the outskirts of galaxies, which
is suggestive of the inside-out picture of galaxy formation.

The dotted lines in Figure \ref{fig:redshift_bias_LBGLAE}
show evolutionary tracks of dark halos for galaxy-conserving models.
The galaxy-conserving model
assumes that the motion of galaxies is purely driven by
gravity, and that merging does not take place. In this case,
the bias value of galaxies decreases as the Universe
evolves with time,
\begin{equation}
b_{\rm g}=1+(b_{\rm g}^0-1)/D(z),
\label{eq:galaxy_conserving}
\end{equation}
where $b_{\rm g}^0$ is a bias at $z=0$ \citep{fry1996}.
Under the assumption of the galaxy-conserving
evolution, dark halos of $z=3$ LAEs evolve into dark halos
with $b_{\rm g} \simeq 1.1-1.3$ at $z=0$. Because,
in more realistic extended Press-Shechter formalism
(e.g. \citealt{lacey1993}),
average evolutionary tracks
come slightly below those of the galaxy-conserving models
(see, e.g., Figure 13 of \citealt{ichikawa2007}),
dark halos of $z=3$ LAEs probably evolve into
dark halos with $b_{\rm g}\simeq 1$ which host present-day
$\simeq L^*$ galaxies including Milky Way, as suggested
by \citet{gawiser2007}. On the other hand,
in Figure \ref{fig:redshift_bias_LBGLAE},
LAEs at $z\simeq 4-7$ have a significantly 
larger bias than $z=3$ LAEs, and
the increase of the LAE bias measurements toward high-$z$
(nearly along solid lines) is steeper
than those of galaxy-conserving evolution (dotted lines).
The galaxy-conserving model indicates that
dark halos of $z\simeq 4-7$ LAEs evolve into
those with $b_{\rm g}\simeq 1.5-2$ at $z=0$
whose value is higher than those of 
$z=3.1$ LAEs ($b_{\rm g} \simeq 1.1-1.3$),
which implies that descendants of $z\simeq 4-7$ LAEs are
different from those of $z=3$ LAEs. 
On average, the descendants
of $z\simeq 4-7$ LAEs
would be more massive than those of $z=3$ LAEs.
Majority of LAEs at $z\simeq 4-7$ are probably 
not ancestors of Milky Way, but
today's large galaxies more massive than Milky Way,
although there should exist some LAEs that 
become today's $L^*$ galaxies by a probability process
of halo mass build up. These implications from
our clustering results are similar to 
those from theoretical predictions by
\citet{salvadori2010} who claim
that only $\simeq 2$\% of Milky Way progenitors
can be LAEs.

\section{Conclusions}
\label{sec:conclusions}

We have identified 207 LAEs at $z=6.6$
in the 1 deg$^2$ area of SXDS field down to 
$L \gtrsim 2.5 \times 10^{42}$ erg s$^{-1}$
and $EW \gtrsim 14$ \AA\ by deep and wide-field
narrow-band and broad-band imaging of Subaru/Suprime-Cam.
Nineteen Ly$\alpha$ lines are confirmed by 
the high-quality Keck/DEIMOS spectra,
and none of interlopers have been found in our sample
through our extensive deep spectroscopy campaign
as well as the SXDS project spectroscopy
consisting of 3,233 objects at $z=0-6$.
We have obtained the Ly$\alpha$ LF, ACF, and Ly$\alpha$ line profiles 
to constrain cosmic reionization and 
early galaxy formation
with the aid of recent theoretical model predictions.
We have also derived ACFs of LAEs at $z=3.1-5.7$ in the SXDS field
to find the evolutionary trend of LAE clustering from $z\sim 3-7$
in the framework of the $\Lambda$CDM model.
The major results of our study are summarized below.

1. Our Ly$\alpha$ LF of $z=6.6$ LAEs shows
the best-fit Schechter parameters of
$\phi^*=8.5_{-2.2}^{+3.0}\times10^{-4}$Mpc$^{-3}$ and
$L_{\rm Ly\alpha}^*=4.4_{-0.6}^{+0.6}\times10^{42}$ erg s$^{-1}$
with a fixed $\alpha=-1.5$, where the errors
include uncertainties of statistics and cosmic variance.
The combination of statistics and cosmic variance
errors presents scatters of number density measurements 
up to a factor of $\simeq 10$ among 5 subfields of $\sim 0.2$ deg$^2$ areas,
although the typical scatters of the subfield number densities
are not far beyond the errors of Poisson statistics.
Comparing this LAE LF at $z=6.6$ with the one at $z=5.7$,
we find that the Ly$\alpha$ LF decreases from $z=5.7$ to
$6.6$ at the $\gtrsim 90$\% confidence level.
A more dominant decrease of $L^*$ (luminosity evolution) than 
$\phi^*$ (number evolution)
is preferable. The decrease of Ly$\alpha$ LF from $z=5.7$ to $6.6$ is 
$\simeq 30$\% in the case of pure luminosity evolution.
Note that this 30\%-luminosity decrease is too small
to be identified by the previous studies, due to their large
uncertainties from small statistics and cosmic variance.

2. We have identified a significant angular-correlation signal
for our $z=6.6$ LAE sample. This is the detection of 
clustering signal for the most distant galaxies, to date.
The correlation length and bias are 
$r_0= 2-5$ h$^{-1}_{100}$ Mpc and
bias of $b=3-6$, 
respectively. There is no sudden boost
of clustering amplitude given by cosmic reionization at $z=6.6$.
In the framework of $\Lambda$CDM models,
the average hosting dark halo mass inferred from clustering 
is $10^{10}-10^{11}M_\odot$. The duty cycle of LAE population,
a product of star-formation and Ly$\alpha$ emitting
duty cycles,
is a few $0.1$ to a few percent, roughly $\sim 1$\%, 
although the constraint on the duty cycle
involves large uncertainties because of
the bias estimate errors and the fairly flat
relation between number density and bias of halos 
from the $\Lambda$CDM model.

3. Based on our high quality DEIMOS spectra,
we have found that most of Ly$\alpha$ emission lines
present a clear asymmetric profile with the average
FWHM velocity width of $251\pm 16$ km s$^{-1}$ (direct measurement)
at $\log L({\rm Ly\alpha})\simeq 42.6-43.6$, 
and that the average FWHM velocity width does not largely
evolve from $z=5.7$ to $6.6$ beyond errors of our 
$\simeq 40$ km s$^{-1}$, which is dominated by 
the error of our reference $z=5.7$ LAE.
There is no anti-correlation between
Ly$\alpha$ luminosity and velocity width.
If our spectroscopic sample is not biased, 
Ly$\alpha$ velocity width positively correlates 
with Ly$\alpha$ luminosity at $z=6.6$ 
in the luminosity range of
$\log L({\rm Ly\alpha})\simeq 42.6-43.6$.
We identify a knee feature in a blue tail of 
Ly$\alpha$ line in our composite spectra
of LAEs at $z=6.6$ as well as $5.7$, which cannot be explained 
by statistical and systematic instrumental errors.
These knee features, if true, would be important
for understanding dynamics, UV radiation field, and structure
of LAEs.

4. We compare evolution of Ly$\alpha$ LF, clustering,
and Ly$\alpha$ line profiles
from our observations with various reionization models
including analytic, semi-analytic, and radiative transfer
models. 
Although there would exist a small $\simeq 20$\%-decrease
of IGM transmission from $z=5.7$ to $6.6$ due to
cosmic reionization, 
the comparisons of all models and observational 
quantities reach the same conclusion that 
hydrogen IGM is not highly neutral 
at $z=6.6$. The upper limit of neutral fraction is 
$x_{\rm HI}\lesssim 0.2\pm 0.2$ from our Ly$\alpha$ LF evolution and
$x_{\rm HI}\lesssim 0.5$ from our clustering evolution
between $z=5.7$ and $6.6$. A fully neutral universe,
$x_{\rm HI}=1$, at $z=6.6$ is ruled out by no large evolution
of Ly$\alpha$ velocity width from $z=5.7$ to $6.6$
and by no anti-correlation between
Ly$\alpha$ luminosity and velocity width
at $z=6.6$. All of these reionization tests with
Ly$\alpha$ LFs, clustering, and line profiles,
agree that a neutral hydrogen fraction of IGM
is not high at $z=6.6$.
Our strongest constraint from Ly$\alpha$ LF,
$x_{\rm HI}\lesssim 0.2\pm 0.2$, 
implies that the major reionization process 
took place early, at $z\gtrsim 7$.

5. Calculating ACFs, bias, and 
hosting dark halos of low-$z$ LAEs of SXDS at $z=3.1-5.7$
in the same manner as those of our $z=6.6$ LAEs, 
we find that hosting dark halo masses
stay at the similar value of $10^{11\pm 1}M_\odot$ over $z=3-7$.
It implies that LAEs are galaxies at the evolutionary stage
for all or some type of galaxies whose dark halos
have reached a mass of $\sim 10^{11\pm 1}M_\odot$.
Because a halo mass of typical ($\simeq L*$) LBGs is 
$\sim 10^{12\pm 1}M_\odot$, about 
one order of magnitude larger than that of LAEs,
there is a possibility that some galaxies 
become LAEs with a halo mass of $\sim 10^{11}M_\odot$
in the $\sim 1$\% duty cycle,
then evolve into typical LBGs with a halo mass of 
$\sim 10^{12}M_\odot$ 
via mass assembly such as mergers and accretion.
This scenario consistently explains two observational
results of LBGs and LAEs; the different LF evolutionary trend
between LBGs and LAEs (e.g. \citealt{ouchi2008}),
and the deficit of strong Ly$\alpha$ emitters
among UV bright population
(e.g. \citealt{ando2006}).

\acknowledgments
We thank Renyue Cen,
Tirthankar Roy Choudhury,
Mark Dijkstra, Andrea Ferrara, 
Ilian Iliev , Masakazu Kobayashi,
Matt McQuinn, Takashi Murayama,
and Xiaohui Fan for providing
their data. Especially, we 
appreciate the efforts of Mark Dijkstra
and Matt McQuinn who responded
to a number of our requests.
M.O. and K.S. acknowledge
Cedric Lacey
and 
Orsi Alvaro
for their helpful comments
on clustering analysis.
M.O. thanks Richard Ellis for
his encouragement to publish
these results as early as possible.
M.O. is grateful to comments
from Kentaro Aoki, Richard Ellis, David Sobral,
and Eros Vanzella, and useful discussions
with Renyue Cen, Mark Dijkstra, Eric Gawiser,
Esther Hu, Matt McQuinn, James Rhoads, 
Alice Shapley, Haojing Yan, and Zheng Zheng
at the Ly$\alpha$ Emitter Workshop, Ohio State
University held in April 26-27, 2010.
M.O. has been supported via Carnegie Fellowship.

{\it Facilities:} \facility{Subaru (Suprime-Cam) KeckII (DEIMOS)}

\end{document}